\documentclass[a4paper]{JHEP3}
\usepackage{amssymb}
\usepackage{amsmath,amssymb,graphicx}
\usepackage{bbm}
\usepackage{amsfonts}
\usepackage{tipa}
\usepackage{mathrsfs}

\title{A Model of Fermion Masses and Mixings Triggered by Family Problem in Warped Extra Dimensions}

\author{Zhi-Qiang Guo and Bo-Qiang
Ma\thanks{Email:~mabq@phy.pku.edu.cn}  \\
School of Physics and State Key Laboratory of Nuclear Physics and
Technology, Peking University, Beijing 100871, China}

\abstract{We suggest a model which addresses both the fermion mass
hierarchy problem and the family problem in two-layer warped extra
dimensions. In this model, 3 family fermions in 4 dimensions (4D)
generate from 1 family in two-layer warped 6D by two step
Kluza-Klein decompositions. The mass hierarchies are produced by the
exponential behavior of 4D fermion zero mode profiles. The mixings
and masses of fermions are closely related to the family problem. By
adjusting parameters in this model, the numerical results can be
very close to the experimental data both in the quark sector and in
the lepton sector. In the lepton sector, we suppose neutrinos to be
Dirac ones. The neutrino masses of sub-$\mathrm{eV}$ scale are
obtained, and also CP violation emerges in the lepton sector.}


\begin{document}

\section{Introduction}\label{sec:1}

The Standard Model~(SM) is very successful and has been tested by
experiments with high precision. However, there still exist puzzles
in SM from theoretical perspectives. Two famous ones of them are the
family problem and the flavor hierarchy problem. In SM, 3 families
of quarks and leptons have similar gauge interactions and it seems
that two heavier generations replicate characters of the lightest
generation; while the masses of quarks and charged leptons have the
obvious hierarchical structure. The hierarchical structure also
exists in the Cabibbo-Kobayashi-Maskawa~(CKM) mixing matrix. SM
accommodates 3 family fermions and their phenomenologies by
adjusting the Yukawa couplings. It does not supply an interpretation
for the origin of 3 families and the hierarchy structure of flavor
parameters.

There have been several different approaches to address these two
problems. A natural and popular way is family symmetry. Froggatt and
Nielsen \cite{Froggatt:1979} suggested a horizontal $U(1)$ symmetry
to understand the hierarchical fermion mass structure; while
continuous non-Abelian family symmetry, like $SU(3)$, can also lead
to very promising results both in the quark sector and in the lepton
sector \cite{King:2002}. Recently, triggered by the data from
neutrino experiments, discrete non-Abelian family symmetries have
attracted many attentions. It is found that the $A_4$ family
symmetry \cite{Ma:2001} can produce the famous tri-bimaximal lepton
mixing matrix \cite{Harrison:2002}, which is a good approximation to
the best fit value of the neutrino experimental data
\cite{Gonzalez-Garcia:2008}.

Different from the family symmetry approach\footnote{However, for
symmetry approach in extra dimensional framework, see
\cite{Haba:2006}.}, the family and flavor problems are addressed
from new approaches in extra dimensional framework. Due to the
exponential behavior of fermion profiles in extra dimensions
\cite{Arkani-Hamed:2000,Grossman:2000}, the mass hierarchy of
fermions can be produced by parameters of the same order naturally.
While family problem also gets new interpretations in extra
dimensions. Several groups of authors show that 3 families of SM in
4D can originate from 1 family in 6D
\cite{Libanov:2000,Neronov:2000} by making an appropriate gauge
background or choosing a metric of special structure. Although there
have been many progresses in these directions, there is still no a
well celebrated model to address the family and flavor problems in
the community.

In this paper, we focus on the warped extra dimension approach to
address the family problem and the flavor problem. In the
Randall-Sundrum warped 5D spacetime \cite{Randall:1999}, let
fermions to propagate in the bulk. The profiles of fermion zero
modes can be of the exponential behavior, which depend on the 5D
bulk mass parameters. Due to this character, the fermion mass
hierarchy can be reproduced by the 5D bulk mass parameters of the
same order. This approach supplies a beautiful geometrical
interpretation to the flavor hierarchy problem. In the paper
\cite{Guo:2008}, we attempted to understand the origin of the same
order 5D bulk mass parameters. The origin of the 5D bulk mass
parameters has close relation to the family problem, because one
mass parameter can stand for one family. We suggested a two-layer
warped 6D spacetime, and begin with 1 family fermion in the 6D bulk.
We reduce the spacetime from 6D to 5D at the first step. As a
result, the 5D mass parameters emerge as eigenvalues of a 1D
Schr\"{o}dinger-like equation~(or Kluza-Klein~(KK) modes in 5D). The
spacetime metric can be chosen such that only 3 eigenvalues are
permitted. So in this setup, 3 families in 5D can originate from 1
family in 6D. When we further reduce the 5D spacetime to the
physical 4D at the second step, the zero modes of 3 family fermions
in 5D produce 3 family fermions in 4D. By coupling with Higgs field
in 4D, these zero modes get masses and produce the 3 family fermions
in SM. According to these considerations, we can imagine that both
the family problem and the flavor hierarchy problem can be addressed
in such an approach. 1 family fermion in 6D produces 3 families in
5D. The 5D mass parameters are of the same order. The same order 5D
mass parameters further produce hierarchical structure of fermions
in 4D due to the exponential behavior of zero mode profiles. In the
following part of this paper, we construct a specific model along
the above ideas. We find that the numerical results of this model
can be very close to the experimental data by adjusting parameters
in this model. The mass hierarchies of quarks and charged leptons
are produced. Supposing the neutrinos to be the Dirac ones, the
small neutrino masses are obtained. The quark mixing matrix and the
lepton mixing matrix are also very close to the realistic CKM matrix
and Petrov-Maki-Nakagawa-Sakata~(PMNS) matrix respectively.

This paper is organized as follows. In section \ref{sec:2}, we
simply introduce the model suggested in \cite{Guo:2008}, and develop
it to be a more realistic one. We notice several treatments which
are different from that in our previous paper \cite{Guo:2008} in
section \ref{sec:2}. In section \ref{sec:3}, we construct a model
along the considerations in the introduction, and adjust the
parameters of this model to show that the numerical results can be
very close to the experimental data both in the quark sector and in
the lepton sector. In section \ref{sec:4}, we give some analytical
treatments about our model. By these analytical treatments, we show
that there are some common concise structures shared by quarks and
leptons. We also make some further discussions about this model in
section \ref{sec:4}. We make summaries in section \ref{sec:5}.
Several appendices are added for a clear understanding of the paper.

\section{Preliminaries for model building}\label{sec:2}

In this section, we introduce some necessary tools for the future
model building in section \ref{sec:3}. In subsection \ref{sec:2.1},
following the discussions in our previous paper \cite{Guo:2008}, we
introduce the basic setup to show that how several fermion families
in 5D can originate from 1 family in 6D by KK decomposition in
two-layer warped extra dimensions. The critical point is how we can
get finite KK modes while there are infinite KK modes in the usual
KK decomposition. In subsection \ref{sec:2.2}, we introduce an
example in the popular quantum mechanics textbook to show that
finite bound KK modes can be obtained. This simple example helps us
to understand the problem more clearly. In subsection \ref{sec:2.3},
we make more discussions about the characters of massive KK modes,
and analyze an example which will be used in the model building in
section \ref{sec:3}. In subsection \ref{sec:2.4}, we reduce the 5D
action to the 4D one, and display the necessary results for model
building. In subsection \ref{sec:2.5}, we discuss a problem which
makes the basic setup in subsection \ref{sec:2.1} to be not
realistic. We further suggest a new setup to bypass the problem.
This new setup will be used in the model building of section
\ref{sec:3}, instead of the basic setup in subsection \ref{sec:2.1}.
In addition, we notice several treatments about the setup which
differ from that in our previous paper \cite{Guo:2008}. We also
notice these different treatments in several footnotes. The model
building in section \ref{sec:3} largely depends on the discussions
in this section.

\subsection{Introduction to the basic setup}\label{sec:2.1}

Following the discussions in \cite{Guo:2008}, we consider a 6D
spacetime metric with the special two-layer warped structure,
\begin{eqnarray}
\label{introduce} ds^2 = B(z)^2\left[A(y)^2\left( \eta_{\mu \nu}
dx^{\mu} dx^{\nu} + d y^2\right)+ d z^2\right].
\end{eqnarray}
We choose $\eta_{\mu \nu}=\mathrm{diag}(-1,1,1,1)$, and suppose that
the two extra dimensions are both intervals.

A massive Dirac fermion in this spacetime has the action
\begin{eqnarray}
\label{action1} S = \int d^4 x dy dz \sqrt{-g} \left\{ \frac{i}{2}
\left[\bar{\Psi} e_a^M \Gamma^a \nabla_M \Psi-\nabla_M
\bar{\Psi}e_a^M \Gamma^a \Psi \right] -i~m \bar{\Psi} \Psi\right\},
\end{eqnarray}
where $e_a^M$ is the sechsbien, and $\nabla_M=\partial_M +
\frac{1}{2} \omega^{ab}_M
\Gamma_{ab},\Gamma_{ab}=\frac{1}{4}[\Gamma_a,\Gamma_b]$ is the
covariant derivative of spinor in curved spacetime. $a$ and
$M=0,1,2,3,5,6$ stand for the flat indices and the curved indices in
the tetrads respectively. $m$ is a real number to ensure that the
action Eq.~(\ref{action1}) is hermitian. The gamma matrix
representations are as follows,
\begin{eqnarray}
\label{Eq3} \Gamma^\mu &=& \left(
\begin{array}{cc}
0 & \gamma^\mu \\
\gamma^\mu & 0
\end{array}
\right), ~\Gamma^5 = \left(
\begin{array}{cc}
0 & \gamma^5 \\
\gamma^5 & 0
\end{array}
\right), ~\Gamma^6 = \left(
\begin{array}{cc}
\mathbf{1}_4 & 0 \\
0 & -\mathbf{1}_4
\end{array}
\right), ~\Gamma^7 = \left(
\begin{array}{cc}
0 & -i\mathbf{1}_4 \\
i\mathbf{1}_4 & 0
\end{array}
\right),\nonumber \\
\gamma^0 &=& \left(
\begin{array}{cc}
0 & -\mathbf{1}_2 \\
\mathbf{1}_2 & 0
\end{array}
\right), ~\gamma^j = \left(
\begin{array}{cc}
0 & \sigma^j \\
\sigma^j & 0
\end{array}
\right), ~\gamma^5 = \left(
\begin{array}{cc}
\mathbf{1}_2 & 0 \\
0 & -\mathbf{1}_2
\end{array}
\right), ~~j=1,2,3,
\end{eqnarray}
where $\mu=0,1,2,3$ and $\sigma^j$ are the Pauli matrices.

As a first step, we reduce the 6D action Eq.~(\ref{action1}) to the
5D one by KK decompositions. Rewrite the 6D 8-component spinor with
two 4-component spinors and expand these 4-component spinors with 5D
fields as follows,
\begin{eqnarray}
\label{solution9}
\Psi=\left(
\begin{array}{c}
\chi_{1}\\
\chi_{2}
\end{array}\right),~~
\chi_1(x^\mu,y,z)= \sum_n \widehat{F}_n(z)\, \psi_{n}
(x^\mu,y),~\chi_2(x^\mu,y,z)= \sum_n \widehat{G}_n(z)\, \psi_{n}
(x^\mu,y).
\end{eqnarray}
Suppose that $\widehat{F}_n(z)$ and $\widehat{G}_n(z)$ subjugated to
the conditions,
\begin{eqnarray}
\label{solution10}
\left(\frac{d}{d z}+\frac{5}{2} B^{-1}B_z\right)\widehat{F}_n(z)-m B \widehat{F}_n(z) + \lambda_n \widehat{G}_n(z)=0,\\
\label{solution11} \left(\frac{d}{d z}+\frac{5}{2}
B^{-1}B_z\right)\widehat{G}_n(z)+m B \widehat{G}_n(z) - \lambda_n
\widehat{F}_n(z)=0,
\end{eqnarray}
where $B_z=\frac{d B(z)}{dz}$. With the above conditions, the 6D
action Eq.~(\ref{action1}) is reduced to
\begin{eqnarray}
\label{norm1} S &=&\int d^4 x dy~K_{mn}\left\{\frac{i}{2}
A^4\left[\bar{\psi}_m\gamma^5\partial_5\psi_n-\partial_5\bar{\psi}_m\gamma^5\psi_n+\bar{\psi}_m\gamma^{\mu}\partial_{\mu}\psi_n-\partial_{\mu}\bar{\psi}_m\gamma^{\mu}\psi_n\right]\right\}\nonumber\\
&-&\int d^4 x dy~M_{mn}A^5 i\bar{\psi}_m {\psi}_n, \\
\label{norm2} K_{mn}&=& \int dz B^5
\left(\widehat{F}_{m}^{\ast}\widehat{F}_n+\widehat{G}_{m}^{\ast}\widehat{G}_n\right)=\int
dz\left({F}_{m}^{\ast}{F}_n+{G}_{m}^{\ast}{G}_n \right),\\
\label{norm3} M_{mn}&=&\int dz B^5\left[
\left(\widehat{F}_{m}^{\ast}\widehat{F}_n+\widehat{G}_{m}^{\ast}\widehat{G}_n
\right)\frac{\lambda_{m}^{\ast}+\lambda_n}{2}\right]=\int dz
\left[\left({F}_{m}^{\ast}{F}_n+{G}_{m}^{\ast}{G}_n\right)\frac{\lambda_{m}^{\ast}+\lambda_n}{2}\right],
\end{eqnarray}
where we have defined the transformations
\begin{eqnarray}
\label{solution12}
\widehat{F}_n(z)=B(z)^{-5/2}F_n(z),~\widehat{G}_n(z)=B(z)^{-5/2}
G_n(z).
\end{eqnarray}
The above equations are satisfied for all KK modes, including zero
mode and massive modes. By the transformations
Eq.~(\ref{solution12}), equations (\ref{solution10}) and
(\ref{solution11}) can be simplified to
\begin{eqnarray}
\label{solution13}
\left(\frac{d}{d z}- m B\right)F_n(z) + \lambda_n G_n(z)=0,\\
\label{solution14} \left(\frac{d}{d z}+ m B \right)G_n(z) -
\lambda_n F_n(z)=0.
\end{eqnarray}

Our purpose is to obtain the conventional effective 5D action
\begin{eqnarray}
\label{action3} S_{5eff}&=&\sum_n\int d^4 x dy\left\{\frac{i}{2}
A^4\left[\bar{\psi}_n\gamma^5\partial_5\psi_n-\partial_5\bar{\psi}_n\gamma^5\psi_n+\bar{\psi}_n\gamma^{\mu}\partial_{\mu}\psi_n-\partial_{\mu}\bar{\psi}_n\gamma^{\mu}\psi_n\right]\right\}\nonumber\\
&-&\sum_n \int d^4 x dy A^5 i\lambda_{n}\bar{\psi}_n {\psi}_n.
\end{eqnarray}
We consider two cases:

Case~(I): The different KK modes are orthogonal, that is, the
normalization conditions
\begin{eqnarray}
\label{norm2a} K_{mn}=\int dz
\left({F}_{m}^{\ast}{F}_n+{G}_{m}^{\ast}{G}_n\right)=\delta_{mn}
\end{eqnarray}
are satisfied. In these conditions, the 6D action Eq.~(\ref{norm1})
reduce to be the 5D action Eq.~(\ref{action3}) naturally.

Case~(II): The second case is that the normalization conditions
Eq.~(\ref{norm2a}) are not satisfied. In this case, $K$ and $M$ are
both matrices. It seems that we can not obtain the conventional
effective 5-dimensional action Eq.~(\ref{action3}) at first sight.
However, if the number of KK modes is finite and the matrix $K$ is
positive-definite, we can redefine the fermion fields to obtain an
action, which has the same form with that of Eq.~(\ref{action3}).
The difference is that the eigenvalues $\lambda_n$ are modified to
different values. The redefinition process can be done in two steps.

Step~(1): Decompose the hermitian matrix $K$ as follows,
\begin{eqnarray}
\label{diag1} K&=&V^{\dagger}\Lambda V=H^{\dagger}H,~~H=\sqrt{\Lambda} V,\\
\Lambda&=&\mathrm{diag}\left(\Lambda_1,\Lambda_2,\cdots,\Lambda_n\right),\nonumber\\
\sqrt{\Lambda}&=&\mathrm{diag}(\sqrt{\Lambda_1},\sqrt{\Lambda_2},\cdots,\sqrt{\Lambda_n}).\nonumber
\end{eqnarray}
In the above expressions, $\Lambda_i>0,~i=1,2,\cdots,n$, as we have
supposed that $K$ is positive-definite. Redefine $\psi_n$ as
\begin{eqnarray}
\label{diag4} \widetilde{\psi}_m=H_{mn}\psi_n,
\end{eqnarray}
then in the new basis $\widetilde{\psi}_n$, $M$ becomes
\begin{eqnarray}
\label{diag5} \widetilde{M}=(H^{-1})^{\dagger}M H^{-1}.
\end{eqnarray}
After this step, the kinetic term of Eq.~(\ref{norm1}) becomes the
conventional form as that in Eq.~(\ref{action3});

Step~(2):  Obviously $\widetilde{M}$ is a hermitian matrix. We can
diagonalize this new matrix as
\begin{eqnarray}
\label{diag6} \widetilde{M}&=&U^{\dagger}\Delta U,\\
\Delta&=&\mathrm{diag}(\widehat{\lambda}_1,\widehat{\lambda}_2,\cdots,\widehat{\lambda}_n).\nonumber
\end{eqnarray}
Redefine also the new basis $\widehat{\psi}_n$,
\begin{eqnarray}
\label{diag5-1} \widehat{\psi}_m=U_{mn}\widetilde{\psi}_n.
\end{eqnarray}
We can obtain the action
\begin{eqnarray}
\label{action3a} \widehat{S}_{5eff}&=&\sum_n\int d^4 x
dy\left\{\frac{i}{2}
A^4\left[\bar{\widehat{\psi}}_n\gamma^5\partial_5\widehat{\psi}_n-\partial_5\bar{\widehat{\psi}}_n\gamma^5\widehat{\psi}_n+\bar{\widehat{\psi}}_n\gamma^{\mu}\partial_{\mu}\widehat{\psi}_n-\partial_{\mu}\bar{\widehat{\psi}}_n\gamma^{\mu}\widehat{\psi}_n\right]\right\}\nonumber\\
&-&\sum_n \int d^4 x dy A^5
i\widehat{\lambda}_{n}\bar{\widehat{\psi}}_n {\widehat{\psi}}_n,
\end{eqnarray}
which has the same form with the action Eq.~(\ref{action3}). The
total operations on the fermion fields equal to
\begin{eqnarray}
\label{diag5-2}
\widehat{\psi}&=&U\widetilde{\psi}=UH\psi,\\
\label{diag5-3}
\mathrm{or}~~\psi&=&H^{-1}\widetilde{\psi}=H^{-1}U^{\dagger}\widehat{\psi},
\end{eqnarray}
where we have omitted the subscripts.

In the above, beginning with the 6D action Eq.~(\ref{action1}), we
obtain the 5D action Eq.~(\ref{action3}) by KK decompositions. We
can interpret this process as follows: the action
Eq.~(\ref{action1}) stands for 1 fermion family in 6D; while the
action Eq.~(\ref{action3}) or Eq.~(\ref{action3a}) can stand for
several fermion families in 5D if we can restrict the number of the
KK modes to be finite. The masses of these KK modes are determined
by the equations (\ref{solution13}) and (\ref{solution14}).
Moreover, it is not necessary to require that these masses have the
hierarchical structure. These 5D bulk masses of the same order are
enough to produce the hierarchical structure of 4D fermions
according to the works in \cite{Grossman:2000}. Therefore, the key
point is how we can get finite KK modes. This issue is the theme of
the following subsection.

\subsection{An example for finite bound KK states from 1D quantum mechanics}\label{sec:2.2}

Now we analyze the solutions of equations (\ref{solution13}) and
(\ref{solution14}). For a zero mode ($\lambda=0)$, these equations
decouple and are easy to be solved. The solutions are given by
\begin{eqnarray}
\label{ZeroMode} F_{0}(z)=\frac{1}{\sqrt{l
N_0}}\mathrm{exp}\left(\int_{z_0}^{z}mB(\zeta)d\zeta\right)~\mathrm{or}~0,
G_{0}(z)=\frac{1}{\sqrt{l\widetilde{N}_0}}\mathrm{exp}\left(-\int_{z_0}^{z}mB(\zeta)d\zeta\right)~\mathrm{or}~0.~
\end{eqnarray}
We introduce $l$ of the length dimension in order to make the
normalization constants to be dimensionless. For massive modes, we
can combine the first order differential equations to obtain second
order equations
\begin{eqnarray}
\label{solution15} \frac{d^2}{d z^2}F_n(z)
+\left[-m B_z-m^2 B^2\right]F_n(z)+\lambda_n^2 F_n(z)=0,\\
\label{solution16} \frac{d^2}{d z^2}G_n(z)+\left[m B_z-m^2
B^2\right]G_n(z)+\lambda_n^2 G_n(z)=0.
\end{eqnarray}
Rewriting them in another form, we see that they are similar to the
1D Schr\"{o}dinger equations
\begin{eqnarray}
\label{solution17} -\frac{d^2}{d z^2}F_n(z)
+V(z)F_n(z)&=&\lambda_n^2 F_n(z),\\
\label{solution18} -\frac{d^2}{d z^2}G_n(z)+\widetilde{V}(z) G_n(z)
&=&\lambda_n^2 G_n(z),
\end{eqnarray}
with potentials
\begin{eqnarray}
\label{solution19} V(z)=m B_z+m^2 B^2,~\widetilde{V}(z)=-m B_z+m^2
B^2,
\end{eqnarray}
where $B_z=\frac{dB(z)}{dz}$. In order to obtain the effective 5D
action Eq.~(\ref{action3}), we should require that the integrands in
Eq.~(\ref{norm2}) and Eq.~(\ref{norm3}) are finite. These
requirements can be satisfied if the KK modes are the bound states
of Schr\"{o}dinger equations (\ref{solution17}) and
(\ref{solution18}). So the problem how we can get finite KK modes
transforms to the problem how we can get finite bound states of the
1D Schr\"{o}dinger equations (\ref{solution17}) and
(\ref{solution18}).

There is a simple example \cite{Schiff:1968} in 1D quantum mechanics
which has finite bound states. It is the square potential well of
finite depth and width. The potential $V(z)$ is given by
\begin{eqnarray}
\label{well} V(z)=
\begin{cases}
0,~~~&\mid z\mid< \frac{a}{2}, \\
 V,~~~&\mid z\mid> \frac{a}{2},
\end{cases}
\end{eqnarray}
where $V>0$. The Schr\"{o}dinger equation is
\begin{eqnarray}
\label{well-sch}
-\frac{{\hbar}^2}{2m}\frac{d^2}{dz^2}u(z)+V(z)u(z)=Eu(z).
\end{eqnarray}
The existence of bound states requires that $E<V$. We first consider
the case $E>0$. The solutions of Eq.~(\ref{well-sch}) can be
classified by parity.

The solution of odd parity is given by
\begin{eqnarray}
\label{odd} u(z)=
\begin{cases}
-C\exp(\beta z),~~~&z <-\frac{a}{2},\\
B\sin(\alpha z),~~~&\mid z\mid < \frac{a}{2},\\
C\exp(-\beta z),~~~&z > \frac{a}{2},
\end{cases}
\end{eqnarray}
where
$\alpha=\sqrt{\frac{2mE}{{\hbar}^2}},~\beta=\sqrt{\frac{2m(V-E)}{{\hbar}^2}}$.
The continuity of $u(z)$ and $\frac{du(z)}{dz}$ at $z=\pm
\frac{a}{2}$ requires that
\begin{eqnarray}
\label{odd1} -\xi \cot\xi&=&\eta,\\
\label{odd2} \xi^2+\eta^2&=&\frac{mV}{2{\hbar}^2}a^2,
\end{eqnarray}
where $\xi=\alpha \frac{a}{2},~\eta= \beta\frac{a}{2}$.

For the solution of even parity, we obtain
\begin{eqnarray}
\label{even} u(z)=
\begin{cases}
C\exp(\beta z),~~~&z <-\frac{a}{2},\\
B\cos(\alpha z),~~~&\mid z\mid < \frac{a}{2},\\
C\exp(-\beta z),~~~&z > \frac{a}{2}.
\end{cases}
\end{eqnarray}
The continuity conditions require that
\begin{eqnarray}
\label{even1} \xi \tan \xi&=&\eta,\\
\label{even2} \xi^2+\eta^2&=&\frac{mV}{2{\hbar}^2}a^2.
\end{eqnarray}
The eigenvalues are determined by equations (\ref{odd1}),
(\ref{odd2}), (\ref{even1}) and (\ref{even2}). Their number is
finite\footnote{For details, see \cite{Schiff:1968}.}, and the
condition that there exist the very $n$ eigenvalues is that
\begin{eqnarray}
\label{Ncon}
 \left(\frac{(n-1)\pi}{2}\right)^2<\frac{m
V}{2\hbar^2}a^2<\left(\frac{n\pi}{2}\right)^2.
\end{eqnarray}
From Eq.~(\ref{Ncon}), we know that the number of eigenvalues is
restricted by the depth and the width of the square potential well.

In the above, we have discussed the situation $E>0$. However, in the
process of KK decomposition, it is not necessary to require that
$E>0$. We should also discuss the situation $E<0$. We can obtain
solutions of $E<0$ by replacing $\alpha$ and $\xi$ with
$i\tilde{\alpha}$ and $i\tilde{\xi}$ in the solutions of $E>0$. We
display only the conditions that determine the eigenvalues here. For
the odd parity solution, we have
\begin{eqnarray}
\label{odd1-1} -\tilde{\xi} \coth \tilde{\xi}&=&\eta,\\
\label{odd2-1} -{\tilde{\xi}}^2+\eta^2&=&\frac{m V}{2{\hbar}^2}a^2,
\end{eqnarray}
and for the even parity solution, we have
\begin{eqnarray}
\label{even1-1} -\tilde{\xi} \tanh \tilde{\xi}&=&\eta,\\
\label{even2-1} -{\tilde{\xi}}^2+\eta^2&=&\frac{m V}{2{\hbar}^2}a^2,
\end{eqnarray}
where
$\tilde{\alpha}=\sqrt{\frac{-2mE}{{\hbar}^2}},~\tilde{\xi}=\tilde{\alpha}
\frac{a}{2}$; while $\beta$ and $\eta$ keep invariant. It is obvious
there are no solutions that satisfy the above equations.

In summary, there are finite bound states in the square potential
well. This is a simple example to show that it is possible to obtain
finite bound states by choosing the potential properly. We are also
interested in what kind of metric $B(z)$ can make the square
potential well. Replacing $V(z)$ with the square potential well in
Eq.~(\ref{solution19}), we obtain
\begin{eqnarray}
\label{forbz} V(z)=m B_z+m^2 B^2=\begin{cases}
0,~~~&\mid z\mid< \frac{a}{2}, \\
 V,~~~&\mid z\mid> \frac{a}{2}.
\end{cases}
\end{eqnarray}
Solving it for $B(z)$, we obtain
\begin{eqnarray}
\label{forbz-1}
B(z)=
\begin{cases}
\frac{1}{mz+C},~~~&\mid z\mid< \frac{a}{2}, \\
 \frac{\sqrt{V}}{m}\tanh(\sqrt{V} z+\tilde{C}),~~~&\mid z\mid>
 \frac{a}{2},
\end{cases}
\end{eqnarray}
where $C$ and $\tilde{C}$ are constants. They can be determined by
boundary conditions and continuity conditions. This metric is very
similar to that in the papers \cite{Gregory:2000}, where the authors
opened up the extra dimension at infinity to address the
cosmological acceleration problem. For another example for finite
bound solutions in different background, see \cite{Afonin:2009}.

\subsection{General characters of massive modes and an example for model building}\label{sec:2.3}

In the above subsection, we give an example in 1D quantum mechanics
to show that it is possible to obtain finite bound modes. In this
subsection, we give another example which also has finite bound
modes. This example will be used in the model building of section
\ref{sec:3}. Before doing that, we analyze some general characters
of equations (\ref{solution13}) and (\ref{solution14}).

The zero mode solution of  equations (\ref{solution13}) and
(\ref{solution14}) has been given in Eq.~(\ref{ZeroMode}). The
massive mode solutions are determined by the second order equations
(\ref{solution17}) and (\ref{solution18}). From equations
(\ref{solution17}) and (\ref{solution18}), we see that eigenvalues
$\lambda_n$ emerge with the form $\lambda_n^2$; so for $\pm
\lambda_n$, equations (\ref{solution17}) and (\ref{solution18}) are
both satisfied. In fact, we can infer from equations
(\ref{solution13}) and (\ref{solution14}) that if the pair
$(F_n,G_n)$ is a solution of equations (\ref{solution13}) and
(\ref{solution14}) corresponding to the eigenvalue $\lambda_n$, then
another
 pair $(F_n,-G_n)$ is also a solution of equations
(\ref{solution13}) and (\ref{solution14}) corresponding to the
eigenvalue $-\lambda_n$. Therefore, the massive solutions always
emerge in pairs\footnote{In our previous paper \cite{Guo:2008}, we
only keep the zero mode and the positive massive modes, and omit the
negative massive modes.}. We denote these pair solutions explicitly
as follows,
\begin{eqnarray}
\label{pair} \lambda_n \longrightarrow
\begin{cases}
F_n \\
G_n
\end{cases},~~~~~-\lambda_n \longrightarrow
\begin{cases}
F_n \\
-G_n
\end{cases}.
\end{eqnarray}
Therefore, if these massive modes are discrete, we obtain $(2n+1)$
modes in all, where $n=0,1,2,3,\cdots$, including the zero mode and
massive modes. Supposing that there only exist a pair of massive
modes, we can write the spectrum of equations (\ref{solution13}) and
(\ref{solution14}) according to above discussions as
\begin{eqnarray}
\label{spectrum1} \mathrm{spectrum~of}~\lambda \longrightarrow
\begin{cases}
\lambda_1 \\
0\\
-\lambda_1 \\
\end{cases}.
\end{eqnarray}

Now we analyze an example which will be used in the model building
of section \ref{sec:3}. As in our previous paper \cite{Guo:2008}, we
choose the metric $B(z)$ to be
\begin{eqnarray}
\label{suppose1} B(z)=s \frac{e^{\omega z}+a}{e^{\omega
z}+b},~~s,a,b,\omega>0.
\end{eqnarray}
We choose the extent of $z$ to be the semi-infinite interval
$[R,~\infty)$. This metric is similar to that we obtained in
Eq.~(\ref{forbz-1}). We implicitly suppose $m>0$ in following
discussions, unless we announce it explicitly.

For this metric, the normalizable zero mode solution is given by
\begin{eqnarray}
\label{ZeroMode1}
F_{0}(z)=0,~G_{0}(z)=\frac{\sqrt{\omega}}{\sqrt{N_0}}\left(\frac{e^{\omega
z}}{b}\right)^{-\frac{m}{\omega}s\frac{a}{b}}\left(\frac{e^{\omega
z}}{b}+1\right)^{\frac{m}{\omega}s(\frac{a}{b}-1)}.
\end{eqnarray}
For massive modes, Eq.~(\ref{solution17}) can be solved by
hypergeometrical functions,
\begin{eqnarray}
\label{suppose1-2} F(z)&=&C_1 e^{-\mu \omega z}(e^{\omega
z}+b)^{\mu-\nu}
\mathrm{hypergeom}\left(\rho-\mu+\nu,1-\rho-\mu+\nu;1-2\mu,\frac{e^{\omega
z}}{e^{\omega z}+b}\right) \nonumber \\&+&C_2 e^{\mu \omega
z}(e^{\omega z}+b)^{-\mu-\nu}
\mathrm{hypergeom}\left(\rho+\mu+\nu,1-\rho+\mu+\nu;1+2\mu,\frac{e^{\omega
z}}{e^{\omega z}+b}\right),~~
\end{eqnarray}
where $\rho=\frac{m}{\omega}s(1-\frac{a}{b})$,
$\mu=\sqrt{\left(\frac{m}{\omega}s\right)^2\left(\frac{a}{b}\right)^2-\left(\frac{\lambda}{\omega}\right)^2}$,
and
$\nu=\sqrt{\left(\frac{m}{\omega}s\right)^2-\left(\frac{\lambda}{\omega}\right)^2}$.
$C_1$ and $C_2$ are constants. We have omitted the subscript $n$
explicitly. We only give the solution for $F(z)$ here. The solution
for $G(z)$ can be determined by $F(z)$ through
Eq.~(\ref{solution13}) or by Eq.~(\ref{solution18}) directly. As
discussed in \cite{Guo:2008}, in order to make the solution to be
finite when $z\rightarrow\infty$, the hypergeometrical series must
be cut off to be a polynomial. This cut off can be completed by four
different kinds of choices. We can cut off the series after the
coefficient $C_1$ by conditions,
\begin{eqnarray}
\label{cut1-a} \rho-\mu+\nu&=&-n,\\
\label{cut1-b} \mathrm{or} ~~~1-\rho-\mu+\nu&=&-n,
\end{eqnarray}
in which $n=0,1,2,3,\cdots$. Otherwise we can cut off the series
after the coefficient $C_2$ by conditions,
\begin{eqnarray}
\label{cut2-a} \rho+\mu+\nu&=&-n,\\
\label{cut2-b} \mathrm{or} ~~~1-\rho+\mu+\nu&=&-n.
\end{eqnarray}
The four equations (\ref{cut1-a}), (\ref{cut1-b}), (\ref{cut2-a})
and (\ref{cut2-b}) are not necessary to be satisfied at the same
time. Anyone of them is enough to cut off the series. So we can
obtain four kinds of solutions corresponding to these four different
cut off ways generally\footnote{In our previous paper
\cite{Guo:2008}, we only consider the solutions corresponding to the
cut off condition Eq.~(\ref{cut1-b}). The other three kinds of
solutions are omitted.}. The next important task is to analyze
whether the number of eigenvalues determined by these equations are
finite. We discuss this problem in appendix \ref{sec:A}. In this
subsection, we only display the results.

In this paper, we are interested in the situation $\frac{a}{b}>1$.
According to discussions in appendix \ref{sec:A}, we can have two
different ways (\ref{cut1-a}) and (\ref{cut1-b}) to cut off the
series. We might obtain two kinds of solutions corresponding to
these two cut off ways. While equations (\ref{cut2-a}) and
(\ref{cut2-b}) have no solutions, so the corresponding cut off ways
do not work. As we analyzed in appendix \ref{sec:A}, $n$ must be
finite in equations (\ref{cut1-a}) and (\ref{cut1-b}). The extents
of $n$ are given by equations (\ref{cut1-a-n-ex}) and
(\ref{cut1-b-n-ex}). For convenience of reading, we copy equations
(\ref{cut1-a-n-ex}) and (\ref{cut1-b-n-ex}) with new sequence
numbers here.
\begin{eqnarray}
\label{cut1-a-n-exc}
\frac{m}{\omega}s\left(\frac{a}{b}-1\right)<&n&\leq
\frac{m}{\omega}s\left[\sqrt{\left(\frac{a}{b}\right)^2-1}+\left(\frac{a}{b}-1\right)\right]~~~~\mathrm{for~~(\ref{cut1-a})},\\
\label{cut1-b-n-exc}
-\frac{m}{\omega}s\left(\frac{a}{b}-1\right)<&n+1&\leq
\frac{m}{\omega}s\left[\sqrt{\left(\frac{a}{b}\right)^2-1}-\left(\frac{a}{b}-1\right)\right]~~~~\mathrm{for~~(\ref{cut1-b})}.
\end{eqnarray}
From (\ref{cut1-a-n-exc}) and (\ref{cut1-b-n-exc}), we know that the
number of eigenvalues is determined by a pair of parameters
$(\frac{a}{b},~\frac{m}{\omega}s)$. We can choose the values of this
pair of parameters to make the very two massive modes left. Note
that conditions (\ref{cut1-a-n-exc}) and (\ref{cut1-b-n-exc}) are
not necessary to be satisfied at the same time. They are different
restrictions for two kinds of cut off ways respectively.

The values of $(\frac{a}{b},~\frac{m}{\omega}s)$ construct a 2D
plane. A point set in this 2D plane restricts $n$ to be a specific
value. We give a point set of $(\frac{a}{b},~\frac{m}{\omega}s)$
determined by following equations
\begin{eqnarray}
\label{cut1-a-pc} \frac{m}{\omega}s\left(\frac{a}{b}-1\right)<1,\\
\label{cut1-b-pc}
1<\frac{m}{\omega}s\left[\sqrt{\left(\frac{a}{b}\right)^2-1}+\left(\frac{a}{b}-1\right)\right]<2,\\
\label{cut1-c-pc}
\frac{m}{\omega}s\left[\sqrt{\left(\frac{a}{b}\right)^2-1}-\left(\frac{a}{b}-1\right)\right]<1.
\end{eqnarray}
Conditions (\ref{cut1-a-pc}) and (\ref{cut1-b-pc}) imply that only
$n=1$ is permitted in Eq.~(\ref{cut1-a-n-exc}); so there is one
solution corresponding to this kind of cut off. The condition
(\ref{cut1-c-pc}) implies that Eq.~(\ref{cut1-b-n-exc}) is
impossible; so there is no solutions corresponding to this kind of
cut off. As discussed in the above, the massive modes always emerge
in pair. So this parameter set is enough to make a pair massive
modes. Together with the zero mode, we obtain the very 3 modes in
all. For the future model building, we give solutions for these
modes explicitly in appendix \ref{sec:B}.

\begin{figure}
\begin{center}
\includegraphics[height=6cm]{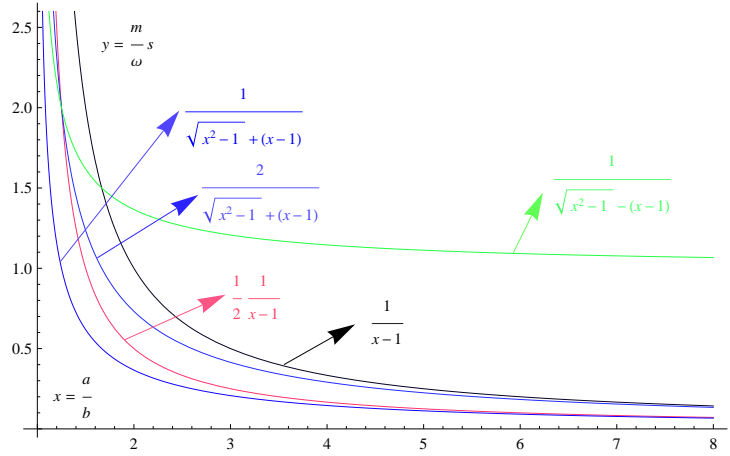}
\end{center}
\caption{\small The parameter set for 3 families.}
\label{fig:paraset}
\end{figure}

The point set determined by equations (\ref{cut1-a-pc}),
(\ref{cut1-b-pc}) and (\ref{cut1-c-pc}) forms a specific area in 2D
plane. This area can be visualized in Figure.~\ref{fig:paraset}.

For $\frac{a}{b}>\frac{5}{4}$, this area is given by the area
between the curve $y=\frac{1}{\sqrt{x^2-1}+(x-1)}$ and the curve
$y=\frac{2}{\sqrt{x^2-1}+(x-1)}$; while for
$\frac{a}{b}<\frac{5}{4}$, this area is given by the area between
the curve $y=\frac{1}{\sqrt{x^2-1}+(x-1)}$ and the curve
$y=\frac{1}{\sqrt{x^2-1}-(x-1)}$. However, the points at the curve
$y=\frac{1}{2}\frac{1}{x-1}$ must be excluded. As we discussed in
appendix \ref{sec:A}, in this curve, the equation
\begin{eqnarray}
\label{lambda-zero-1} 1=2\frac{m}{\omega}s\left(\frac{a}{b}-1\right)
\end{eqnarray}
is satisfied. According to Eq.~(\ref{lambda-zero}), when
Eq.~(\ref{lambda-zero-1}) is satisfied, $\lambda=0$ for
Eq.~(\ref{cut1-a}). So in this situation, $n=1$ gives a massless
solution but not a massive solution. It is not a new solution and it
coincides with the zero mode solution given by
Eq.~(\ref{ZeroMode1}). We should exclude the points at this curve in
order to ensure that we obtain massive solutions.

\subsection{Reducing spacetime from 5D to 4D}\label{sec:2.4}

We have reduced the spacetime form 6D to 5D in subsection
\ref{sec:2.1}. In this subsection, we further reduce the spacetime
form 5D to 4D. We begin with the effective 5D action
(\ref{action3a}). Note that masses $\widehat{\lambda}_{n}$ are real
irrespective of values of $\lambda_{n}$ in equations
(\ref{solution13}) and (\ref{solution14})\footnote{In our previous
paper \cite{Guo:2008}, we require that $\lambda_{n}$ must be real.
It is a superfluous requirement and is not necessary. As in the
discussions in appendix \ref{sec:B}, $\lambda_{n}$ can be pure
imaginary.}, because the induced matrix $\widetilde{M}$ in
(\ref{diag5}) is hermitian. As in the interval approach of RS model,
we choose the extent of $y$ to be a finite interval
$[L,~L^{\prime}]$.

We make KK decompositions by expanding the 5D field $\widehat{\psi}$
with the 4D chiral fields as
\begin{eqnarray}
\label{KK5} \widehat{\psi}=\sum_j
A(y)^{-2}[f_{L}^{(j)}(y)\psi_{L}^{(j)}+f_{R}^{(j)}(y)\psi_{R}^{(j)}],
\end{eqnarray}
where $\psi_{L}=-\gamma^5\psi_{L},~\psi_{R}=\gamma^5\psi_{R}$. In
the above expanding, we have omitted the subscript for
$\widehat{\psi}$. It is understood that we do the similar operation
for each $\widehat{\psi}_n$ in the action Eq.~(\ref{action3a}). The
KK modes satisfy following equations
\begin{eqnarray}
\label{KK5-1}
\left(\frac{d}{d y}+ \hat{\lambda} A \right)f_{L}^{(j)}(y)-\mathbbm{m}^{(j)} f_{R}^{(j)}(y)=0,\\
\label{KK5-2} \left(\frac{d}{d y}-\hat{\lambda} A
\right)f_{R}^{(j)}(y)+\mathbbm{m}^{(j)} f_{L}^{(j)}(y)=0,
\end{eqnarray}
where $\mathbbm{m}^{(j)}$ are eigenvalues.

In this paper, following a popularly adopted approach in papers
\cite{Grossman:2000,Agashe:2005}, we employ the zero mode
approximation~(ZMA) approach, that is, we solve for the fermion bulk
profiles without the brane interaction terms at first and then we
treat the brane interaction terms as a perturbation. In this ZMA
approach, we can choose two groups of consistent boundary conditions
for equations (\ref{KK5-1}) and (\ref{KK5-2}) as follows,
\begin{eqnarray}
\label{KK5-1-bound}
\mathrm{Group~(I):}~~\left(\frac{d}{d y}+ \hat{\lambda} A \right)f_{L}^{(j)}(y)=0,~~f_{R}^{(j)}(y)=0,~~\mathrm{at}~~y=L~~\mathrm{and}~~L^{\prime},\\
\label{KK5-2-bound} \mathrm{Group~(II):}~~\left(\frac{d}{d
y}-\hat{\lambda} A
\right)f_{R}^{(j)}(y)=0,~~f_{L}^{(j)}(y)=0,~~\mathrm{at}~~y=L~~
\mathrm{and}~~L^{\prime}.
\end{eqnarray}
The boundary conditions (\ref{KK5-1-bound}) save the left-handed
zero mode and kill the right-handed one; while the boundary
conditions (\ref{KK5-2-bound}) do the opposite.

In the model building in section \ref{sec:3}, we will choose the
metric factor $A(y)$ to be
\begin{eqnarray}
\label{ansatz5} A(y)=\frac{L}{y},
\end{eqnarray}
that is, the RS spacetime. By solving the bulk equations for this
metric, we obtain the normalized left-handed zero mode profile and
the right-handed one as
\begin{eqnarray}
\label{KK5-1-zero}
f_{L}^{(0)}(y)&=&\frac{1}{\sqrt{L}}\sqrt{\frac{2c-1}{1-\epsilon^{2c-1}}}\left(\frac{y}{L}\right)^{-c},\\
\label{KK5-2-zero}
f_{R}^{(0)}(y)&=&\frac{1}{\sqrt{L}}\sqrt{\frac{-2c-1}{1-\epsilon^{-2c-1}}}\left(\frac{y}{L}\right)^{c},
\end{eqnarray}
where $c=\hat{\lambda}L,~~\epsilon=\frac{L}{L^{\prime}}$, and we
suppose that $\epsilon\ll 1$. In this paper, we will also use the
canonical normalized zero mode profiles and their values at
$y=L^{\prime}$. Their values at $y=L^{\prime}$ are given by
\begin{eqnarray}
\label{KK5-1-zero-c}
\hat{f}_{L}^{(0)}&=&\sqrt{\frac{2c-1}{1-\epsilon^{2c-1}}}\epsilon^{c-\frac{1}{2}},\\
\label{KK5-2-zero-c}
\hat{f}_{R}^{(0)}&=&\sqrt{\frac{-2c-1}{1-\epsilon^{-2c-1}}}\epsilon^{-c-\frac{1}{2}},
\end{eqnarray}
where $\hat{f}_{L,R}^{(0)}(y)=y^{\frac{1}{2}}f_{L,R}^{(0)}(y)$. For
discussions in the next subsection, it is useful to find the
asymptotic behavior of these profiles,
\begin{eqnarray}
\label{KK5-1-zero-asy} \hat{f}_{L}^{(0)}\thicksim
\begin{cases}
\sqrt{1-2c},&\mathrm{for}~~c< \frac{1}{2}, \\
 \sqrt{\frac{1}{-\ln{\epsilon}}},~&\mathrm{for}~~c\rightarrow\frac{1}{2},\\
\sqrt{2c-1}~\epsilon^{c-\frac{1}{2}},~&\mathrm{for}~~c> \frac{1}{2}.
\end{cases}~~~\hat{f}_{R}^{(0)}\thicksim\begin{cases}
\sqrt{1+2c},&\mathrm{for}~-c< \frac{1}{2}, \\
 \sqrt{\frac{1}{-\ln{\epsilon}}},~&\mathrm{for}~-c\rightarrow\frac{1}{2},\\
\sqrt{-1-2c}~\epsilon^{-c-\frac{1}{2}},~&\mathrm{for}~-c>
\frac{1}{2}.
\end{cases}
\end{eqnarray}
We have supposed that $\epsilon\ll 1$ in above expressions.

In the above, we discuss characters of zero modes. In the following
we discuss massive modes. For massive modes, equations (\ref{KK5-1})
and (\ref{KK5-2}) can be combined to be the second order equations
\begin{eqnarray}
\label{KK5-1-massive}
\frac{d^2}{d y^2}f_{L}^{(j)}(y)+[\hat{\lambda} A_y-\hat{\lambda}^2 A^2 ]f_{L}^{(j)}(y)+\mathbbm{m}^{(j)2} f_{L}^{(j)}(y)=0,\\
\label{KK5-2-massive} \frac{d^2}{d
y^2}f_{R}^{(j)}(y)+[-\hat{\lambda} A_y-\hat{\lambda}^2 A^2
]f_{R}^{(j)}(y)+\mathbbm{m}^{(j)2} f_{R}^{(j)}(y)=0,
\end{eqnarray}
where $A_y=\frac{d A(y)}{d y}$. For the metric (\ref{ansatz5}),
equations (\ref{KK5-1-massive}) and (\ref{KK5-2-massive}) can be
solved by Bessel functions. Following the standard procedure of
analyzing Sturm-Liouville equations, we can obtain from equations
(\ref{KK5-1-massive}) and (\ref{KK5-2-massive}) that
\begin{eqnarray}
\label{KK5-1-massive-ortho}
\left[\mathbbm{m}^{(j)2}-(\mathbbm{m}^{(i)2})^{\ast}\right]\int_L^{L^{\prime}}
dy\left(f_{L}^{(i)\ast}f_{L}^{(j)}\right)
&=&\left[\left(f_{L}^{(j)}\frac{d}{d
y}f_{L}^{(i)\ast}-f_{L}^{(i)\ast}\frac{d}{d y}f_{L}^{(j)}\right)\right]\bigg{|}_{L}^{L'},\\
\label{KK5-2-massive-ortho}
\left[\mathbbm{m}^{(j)2}-(\mathbbm{m}^{(i)2})^{\ast}\right]\int_L^{L^{\prime}}
dy\left(f_{R}^{(i)\ast}f_{R}^{(j)}\right)
&=&\left[\left(f_{R}^{(j)}\frac{d}{d
y}f_{R}^{(i)\ast}-f_{R}^{(i)\ast}\frac{d}{d
y}f_{R}^{(j)}\right)\right]\bigg{|}_{L}^{L'}.
\end{eqnarray}
The above equations are derived from equations (\ref{KK5-1-massive})
and (\ref{KK5-2-massive}) for massive modes, but it also applies for
zero modes. The boundary condition (\ref{KK5-1-bound}) ensures that
left-handed and right-handed KK modes satisfy orthogonal conditions
\begin{eqnarray}
\label{KK5-1-massive-norm} \int_L^{L^{\prime}}
dy\left(f_{L}^{(i)\ast}f_{L}^{(j)}\right)=\delta^{ij},~~\int_L^{L^{\prime}}
dy\left(f_{R}^{(i)\ast}f_{R}^{(j)}\right)=\delta^{ij}.
\end{eqnarray}
While the boundary condition (\ref{KK5-2-bound}) also makes
orthogonal conditions (\ref{KK5-1-massive-norm}) satisfied.

\subsection{A more realistic setup for model building}\label{sec:2.5}
According to discussions in subsection \ref{sec:2.3} subsection
\ref{sec:2.4}, we find a problem which prevents us to construct a
realistic model. This problem arises from the following
contradiction. On one side, from the asymptotic expressions of the
zero mode profiles (\ref{KK5-1-zero-asy}) in subsection
\ref{sec:2.4}, we know that in order to produce fermion mass
hierarchies, the zero mode profiles should be of exponential
behavior. According to Eq.~(\ref{KK5-1-zero-asy}), because
$\epsilon\ll 1$, we need positive parameters $c>\frac{1}{2}$ for
left-handed zero mode profile, and negative parameters
$-c>\frac{1}{2}$ for right-handed zero mode profile. On the other
side, the parameter $c$ comes from the 5D mass $\hat{\lambda}$. We
have found in subsection \ref{sec:2.3} that the spectrum of
$\hat{\lambda}$ consists of one zero mode and two massive modes
$\pm\hat{\lambda}$ in pairs\footnote{In subsection \ref{sec:2.3}, we
show that the spectrum of $\lambda$ consists of one zero mode and
two massive modes $\pm\lambda$ in pairs. However, we can prove that
this results also applies to $\hat{\lambda}$. We will give the
proofs in subsection \ref{sec:4.1}. Or one can employ the example in
subsection \ref{sec:2.3} to check it with numerical method
directly.}. For left-handed zero modes, $\hat{\lambda}=0$ and
$-\hat{\lambda}$ prevent the profiles to have exponential behavior;
while for right-handed zero modes, $\hat{\lambda}=0$ and
$\hat{\lambda}$ prevent them to have exponential behavior. This
contradiction constructs an obstacle to produce the fermion mass
hierarchies. In this subsection, we construct a new setup to bypass
this problem.

Instead of the action (\ref{action1}), we suggest a new bulk action
as follows,
\begin{equation}
\label{action-new} S = \int d^4 x dy dz \sqrt{-g} \left\{
\frac{i}{2} \left[\bar{\Psi}e_a^N \Gamma^a \nabla_N \Psi - \nabla_N
\bar{\Psi}e_a^N \Gamma^a \Psi \right] -i~m \bar{\Psi}
\Psi-\mathcal{M}\bar{\Psi}\Gamma^7e^6_a\Gamma^a\Psi\right\},
\end{equation}
where $\Gamma^7$ is defined by Eq.~(\ref{Eq3}). Our metric Ansatz
(\ref{introduce}) keeps invariant. $\mathcal{M}$ is real to ensure
that this new action is hermitian. Comparing it with the action
(\ref{action1}), we add a new term. This new term has a strange
form, as it breaks the 6D local lorentz invariance obviously. Here
we regard this new action as an effective one to bypass the problem
discussed above. We will discuss the possible origin of this new
action in section \ref{sec:4}.

Making the KK decompositions as in subsection \ref{sec:2.1},
equations (\ref{solution13}) and (\ref{solution14}) are modified to
be
\begin{eqnarray}
\label{solution13-new} \left(\frac{d}{d z}- m B\right)F_n(z) +
(\lambda_n-\mathcal{M}) G_n(z)=0,\\
\label{solution14-new} \left(\frac{d}{d z}+ m B \right)G_n(z) -
(\lambda_n-\mathcal{M}) F_n(z)=0.
\end{eqnarray}
The induced 5D action is the same with that in equations
(\ref{norm1}), (\ref{norm2}) and (\ref{norm3}). The differences are
that the KK modes and the eigenvalues $\lambda_n$ determined by the
new equations (\ref{solution13-new}) and (\ref{solution14-new})
instead of equations (\ref{solution13}) and (\ref{solution14}). By
defining new variables
\begin{eqnarray}
\label{new-var} \triangle\lambda_n=\lambda_n-\mathcal{M},
\end{eqnarray}
we obtain
\begin{eqnarray}
\label{sol13-new-var}
\left(\frac{d}{d z}- m B\right)F_n(z) + \triangle\lambda_n G_n(z)=0,\\
\label{sol14-new-var} \left(\frac{d}{d z}+ m B \right)G_n(z) -
\triangle\lambda_n F_n(z)=0.
\end{eqnarray}
These new equations have the same form with equations
(\ref{solution13}) and (\ref{solution14}). So if we replace
$\lambda_n$ with $\triangle\lambda_n$ in equations
(\ref{solution13}) and (\ref{solution14}), the results that we
obtain in subsections \ref{sec:2.2}-\ref{sec:2.3}, appendix
\ref{sec:A} and appendix \ref{sec:B} apply here. Of course, we
should notice that $\lambda_n$ in Eq.~(\ref{new-var}) emerge in
actions (\ref{norm1}), (\ref{norm2}) and (\ref{norm3}) but not the
new variables $\triangle\lambda_n$. We will give more details in
section \ref{sec:4}. There is a zero mode solution corresponding to
$\triangle\lambda=0$, and the massive mode solutions also emerge in
pairs. A pair of massive modes for $\triangle\lambda_n$ are given by
\begin{eqnarray}
\label{pair-real} \triangle\lambda_n \longrightarrow
\begin{cases}
F_n \\
G_n
\end{cases},~~~~~-\triangle\lambda_n \longrightarrow
\begin{cases}
F_n \\
-G_n
\end{cases}.
\end{eqnarray}
The spectrum of $\triangle\lambda$ will be the same with that of
$\lambda$ given by Eq.~(\ref{spectrum1}). While by
Eq.~(\ref{new-var}), the spectrum formula (\ref{spectrum1}) of
$\lambda_n$ now changes to be
\begin{eqnarray}
\label{spectrum2} \mathrm{spectrum~of}~\triangle\lambda
\longrightarrow
\begin{cases}
\triangle\lambda_1 \\
0\\
-\triangle\lambda_1 \\
\end{cases}~\underrightarrow{\lambda_n=\triangle\lambda_n+\mathcal{M}}~\mathrm{spectrum~of}~\lambda
\longrightarrow \begin{cases}
\triangle\lambda_1+\mathcal{M} \\
\mathcal{M}\\
-\triangle\lambda_1+\mathcal{M}\\
\end{cases}.~~
\end{eqnarray}

By the new spectrum Eq.~(\ref{spectrum2}), we can bypass the problem
discussed at the onset of this subsection. From equations
(\ref{sol13-new-var}) and (\ref{sol14-new-var}), we know that
$\triangle\lambda_n$ depends on the mass parameter $m$ and the
metric $B(z)$. So just like we discussed in subsection
\ref{sec:2.3}, the number and the spectrum of $\triangle\lambda_n$
are completely determined by $m$ and parameters in $B(z)$. The
number and the spectrum of $\triangle\lambda$ are irrelevant to the
new parameter $\mathcal{M}$. We can adjust $\mathcal{M}$ to change
the spectrum of the final eigenvalues $\lambda$. If the value of
$\mathcal{M}$ is a large positive number, then the spectrum of
$\lambda$ can be all positive by Eq.~(\ref{spectrum2}), which are
expected for left-handed zero modes. We can also make the spectrum
of $\lambda$ to be all negative by choosing small negative value of
$\mathcal{M}$, which are expected for right-handed zero modes. As we
will discuss in subsection \ref{sec:4.1}, the new spectrum
Eq.~(\ref{spectrum2}) for $\lambda$ also applies to $\hat{\lambda}$.
The above discussions for $\lambda$ also apply to $\hat{\lambda}$.
These features bypass the problem at the beginning of this
subsection.

In the model building of section \ref{sec:3}, we will adopt this new
setup in this subsection. Compared with the setup in subsection
\ref{sec:2.1}, it is not more difficult to analyze this new setup.
We only define the new variable by Eq.~(\ref{new-var}). All other
analysis in subsection \ref{sec:2.2}, subsection \ref{sec:2.3},
appendix \ref{sec:A} and appendix \ref{sec:B} apply here.

\section{A model of fermion mass and mixing}\label{sec:3}
In subsection \ref{sec:2.1}, we discuss how several families in 5D
can generate from 1 family in 6D. In subsection \ref{sec:2.5}, we
suggest a new setup. By this new setup, we saw that the fermion mass
hierarchies in 4D can be produced when we further reduce the
spacetime from 5D to 4D. Therefore, we hope that fermion mass
hierarchy problem and family problem can be both addressed in such
an approach. In such an approach, the mass hierarchy problem will
closely relate to the family problem. In this section, we construct
a specific model along these discussions. In subsection
\ref{sec:3.1}, we introduce the model in the 6D bulk spacetime. In
order to construct a realistic scenario, we make two assumptions in
the model building. In subsection \ref{sec:3.2}, in order to obtain
a 4D effective theories, we employ two step KK decompositions to
reduce the spacetime from 6D to 4D. At the first step, we reduce the
spacetime from 6D to 5D, then we further reduce the spacetime from
5D to 4D. After deriving the 4D effective theories, we further give
numerical results in subsection \ref{sec:3.3}. In subsection
\ref{sec:3.3.1}, we give numerical results in quark sector. In
subsection \ref{sec:3.3.2}, supposing neutrinos to be Dirac ones, we
apply this model to the lepton sector. In both sectors, the
numerical results can be very close to the experimental data. We
also give brief comments about the numerical results in subsection
\ref{sec:3.3.3}.

\subsection{Model building}\label{sec:3.1}
We begin with 6D spacetime with the two-layer warped metric
\begin{eqnarray}
\label{ansatz-model} ds^2 = g_{MN}dx^Mdx^N=B(z)^2\left[A(y)^2\left(
\eta_{\mu \nu} dx^{\mu} dx^{\nu} + d y^2\right)+ d z^2\right].
\end{eqnarray}
We also choose the metric factors to be
\begin{eqnarray}
\label{suppose1-model} B(z)=s \frac{e^{\omega z}+a}{e^{\omega
z}+b},~~~A(y)=\frac{L}{y}.
\end{eqnarray}
The parameters in this metric will be designated when we give
numerical results in subsection \ref{sec:3.3}. As discussed in
section \ref{sec:2}, we choose the two extra dimensions are both
intervals. The extent of $z$ is a semi-infinite interval
$[R,~\infty]$; while the extent of $y$ is a finite interval
$[L,~L^{\prime}]$. By these choices, the spacetime is sandwiched by
3 co-dimension 1 4-branes sited at $z=R$, $y=L$ and $y=L^{\prime}$.
We also introduce two co-dimension 2 3-branes sited at the brane
intersections $(z=R,~y=L)$ and $(z=R,~y=L^{\prime})$. They can be
dubbed UV-brane and IR-brane respectively. We will only designate
the field contents on IR-brane sited at $(z=R,~y=L^{\prime})$, while
the field contents on UV-brane are omitted in the following
discussions. The metric on these 3-branes are given by the induced
metric
\begin{eqnarray}
\label{metric-ind-1}
g^{\mathrm{hid}}_{\mu\nu}(x)=g_{\mu\nu}(x,R,L),\\
\label{metric-ind-2}
g^{\mathrm{vis}}_{\mu\nu}(x)=g_{\mu\nu}(x,R,L^{\prime}).
\end{eqnarray}

We discuss the quark sector at first. We introduce the quark field
contents as
\begin{eqnarray}
\label{quark}
\mathbb{Q}=\left(\begin{array}{c}
   \mathbb{U}\\
   \mathbb{D}
\end{array}\right)
,~~~~\mathcal{U},~~~~\mathcal{D}.
\end{eqnarray}
They transform under the gauge group $\mathrm{SU}(3)_c\times
\mathrm{SU}(2)_L\times\mathrm{U}(1)_Y$ as
$\mathbb{Q}=(3,2)_{+1/6},~\mathcal{U}=(3,1)_{+2/3},~\mathcal{D}=(3,1)_{-1/3}$.
These fields are 6D fields. Note that there are no family indices
for them, that is, we introduce only 1 family in the 6D bulk. Their
actions are given by
\begin{eqnarray}
\label{action-Q} S_{\mathbb{Q}}&=&\int d^4 x dy dz \sqrt{-g}
\bigg{\{} \frac{i}{2} \bigg{[}\bar{\mathbb{Q}} e_a^M \Gamma^a
\nabla_M \mathbb{Q}-\nabla_M \bar{\mathbb{Q}}e_a^M \Gamma^a
\mathbb{Q} \bigg{]}-im_{\mathbb{Q}}
\bar{\mathbb{Q}}\mathbb{Q}\bigg{\}}\nonumber\\
&-&\int d^4 x dy dz \sqrt{-g}\mathcal{M}_{\mathbb{Q}}\bar{\mathbb{Q}}\Gamma^7e^6_a\Gamma^a\mathbb{Q},\\
\label{action-U} S_{\mathcal{U}}&=& \int d^4 x dy dz \sqrt{-g}
\bigg{\{} \frac{i}{2} \bigg{[}\bar{\mathcal{U}} e_a^M \Gamma^a
\nabla_M \mathcal{U}-\nabla_M \bar{\mathcal{U}}e_a^M \Gamma^a
\mathcal{U} \bigg{]} -im_{\mathcal{U}}
\bar{\mathcal{U}} \mathcal{U}\bigg{\}}\nonumber\\
&-&\int d^4 x dy dz \sqrt{-g}\mathcal{M}_{\mathcal{U}}\bar{\mathcal{U}}\Gamma^7e^6_a\Gamma^a \mathcal{U},\\
\label{action-D} S_{\mathcal{D}}&=&\int d^4 x dy dz \sqrt{-g}
\bigg{\{} \frac{i}{2} \bigg{[}\bar{\mathcal{D}} e_a^M \Gamma^a
\nabla_M \mathcal{D}-\nabla_M \bar{\mathcal{D}}e_a^M \Gamma^a
\mathcal{D} \bigg{]} -im_{\mathcal{D}} \bar{\mathcal{D}}
\mathcal{D}\bigg{\}}\nonumber\\
&-&\int d^4 x dy dz
\sqrt{-g}\mathcal{M}_{\mathcal{D}}\bar{\mathcal{D}}\Gamma^7e^6_a\Gamma^a
\mathcal{D}.
\end{eqnarray}
As in the usual field theory, we introduce the interactions between
fermion field and gauge filed by requiring that actions
(\ref{action-Q}), (\ref{action-U}) and (\ref{action-D}) are
invariant under the 6D local gauge transformation
\begin{eqnarray}
\label{gauge-6D} \mathbb{Q}&\longrightarrow&
e^{-i[\hat{g}\hat{\theta}_j(x,y,z)\frac{\sigma^j}{2}+g^{\prime}\tilde{\theta}(x,y,z)]}\mathbb{Q},\\\nonumber
\mathcal{U}&\longrightarrow&
e^{-ig^{\prime}\tilde{\theta}(x,y,z)}\mathcal{U},~~
\mathcal{D}\longrightarrow
e^{-ig^{\prime}\tilde{\theta}(x,y,z)}\mathcal{D},
\end{eqnarray}
where $\frac{\sigma^j}{2}$ are the generators of the gauge group
$\mathrm{SU}(2)_L$\footnote{In this paper, we focus on the
electroweak interaction sector, so the color indices are implicit.
We can introduce the color interaction just like that we do for
electroweak interaction.}. However, we do not employ this 6D local
gauge transformation (\ref{gauge-6D}) in this paper. Instead of it,
we assume that the fermion actions are only invariant under the 4D
local gauge transformation
\begin{eqnarray}
\label{gauge-4D} \mathbb{Q}&\longrightarrow&
e^{-i[\hat{g}\hat{\theta}_j(x)\frac{\sigma^j}{2}+g^{\prime}\tilde{\theta}(x)]}\mathbb{Q},\\\nonumber
\mathcal{U}&\longrightarrow&
e^{-ig^{\prime}\tilde{\theta}(x)}\mathcal{U},~~
\mathcal{D}\longrightarrow
e^{-ig^{\prime}\tilde{\theta}(x)}\mathcal{D}.
\end{eqnarray}
Here the gauge parameters only depend on the 4D coordinates.
Requiring the invariance under transformations Eq.~(\ref{gauge-4D}),
we obtain the interaction terms
\begin{eqnarray}
\label{gauge-int-Q} S_{\mathbb{Q}\mathrm{int}}&=&-\int d^4 x dy dz
\sqrt{-g} \left\{\bar{\mathbb{Q}}e_a^\mu
\Gamma^a\bigg{[}\hat{g}W^j_\mu(x)\frac{\sigma^j}{2}+g^{\prime}B_\mu(x)\bigg{]}\mathbb{Q}\right\},\\
\label{gauge-int-U} S_{\mathcal{U}\mathrm{int}}&=&-\int d^4 x dy dz
\sqrt{-g} \left\{\bar{\mathcal{U}}e_a^\mu
\Gamma^a[g^{\prime}B_\mu(x)]\mathcal{U}\right\},\\
\label{gauge-int-D} S_{\mathcal{D}\mathrm{int}}&=&-\int d^4 x dy dz
\sqrt{-g} \left\{\bar{\mathcal{D}}e_a^\mu
\Gamma^a[g^{\prime}B_\mu(x)]\mathcal{D}\right\},
\end{eqnarray}
where $\mu=0,1,2,3$. Because we suppose only the 4D local gauge
invariance, the gauge field components $W_5,~W_6$ and $B_5,~B_6$
relevant to extra dimensions are not necessary to be introduced
here. These gauge fields only depend on the 4D coordinates. We
designate their actions to be the brane actions on the IR-brane
\begin{eqnarray}
\label{action-gauge} S_{\mathrm{gauge}}=\int d^4 x dy dz
\sqrt{-g_{\mathrm{vis}}}
\left\{-g_{\mathrm{vis}}^{\mu\alpha}g_{\mathrm{vis}}^{\nu\beta}\bigg{[}\frac{1}{4}\hat{F}^a_{\mu\nu}\hat{F}^a_{\alpha\beta}+\frac{1}{4}\tilde{F}_{\mu\nu}\tilde{F}_{\alpha\beta}\bigg{]}\delta(y-L^{\prime})\delta(z-R)\right\},
\end{eqnarray}
where $g_{\mathrm{vis}}^{\mu\alpha}$ is given by
Eq.~(\ref{metric-ind-2}).

Note that interaction terms (\ref{gauge-int-Q}), (\ref{gauge-int-U})
and (\ref{gauge-int-D}) are different from conventional brane
interaction terms
\begin{eqnarray}
\label{int-Q-brane} S_{\mathbb{Q}\mathrm{int}}&=&\int d^4 x dy dz
\sqrt{-g_{\mathrm{vis}}}
\left\{\bar{\mathcal{U}}e_{\mathrm{vis}}^{a\mu}
\Gamma_a[g^{\prime}B_\mu(x)]\mathcal{U}\delta(y-L^{\prime})\delta(z-R)\right\},
\end{eqnarray}
where $e_{\mathrm{vis}}^{a\mu},~a,\mu=0,1,2,3$ are the tetrad
determined by the IR-brane metric $g_{\mathrm{vis}}^{\mu\alpha}$.
Interaction terms (\ref{gauge-int-Q}), (\ref{gauge-int-U}) and
(\ref{gauge-int-D}) are introduced as above to ensure that we can
obtain an unitary mixing matrix in the ZMA approach; while the brane
interaction terms like Eq.~(\ref{int-Q-brane}) break the unitarity
of mixing matrix remarkably in our present model. Interaction terms
(\ref{gauge-int-Q}), (\ref{gauge-int-U}) and (\ref{gauge-int-D}) are
our important assumptions in this paper. We can also introduce gauge
fields in the 5D bulk as in the papers \cite{Chang:2000}. The
unitarity of mixing matrix in the ZMA approach is also kept.
However, in this paper, we employ the more simple approach as
introduced in the above.

Now we introduce the Yukawa interactions between fermion fields and
the Higgs fields. We designate them as
\begin{eqnarray}
\label{yukawa-D-brane} S_{\mathcal{D}\mathrm{Yukawa}}&=&\int d^4 x
dy dz \sqrt{-g}
\left\{\frac{f_{\mathcal{D}}}{\Lambda}[\bar{\mathbb{Q}}\mathcal{D}+
\beta_{\mathcal{D}}\bar{\mathbb{Q}}\Gamma^7\mathcal{D}]\Phi(x)\frac{L^{\prime}}{L}\delta(y-L^{\prime})+\mathrm{H.C.}\right\},\\
\label{yukawa-U-brane} S_{\mathcal{U}\mathrm{Yukawa}}&=&\int d^4 x
dy dz \sqrt{-g}
\left\{\frac{f_{\mathcal{U}}}{\Lambda}[\bar{\mathbb{Q}}\mathcal{U}+
\beta_{\mathcal{U}}\bar{\mathbb{Q}}\Gamma^7\mathcal{U}]\tilde{\Phi}(x)\frac{L^{\prime}}{L}\delta(y-L^{\prime})+\mathrm{H.C.}\right\},\\
\label{higgs} \Phi(x)&=&\frac{1}{\sqrt{2}}\left(\begin{array}{c}
   \phi_1(x)+i\phi_2(x)\\
   \phi_3(x)+i\phi_4(x)
\end{array}\right),~\tilde{\Phi}(x)=i\tau^2\Phi^{\ast}(x)=\frac{1}{\sqrt{2}}\left(\begin{array}{c}
   \phi_3(x)-i\phi_4(x)\\
  -\phi_1(x)+i\phi_2(x)
\end{array}\right),~
\end{eqnarray}
where $\Phi(x)$ is a doublet of $\mathrm{SU}(2)_L$. $\Lambda$ is a
constant with mass dimension. The factor $\frac{L^{\prime}}{L}$
emerges because our metric Ansatz Eq.~(\ref{ansatz-model}) has the
conformal form. If we employ the Gauss normal coordinates for the
sub 5D metric, the factor $\frac{L^{\prime}}{L}$ is superfluous. The
parameters $f_{\mathcal{U}}$ and $f_{\mathcal{D}}$ are real numbers
and dimensionless; while $\beta_{\mathcal{U}}$ and
$\beta_{\mathcal{D}}$ can be complex numbers generally. We also
introduce the terms after $\beta_{\mathcal{U}}$ and
$\beta_{\mathcal{D}}$, which break the parity symmetry of the
action. As we will discuss in following subsections, these terms are
the origin of CP violation. If we drop these terms, the effective 4D
actions will be CP invariant measured by the Jarlskog invariant
measure.

Note that the Yukawa interactions are different from the
conventional brane Yukawa interactions
\begin{eqnarray}
\label{yukawa-D-brane-1} S_{\mathcal{D}\mathrm{Brane}}&=&\int d^4 x
dy dz \sqrt{-g_{\mathrm{vis}}}
\left\{\frac{f_{\mathcal{D}}}{\Lambda}[\bar{\mathbb{Q}}\mathcal{D}+
\beta_{\mathcal{D}}\bar{\mathbb{Q}}\Gamma^7\mathcal{D}]\Phi(x)\frac{L^{\prime}}{L}\delta(y-L^{\prime})\delta(z-R)\right\}\nonumber\\
&+&\mathrm{H.C.},\\
\label{yukawa-U-brane-1} S_{\mathcal{U}\mathrm{Brane}}&=&\int d^4 x
dy dz \sqrt{-g_{\mathrm{vis}}}
\left\{\frac{f_{\mathcal{U}}}{\Lambda}[\bar{\mathbb{Q}}\mathcal{U}+
\beta_{\mathcal{U}}\bar{\mathbb{Q}}\Gamma^7\mathcal{U}]\tilde{\Phi}(x)\frac{L^{\prime}}{L}\delta(y-L^{\prime})\delta(z-R)\right\}\nonumber\\
&+&\mathrm{H.C.}.
\end{eqnarray}
We adopt interactions (\ref{yukawa-D-brane}) and
(\ref{yukawa-U-brane}) instead of the brane interactions
(\ref{yukawa-D-brane-1}) and (\ref{yukawa-U-brane-1}). The reasons
are as follows. By the numerical method in section \ref{sec:3.3}, we
found that interactions (\ref{yukawa-D-brane-1}) and
(\ref{yukawa-U-brane-1}) lead to a massless fermion in the ZMA
approach in the 4D effective theories. While interactions
(\ref{yukawa-D-brane}) and (\ref{yukawa-U-brane}) can lead to a
small but non-zero mass fermion. So we regard interactions
(\ref{yukawa-D-brane}) and (\ref{yukawa-U-brane}) as more plausible
choices.

The Higgs field is confined on the 3-brane. Its action is given by
the brane action
\begin{eqnarray}
\label{higgs-action} S_{\mathrm{Higgs}}=\int d^4
x\sqrt{-g_{\mathrm{vis}}}\left\{g_{\mathrm{vis}}^{\mu\nu}D_{\mu}\Phi^{\dagger}D_{\nu}\Phi
-\mu_0(\Phi^{\dagger}\Phi-v_0^2)^2\right\},
\end{eqnarray}
where
$D_{\mu}=\partial_\mu+i\hat{g}W^j_\mu(x)\frac{\sigma^j}{2}+ig^{\prime}B_\mu(x)$
is the gauge covariant derivative.

Now we have completed the model building in the 6D bulk. From the
above, we know that the gauge field and Higgs field contents are the
same with that in SM. The gauge-fixing terms and the ghost fields
can be introduced as in SM. We omit them in this paper.

\subsection{4D effective theories from the 6D bulk model}\label{sec:3.2}
We constructed the 6D bulk model in the last subsection. In this
subsection, we plan to derive 4D effective theories from the 6D bulk
model in subsection \ref{sec:3.1} by KK decompositions. In that 6D
bulk model, the gauge fields and the Higgs fields are confined on
the brane; while the fermion fields propagate in the bulk. So we
only need to reduce the fermion fields from 6D to 4D. The process of
reducing fermion fields and relative problems have been discussed in
details in section \ref{sec:2}. In this subsection, we derive 4D
effective theories following discussions in section \ref{sec:2}. In
subsection \ref{sec:3.2.1}, we give general results when we reduce
the fermion actions from 6D to 5D by KK decompositions. In
subsection \ref{sec:3.2.2}, we discuss the special metric
Eq.~(\ref{suppose1-model}). In this example, as we analyzed in
subsection \ref{sec:2.3}, we can obtain 3 families fermions in 5D by
adjusting parameters in the metric. In subsection \ref{sec:3.2.3},
we further reduce the actions from 5D to 4D. By this step, we can
obtain zero modes in 4D. These zero modes produce the 3 family
fermions in SM by coupling with the Higgs fields on the brane.

\subsubsection{General discussions about KK decompositions from 6D to 5D}\label{sec:3.2.1}

In this subsection, we follow discussions in subsection
\ref{sec:2.1}. According to Eq.~(\ref{solution9}), we expand the 6D
fields $\mathbb{U}$, $\mathbb{D}$, $\mathcal{U}$ and $\mathcal{D}$
with 5D fields as
\begin{eqnarray}
\label{KK-Q-1} \mathbb{U}&=&\left(
\begin{array}{c}
\chi_{\mathbb{U}}^{1}\\
\chi_{\mathbb{U}}^{2}
\end{array}\right),
\chi_{\mathbb{U}}^{1}(x^\mu,y,z)= \sum_n
\widehat{F}^{\mathbb{Q}}_n(z) U_{n}
(x^\mu,y),\chi_{\mathbb{U}}^{2}(x^\mu,y,z)= \sum_n
\widehat{G}^{\mathbb{Q}}_n(z) U_{n} (x^\mu,y),\\
\label{KK-Q-2} \mathbb{D}&=&\left(
\begin{array}{c}
\chi_{\mathbb{D}}^{1}\\
\chi_{\mathbb{D}}^{2}
\end{array}\right),
\chi_{\mathbb{D}}^{1}(x^\mu,y,z)= \sum_n
\widehat{F}^{\mathbb{Q}}_n(z) D_{n}
(x^\mu,y),\chi_{\mathbb{D}}^{2}(x^\mu,y,z)= \sum_n
\widehat{G}^{\mathbb{Q}}_n(z)D_{n} (x^\mu,y),\\
\label{KK-U} \mathcal{U}&=&\left(
\begin{array}{c}
\chi_{\mathcal{U}}^{1}\\
\chi_{\mathcal{U}}^{2}
\end{array}\right),
\chi_{\mathcal{U}}^{1}(x^\mu,y,z)= \sum_n
\widehat{F}^{\mathcal{U}}_n(z) \mathscr{U}_{n}
(x^\mu,y),\chi_{\mathcal{U}}^{2}(x^\mu,y,z)= \sum_n
\widehat{G}^{\mathcal{U}}_n(z) \mathscr{U}_{n} (x^\mu,y),\\
\label{KK-D} \mathcal{D}&=&\left(
\begin{array}{c}
\chi_{\mathcal{D}}^{1}\\
\chi_{\mathcal{D}}^{2}
\end{array}\right),
\chi_{\mathcal{D}}^{1}(x^\mu,y,z)= \sum_n
\widehat{F}^{\mathcal{D}}_n(z) \mathscr{D}_{n}
(x^\mu,y),\chi_{\mathcal{D}}^{2}(x^\mu,y,z)= \sum_n
\widehat{G}^{\mathcal{D}}_n(z) \mathscr{D}_{n} (x^\mu,y).~~
\end{eqnarray}
Note that we expand $\mathbb{U}$ and $\mathbb{D}$ with the same
functions $\widehat{F}^{\mathbb{Q}}_n(z)$ and
$\widehat{G}^{\mathbb{Q}}_n(z)$. Because they are in the same
doublet $\mathbb{Q}$ of $\mathrm{SU}(2)_L$, they have the same bulk
mass parameters by Eq.~(\ref{action-Q}). So they have the same
expanding functions. These expanding modes are determined by
equations
\begin{eqnarray}
\label{modes-6d-1}
\left(\frac{d}{d z}- m_{\Psi} B\right)F^{\Psi}_n(z) + (\lambda^{\Psi}_n-\mathcal{M}_{\Psi}) G^{\Psi}_n(z)=0,\\
\label{modes-6d-2} \left(\frac{d}{d z}+ m_{\Psi} B
\right)G^{\Psi}_n(z) - (\lambda^{\Psi}_n-\mathcal{M}_{\Psi})
F^{\Psi}_n(z)=0,
\end{eqnarray}
where $\Psi$ can stand for $\mathbb{Q}$, $\mathcal{U}$ or
$\mathcal{D}$, and we have employed the definitions
\begin{eqnarray}
\label{def-6d}
\widehat{F}^{\Psi}_n(z)=B(z)^{-5/2}F^{\Psi}_n(z),~\widehat{G}^{\Psi}_n(z)=B(z)^{-5/2}
G^{\Psi}_n(z).
\end{eqnarray}
By these expanding, the 6D action (\ref{action-Q}) for $\mathbb{Q}$
is reduced to be
\begin{eqnarray}
\label{action-Q-5d-1} S_{\mathbb{Q}} &=&\int d^4 x
dy~K^{\mathbb{Q}}_{mn}\left\{\frac{i}{2}
A^4\left[\bar{Q}_m\gamma^5\partial_5 Q_n-\partial_5\bar{Q}_m\gamma^5
Q_n+\bar{Q}_m\gamma^{\mu}\partial_{\mu}Q_n-\partial_{\mu}\bar{Q}_m\gamma^{\mu}Q_n\right]\right\}\nonumber\\
&-&\int d^4 x dy~M^{\mathbb{Q}}_{mn}A^5 i\bar{Q}_m {Q}_n,
~~~{Q}_n=\left(\begin{array}{c}
{U}_n\\
{D}_n
\end{array}\right),\\
\label{action-Q-5d-2} K^{\mathbb{Q}}_{mn}&=&\int dz
\left({F}^{\mathbb{Q}\ast}_{m}{F}^{\mathbb{Q}}_n+{G}^{\mathbb{Q}\ast}_{m}{G}^{\mathbb{Q}}_n
\right),~~M^{\mathbb{Q}}_{mn}=\int dz
\left[\left({F}_{m}^{\mathbb{Q}\ast}{F}^\mathbb{Q}_n+{G}_{m}^{\mathbb{Q}\ast}{G}^\mathbb{Q}_n\right)\frac{\lambda_{m}^{\mathbb{Q}\ast}+\lambda^{\mathbb{Q}}_n}{2}\right],
\end{eqnarray}
where we have combined $U_n$ and $D_n$ into a doublet $Q_n$, because
they have the same expanding functions. The action (\ref{action-U})
for $\mathcal{U}$ is reduced to be
\begin{eqnarray}
\label{action-U-5d-1} S_{\mathcal{U}} &=&\int d^4 x
dy~K^{\mathcal{U}}_{mn}\left\{\frac{i}{2}
A^4\left[\bar{\mathscr{U}}_m\gamma^5\partial_5
\mathscr{U}_n-\partial_5\bar{\mathscr{U}}_m\gamma^5
\mathscr{U}_n+\bar{\mathscr{U}}_m\gamma^{\mu}\partial_{\mu}\mathscr{U}_n-\partial_{\mu}\bar{\mathscr{U}}_m\gamma^{\mu}\mathscr{U}_n\right]\right\}\nonumber\\
&-&\int d^4 x dy~M^{\mathcal{U}}_{mn}A^5 i\bar{\mathscr{U}}_m {\mathscr{U}}_n, \\
\label{action-U-5d-2} K^{\mathcal{U}}_{mn}&=&\int dz
\left({F}^{\mathcal{U}\ast}_{m}{F}^{\mathcal{U}}_n+{G}^{\mathcal{U}\ast}_{m}{G}^{\mathcal{U}}_n
\right),~~M^{\mathcal{U}}_{mn}=\int dz
\left[\left({F}_{m}^{\mathcal{U}\ast}{F}^\mathcal{U}_n+{G}_{m}^{\mathcal{U}\ast}{G}^\mathcal{U}_n\right)\frac{\lambda_{m}^{\mathcal{U}\ast}+\lambda^{\mathcal{U}}_n}{2}\right].
\end{eqnarray}
The action (\ref{action-D}) for $\mathcal{D}$ is reduced to be
\begin{eqnarray}
\label{action-D-5d-1} S_{\mathcal{D}} &=&\int d^4 x
dy~K^{\mathcal{D}}_{mn}\left\{\frac{i}{2}
A^4\left[\bar{\mathscr{D}}_m\gamma^5\partial_5
\mathscr{D}_n-\partial_5\bar{\mathscr{D}}_m\gamma^5
\mathscr{D}_n+\bar{\mathscr{D}}_m\gamma^{\mu}\partial_{\mu}\mathscr{D}_n-\partial_{\mu}\bar{\mathscr{D}}_m\gamma^{\mu}\mathscr{D}_n\right]\right\}\nonumber\\
&-&\int d^4 x dy~M^{\mathcal{D}}_{mn}A^5 i\bar{\mathscr{D}}_m {\mathscr{D}}_n, \\
\label{action-D-5d-2} K^{\mathcal{D}}_{mn}&=&\int dz
\left({F}^{\mathcal{D}\ast}_{m}{F}^{\mathcal{D}}_n+{G}^{\mathcal{D}\ast}_{m}{G}^{\mathcal{D}}_n
\right),~~M^{\mathcal{D}}_{mn}=\int dz
\left[\left({F}_{m}^{\mathcal{D}\ast}{F}^\mathcal{D}_n+{G}_{m}^{\mathcal{D}\ast}{G}^\mathcal{D}_n\right)\frac{\lambda_{m}^{\mathcal{D}\ast}+\lambda^{\mathcal{D}}_n}{2}\right].
\end{eqnarray}

Now we consider the interaction sectors under the KK decompositions.
For the gauge interaction sector (\ref{gauge-int-Q}),
(\ref{gauge-int-U}) and (\ref{gauge-int-D}), by the expanding in
equations (\ref{KK-Q-1})-(\ref{KK-D}), we obtain
\begin{eqnarray}
\label{gauge-int-Q-5d} S_{\mathbb{Q}\mathrm{int}}&=&-\int d^4 x dy
A(y)^4 K^{\mathbb{Q}}_{mn}\left\{\bar{Q}_m\gamma^{\mu}\bigg{[}\hat{g}W^j_\mu(x)\frac{\sigma^j}{2}+g^{\prime}B_\mu(x)\bigg{]}Q_n\right\},\\
\label{gauge-int-U-5d} S_{\mathcal{U}\mathrm{int}}&=&-\int d^4 x dy
A(y)^4 K^{\mathcal{U}}_{mn}\left\{\bar{\mathscr{U}}_m\gamma^{\mu}[g^{\prime}B_\mu(x)]\mathscr{U}_n\right\},\\
\label{gauge-int-D-5d} S_{\mathcal{D}\mathrm{int}}&=&-\int d^4 x dy
A(y)^4
K^{\mathcal{D}}_{mn}\left\{\bar{\mathscr{D}}_m\gamma^{\mu}[g^{\prime}B_\mu(x)]\mathscr{D}_n\right\},
\end{eqnarray}
where matrices $K^\mathbb{Q}$, $K^\mathcal{U}$ and $K^\mathcal{D}$
have been defined by equations (\ref{action-Q-5d-2}),
(\ref{action-U-5d-2}) and (\ref{action-D-5d-2}) respectively. For
the the Yukawa interaction sector (\ref{yukawa-D-brane}) and
(\ref{yukawa-U-brane}), after the KK decompositions, we obtain
\begin{eqnarray}
\label{yukawa-D-brane-5d} S_{\mathcal{D}\mathrm{Yukawa}}&=&\int d^4
x dy A(y)^5\left\{\frac{f_{\mathcal{D}}}{\Lambda}[{\mathcal
{Y}}^{D}_{mn}+
i\beta_{\mathcal{D}}Y^{D}_{mn}]\bar{Q}_m\mathscr{D}_n\Phi(x)\frac{L^{\prime}}{L}\delta(y-L^{\prime})+\mathrm{H.C.}\right\},\\
\label{yukawa-D-brane-5d-M} {\mathcal {Y}}^{D}_{mn}&=&\int dz
B(z)\left({F}^{\mathbb{Q}\ast}_{m}{G}^{\mathcal{D}}_n+{G}^{\mathbb{Q}\ast}_{m}{F}^{\mathcal{D}}_n\right),Y^{D}_{mn}=\int
dz
B(z)\left({F}^{\mathbb{Q}\ast}_{m}{F}^{\mathcal{D}}_n-{G}^{\mathbb{Q}\ast}_{m}{G}^{\mathcal{D}}_n\right),\\
\label{yukawa-U-brane-5d} S_{\mathcal{U}\mathrm{Yukawa}}&=&\int d^4
x dy
A(y)^5\left\{\frac{f_{\mathcal{U}}}{\Lambda}[{\mathcal{Y}}^{U}_{mn}+
i\beta_{\mathcal{U}}Y^{U}_{mn}]\bar{Q}_m\mathscr{U}_n\tilde{\Phi}(x)\frac{L^{\prime}}{L}\delta(y-L^{\prime})+\mathrm{H.C.}\right\},\\
\label{yukawa-U-brane-5d-M} {\mathcal{Y}}^{U}_{mn}&=&\int dz
B(z)\left({F}^{\mathbb{Q}\ast}_{m}{G}^{\mathcal{U}}_n+{G}^{\mathbb{Q}\ast}_{m}{F}^{\mathcal{U}}_n\right),
Y^{U}_{mn}=\int dz
B(z)\left({F}^{\mathbb{Q}\ast}_{m}{F}^{\mathcal{U}}_n-{G}^{\mathbb{Q}\ast}_{m}{G}^{\mathcal{U}}_n\right).~~
\end{eqnarray}

In the above, we obtain 5D effective actions by KK decompositions.
We have not consider the concrete form of the metric $B(z)$. In next
subsection, we choose the  metric $B(z)$ to be that in
Eq.~(\ref{suppose1-model}), and discuss the further simplifications
of the above 5D effective actions.

\subsubsection{Deriving 5D effective theories for finite families}\label{sec:3.2.2}
In subsection \ref{sec:3.2.1}, we obtain the 5D effective actions
for a general metric $B(z)$. In this subsection, we plan to obtain
5D effective actions which include only finite KK modes. As we
analyzed in section \ref{sec:2}, one can obtain finite KK modes by
choose a special form of the metric $B(z)$. In this paper, we choose
the metric $B(z)$ to be that in Eq.~(\ref{suppose1-model}). This
metric has been analyzed in subsection \ref{sec:2.3} and in our
previous paper \cite{Guo:2008}. The results are that we can cut off
the hypergeometrical series by requiring that it is finite when
$z\rightarrow\infty$. This requirement determines the solutions
uniquely up to the normalization constants. So we do not have
freedom to imposing boundary conditions at the other boundary $z=R$.
This implies that the normalization conditions Eq.~(\ref{norm2a})
are not be satisfied, and we must change to the case~(II) in
subsection \ref{sec:2.1}. In this case, we should redefine the
fermion fields to obtain the conventional 5D effective actions like
that in Eq.~(\ref{action3a}). As in subsection \ref{sec:2.1}, we can
make these redefinitions in two steps.

Step~(I): At this step, we analyze the kinetic terms of fermion
actions. As the step~(1) in subsection \ref{sec:2.1}, we make the
Cholesky decompositions for matrices $K$ in the kinetic terms as
follows,
\begin{eqnarray}
\label{cholesky-decomp}
K^{\mathbb{Q}}=H^{\mathbb{Q}\dagger}H^{\mathbb{Q}},~~K^{\mathcal
{U}}=H^{\mathcal {U}\dagger}H^{\mathcal {U}},~~K^{\mathcal
{D}}=H^{\mathcal {D}\dagger}H^{\mathcal {D}}.
\end{eqnarray}
One can make Cholesky decomposition for matrix $K$ only when $K$ is
a positive-definite hermitian matrix. In the numerical examples in
subsection \ref{sec:3.3.1} and subsection \ref{sec:3.3.2}, this
condition is satisfied. Redefine the fermion fields as
\begin{eqnarray}
\label{redef-5d}
\widetilde{Q}_m=H^{\mathbb{Q}}_{mn}Q_n,~~\widetilde{\mathscr{U}}_m=H^{\mathcal
{U}}_{mn}\mathscr{U}_n,~~\widetilde{\mathscr{D}}_m=H^{\mathcal
{D}}_{mn}\mathscr{D}_n.
\end{eqnarray}
In these new basis, the kinetic terms become the conventional ones
similar to that of Eq.~(\ref{action3a}); while the mass terms are
modified to be
\begin{eqnarray}
\label{mass-term-redef-1}
\widetilde{M}^{\mathbb{Q}}=(H^{\mathbb{Q}-1})^{\dagger}M
H^{\mathbb{Q}-1},~~\widetilde{M}^{\mathcal{U}}=(H^{\mathcal
{U}-1})^{\dagger}M H^{\mathcal {U}-1},~~\widetilde{M}^{\mathcal
{D}}=(H^{\mathcal {D}-1})^{\dagger}M H^{\mathcal {D}-1}.
\end{eqnarray}
The fermions actions in the new basis are given by
\begin{eqnarray}
\label{action-psi-5d-2} S_{\mathcal{Q}} &=&\int d^4 x
dy~\left\{\frac{i}{2}
A^4\delta_{mn}\left[\bar{\widetilde{\Psi}}_m\gamma^5\partial_5
\widetilde{\Psi}_n-\partial_5\bar{\widetilde{\Psi}}_m\gamma^5
\widetilde{\Psi}_n+\bar{\widetilde{\Psi}}_m\gamma^{\mu}\partial_{\mu}\widetilde{\Psi}_n-\partial_{\mu}\bar{\widetilde{\Psi}}_m\gamma^{\mu}\widetilde{\Psi}_n\right]\right\}\nonumber\\
&-&\int d^4 x dy~\widetilde{M}^{\mathcal{Q}}_{mn}A^5
i\bar{\widetilde{\Psi}}_m {\widetilde{\Psi}}_n,
\end{eqnarray}
where $\mathcal{Q}$ can be $\mathbb{Q}$, $\mathcal{U}$ or $\mathcal
{D}$; while the corresponding $\Psi$ can stand for $Q$,
$\mathscr{U}$ or $\mathscr{D}$ respectively.

For the gauge interaction sector (\ref{gauge-int-Q-5d}),
(\ref{gauge-int-U-5d}) and (\ref{gauge-int-D-5d}), by the
definitions (\ref{redef-5d}), we obtain
\begin{eqnarray}
\label{gauge-int-Q-5d-1} S_{\mathbb{Q}\mathrm{int}}&=&-\int d^4 x dy
A(y)^4 \left\{\bar{\widetilde{Q}}_n\gamma^{\mu}\bigg{[}\hat{g}W^j_\mu(x)\frac{\sigma^j}{2}+g^{\prime}B_\mu(x)\bigg{]}\widetilde{Q}_n\right\},\\
\label{gauge-int-U-5d-1} S_{\mathcal{U}\mathrm{int}}&=&-\int d^4 x
dy
A(y)^4 \left\{\bar{\widetilde{\mathscr{U}}}_n\gamma^{\mu}[g^{\prime}B_\mu(x)]\widetilde{\mathscr{U}}_n\right\},\\
\label{gauge-int-D-5d-1} S_{\mathcal{D}\mathrm{int}}&=&-\int d^4 x
dy A(y)^4
\left\{\bar{\widetilde{\mathscr{D}}}_n\gamma^{\mu}[g^{\prime}B_\mu(x)]\widetilde{\mathscr{D}}_n\right\},
\end{eqnarray}
where the index $n$ is summed. Like matrices in the kinetic terms,
the matrices in this sector become identities by the field
redefinitions in Eq.~(\ref{redef-5d}).

For the Yukawa interaction sector, after redefinitions
(\ref{redef-5d}), we obtain
\begin{eqnarray}
\label{yukawa-D-brane-5d-1} S_{\mathcal{D}\mathrm{Yukawa}}&=&\int
d^4 x dy
A(y)^5\left\{\frac{f_{\mathcal{D}}}{\Lambda}[\widetilde{{\mathcal
{Y}}}^{D}_{mn}+
i\beta_{\mathcal{D}}\widetilde{Y}^{D}_{mn}]\bar{\widetilde{Q}}_m\widetilde{\mathscr{D}}_n\Phi(x)\frac{L^{\prime}}{L}\delta(y-L^{\prime})+\mathrm{H.C.}\right\},\\
\label{yukawa-D-brane-5d-M-1} \widetilde{{\mathcal
{Y}}}^{D}&=&(H^{\mathbb{Q}-1})^{\dagger}{\mathcal
{Y}}^{D}H^{\mathcal {D}-1},~~\widetilde{Y}^{D}=(H^{\mathbb{Q}-1})^{\dagger}Y^{D}H^{\mathcal {D}-1},\\
\label{yukawa-U-brane-5d-1} S_{\mathcal{U}\mathrm{Yukawa}}&=&\int
d^4 x dy
A(y)^5\left\{\frac{f_{\mathcal{U}}}{\Lambda}[\widetilde{{\mathcal
{Y}}}^{U}_{mn}+
i\beta_{\mathcal{U}}\widetilde{Y}^{U}_{mn}]\bar{\widetilde{Q}}_m\widetilde{\mathscr{U}}_n\tilde{\Phi}(x)\frac{L^{\prime}}{L}\delta(y-L^{\prime})+\mathrm{H.C.}\right\},\\
\label{yukawa-U-brane-5d-M-1} \widetilde{{\mathcal
{Y}}}^{U}&=&(H^{\mathbb{Q}-1})^{\dagger}{\mathcal
{Y}}^{U}H^{\mathcal
{U}-1},~~\widetilde{Y}^{U}=(H^{\mathbb{Q}-1})^{\dagger}Y^{U}H^{\mathcal
{U}-1}.
\end{eqnarray}
The Yukawa interaction matrices are modified by the field
redefinitions Eq.~(\ref{redef-5d}).

In the above, we have completed the first step. This step is to make
the kinetic terms of fermion actions to be the conventional ones.
The mass matrices and the interaction sectors are modified
accordingly. Especially, the gauge interaction sector becomes the
flavor universal one, which is important to ensure the unitarity of
the mixing matrix in the ZMA approach.

Step~(II): At the second step, we diagonalize the mass marix in the
action (\ref{action-psi-5d-2}). These matrices are hermitian, as
they are defined in Eq.~(\ref{mass-term-redef-1}). They are
diagonalized as
\begin{eqnarray}
\label{diagQ-5d-2}
\widetilde{M}^{\mathbb{Q}}&=&U^{\mathbb{Q}\dagger}\Delta_{\mathbb{Q}}
U^{\mathbb{Q}},~~
\Delta_{\mathbb{Q}}=\mathrm{diag}(\hat{\lambda}^{\mathbb{Q}}_1,\hat{\lambda}^{\mathbb{Q}}_2,\cdots,\hat{\lambda}^{\mathbb{Q}}_n),\\
\label{diagU-5d-2}
\widetilde{M}^{\mathcal{U}}&=&U^{\mathcal{U}\dagger}\Delta_{\mathcal{U}}
U^{\mathcal{U}},~~
\Delta_{\mathcal{U}}=\mathrm{diag}(\hat{\lambda}^{\mathcal{U}}_1,\hat{\lambda}^{\mathcal{U}}_2,\cdots,\hat{\lambda}^{\mathcal{U}}_n),\\
\label{diagD-5d-2}
\widetilde{M}^{\mathcal{D}}&=&U^{\mathcal{D}\dagger}\Delta_{\mathcal{D}}
U^{\mathcal{D}},~~
\Delta_{\mathcal{D}}=\mathrm{diag}(\hat{\lambda}^{\mathcal{D}}_1,\hat{\lambda}^{\mathcal{D}}_2,\cdots,\hat{\lambda}^{\mathcal{D}}_n).
\end{eqnarray}
Redefining the fields in Eq.~(\ref{action-psi-5d-2}) as
\begin{eqnarray}
\label{field-redef-5d-2}
\widehat{Q}_m=U^{\mathbb{Q}}_{mn}\widetilde{Q}_n,~~\widehat{\mathscr{U}}_m=U^{\mathcal{U}}_{mn}\widetilde{\mathscr{U}}_n,~~
\widehat{\mathscr{D}}_m=U^{\mathcal{D}}_{mn}\widetilde{\mathscr{D}}_n.
\end{eqnarray}
These transformations are unitary. So the kinetic terms keep
invariant; while the mass terms become the diagonal ones. By these
transformations, the action (\ref{action-psi-5d-2}) becomes the
conventional one
\begin{eqnarray}
\label{action-psi-5d-3} S_{\mathcal{Q}} &=&\sum_n\int d^4 x
dy~\left\{\frac{i}{2}
A^4\left[\bar{\widehat{\Psi}}_n\gamma^5\partial_5
\widehat{\Psi}_n-\partial_5\bar{\widehat{\Psi}}_n\gamma^5
\widehat{\Psi}_n+\bar{\widehat{\Psi}}_n\gamma^{\mu}\partial_{\mu}\widehat{\Psi}_n-\partial_{\mu}\bar{\widehat{\Psi}}_n\gamma^{\mu}\widehat{\Psi}_n\right]\right\}\nonumber\\
&-&\sum_n\int d^4 x dy~\hat{\lambda}^{\mathcal{Q}}_{n}A^5
i\bar{\widehat{\Psi}}_n {\widehat{\Psi}}_n.
\end{eqnarray}

For the gauge interaction sector, by the unitary transformations
(\ref{field-redef-5d-2}), we obtain
\begin{eqnarray}
\label{gauge-int-Q-5d-2} S_{\mathbb{Q}\mathrm{int}}&=&-\int d^4 x dy
A(y)^4 \left\{\bar{\widehat{Q}}_n\gamma^{\mu}\bigg{[}\hat{g}W^j_\mu(x)\frac{\sigma^j}{2}+g^{\prime}B_\mu(x)\bigg{]}\widehat{Q}_n\right\},\\
\label{gauge-int-U-5d-2} S_{\mathcal{U}\mathrm{int}}&=&-\int d^4 x
dy
A(y)^4 \left\{\bar{\widehat{\mathscr{U}}}_n\gamma^{\mu}[g^{\prime}B_\mu(x)]\widehat{\mathscr{U}}_n\right\},\\
\label{gauge-int-D-5d-2} S_{\mathcal{D}\mathrm{int}}&=&-\int d^4 x
dy A(y)^4
\left\{\bar{\widehat{\mathscr{D}}}_n\gamma^{\mu}[g^{\prime}B_\mu(x)]\widehat{\mathscr{D}}_n\right\}.
\end{eqnarray}
Because the transformations in Eq.~(\ref{field-redef-5d-2}) are
unitary. They keep universal for the flavors still.

For the Yukawa interaction sector, after redefinitions
Eq.~(\ref{field-redef-5d-2}), we obtain
\begin{eqnarray}
\label{yukawa-D-brane-5d-2} S_{\mathcal{D}\mathrm{Yukawa}}&=&\int
d^4 x dy
A(y)^5\left\{\frac{f_{\mathcal{D}}}{\Lambda}[\widehat{{\mathcal
{Y}}}^{D}_{mn}+
i\beta_{\mathcal{D}}\widehat{Y}^{D}_{mn}]\bar{\widehat{Q}}_m\widehat{\mathscr{D}}_n\Phi(x)\frac{L^{\prime}}{L}\delta(y-L^{\prime})+\mathrm{H.C.}\right\},\\
\label{yukawa-D-brane-5d-Y-2} \widehat{{\mathcal
{Y}}}^{D}&=&U^{\mathbb{Q}}(H^{\mathbb{Q}-1})^{\dagger}{\mathcal
{Y}}^{D}H^{\mathcal {D}-1}U^{\mathcal{D}\dagger},~~\widehat{Y}^{D}=
U^{\mathbb{Q}}(H^{\mathbb{Q}-1})^{\dagger}Y^{D}H^{\mathcal{D}-1}U^{\mathcal{D}\dagger},\\
\label{yukawa-U-brane-5d-2} S_{\mathcal{U}\mathrm{Yukawa}}&=&\int
d^4 x dy
A(y)^5\left\{\frac{f_{\mathcal{U}}}{\Lambda}[\widehat{{\mathcal
{Y}}}^{U}_{mn}+
i\beta_{\mathcal{U}}\widehat{Y}^{U}_{mn}]\bar{\widehat{Q}}_m\widehat{\mathscr{U}}_n\tilde{\Phi}(x)\frac{L^{\prime}}{L}\delta(y-L^{\prime})+\mathrm{H.C.}\right\},\\
\label{yukawa-U-brane-5d-Y-2} \widehat{{\mathcal
{Y}}}^{U}&=&U^{\mathbb{Q}}(H^{\mathbb{Q}-1})^{\dagger}{\mathcal
{Y}}^{U}H^{\mathcal
{U}-1}U^{\mathcal{U}\dagger},~~\widehat{Y}^{U}=U^{\mathbb{Q}}(H^{\mathbb{Q}-1})^{\dagger}Y^{U}H^{\mathcal
{U}-1}U^{\mathcal{U}\dagger}.
\end{eqnarray}
The interaction matrices are modified by the unitary transformations
(\ref{field-redef-5d-2}).

Now we complete the second step. This step makes the fermion mass
terms to be diagonal ones. After this second step, we obtain the
conventional 5D effective fermion action (\ref{action-psi-5d-3}).
The interaction sectors are modified by unitary transformations
(\ref{field-redef-5d-2}) accordingly. The gauge interaction sector
are still universal for flavors after this step.

We make some summaries about this subsection. By choosing the
parameters in the metric, we can obtain finite KK modes. Because of
this requirement, we must consider the normalization conditions
case~(II) in subsection \ref{sec:2.1}. However, through twice
redefinitions of fermion fields, we can also obtain the conventional
5D effective fermion actions. Having obtained the 4D effective
actions for finite KK modes, we can derive 4D effective actions from
these 5D actions in the next subsection.

\subsubsection{4D effective theories from 5D effective theories}\label{sec:3.2.3}

In this subsection, we further reduce the actions from 5D to 4D by
KK decompositions.

We begin with the 5D effective actions obtained in last subsection.
As that in subsection \ref{sec:2.4}, we expand the 5D fields with
the 4D fields as follows
\begin{eqnarray}
\label{expand-4d-Q-u} \widehat{U}_n(x,y)&=&\sum_j
A^{-2}(y)\left[f^{Q(j)}_{n,L}(y)u^{(j)}_{n,L}(x)+f^{Q(j)}_{n,R}(y)u^{(j)}_{n,R}(x)\right],\\
\label{expand-4d-Q-d} \widehat{D}_n(x,y)&=&\sum_j
A^{-2}(y)\left[f^{Q(j)}_{n,L}(y)d^{(j)}_{n,L}(x)+f^{Q(j)}_{n,R}(y)d^{(j)}_{n,R}(x)\right],\\
\label{expand-4d-u} \widehat{\mathscr{U}}_n(x,y)&=&\sum_j
A^{-2}(y)\left[f^{\mathscr{U}(j)}_{n,L}(y)u^{(j)}_{n,L}(x)+f^{\mathscr{U}(j)}_{n,R}(y)u^{(j)}_{n,R}(x)\right],\\
\label{expand-4d-d} \widehat{\mathscr{D}}_n(x,y)&=&\sum_j
A^{-2}(y)\left[f^{\mathscr{D}(j)}_{n,L}(y)d^{(j)}_{n,L}(x)+f^{\mathscr{D}(j)}_{n,R}(y)d^{(j)}_{n,R}(x)\right].
\end{eqnarray}
We give some interpretations about these expanding here. In the
above expanding, the superscript $j$ stands for different KK modes,
while the subscript $n$ can be interpreted as the family index. Note
that they are not summed. We have expanded $\widehat{U}_n(x,y)$ and
$\widehat{D}_n(x,y)$ with the same functions $f^{Q(j)}_{n,L}(y)$ and
$f^{Q(j)}_{n,R}(y)$, because they have the same bulk mass parameters
as shown in last subsection. As in subsection \ref{sec:2.4}, we
require that these expanding functions satisfy equations
\begin{eqnarray}
\label{KK5-1-model}
\left(\frac{d}{d y}+ \hat{\lambda}^{\mathcal{Q}}_n A \right)f_{n,L}^{\Psi(j)}(y)-\mathbbm{m}_n^{\Psi(j)} f_{n,R}^{\Psi(j)}(y)=0,\\
\label{KK5-2-model} \left(\frac{d}{d
y}-\hat{\lambda}^{\mathcal{Q}}_n A
\right)f_{n,R}^{\Psi(j)}(y)+\mathbbm{m}_n^{\Psi(j)}
f_{n,L}^{\Psi(j)}(y)=0,
\end{eqnarray}
where $\mathcal{Q}$ can be $\mathbb{Q}$, $\mathcal{U}$ or $\mathcal
{D}$ as in last subsection; while $\Psi$ stands for $Q$,
$\mathscr{U}$ or $\mathscr{D}$ accordingly. Note that $n$ can be
regarded as the family index here and is not summed. For functions
$f^{Q(j)}_{n,L}(y)$ and $f^{Q(j)}_{n,R}(y)$, we designate the
boundary conditions as in Eq.~(\ref{KK5-1-bound})
\begin{eqnarray}
\label{KK5-1-bound-model} \left(\frac{d}{d y}+
\hat{\lambda}^{\mathbb{Q}}_n A
\right)f_{n,L}^{Q(j)}(y)=0,~~f_{n,R}^{Q(j)}(y)=0,~~\mathrm{at}~~y=L~~\mathrm{and}~~L^{\prime}.
\end{eqnarray}
These boundary conditions save left-handed zero modes while kill
right-handed ones. These left-handed zero modes make the doublet of
$\mathrm{SU}(2)_L$. For functions $f^{\mathscr{U}(j)}_{n,L(R)}(y)$
and $f^{\mathscr{D}(j)}_{n,L(R)}(y)$, we designate the boundary
conditions as in Eq.~(\ref{KK5-2-bound})
\begin{eqnarray}
\label{KK5-2-bound-model-u} \left(\frac{d}{d
y}-\hat{\lambda}^{\mathcal{U}}_n A
\right)f_{n,R}^{\mathscr{U}(j)}(y)=0,~~f_{n,L}^{\mathscr{U}(j)}(y)=0,~~\mathrm{at}~~y=L~~
\mathrm{and}~~L^{\prime},\\
\label{KK5-2-bound-model-d} \left(\frac{d}{d
y}-\hat{\lambda}^{\mathcal{D}}_n A
\right)f_{n,R}^{\mathscr{D}(j)}(y)=0,~~f_{n,L}^{\mathscr{D}(j)}(y)=0,~~\mathrm{at}~~y=L~~
\mathrm{and}~~L^{\prime}.
\end{eqnarray}
These boundary conditions save right-handed zero modes while kill
left-handed ones. These right-handed zero modes make the singlets of
$\mathrm{SU}(2)_L$. As we discussed in subsection \ref{sec:2.4},
these boundary conditions ensure that the expanding functions
satisfy following normalization conditions
\begin{eqnarray}
\label{KK5-1-massive-norm-model} \int_L^{L^{\prime}}
dy\left(f_{n,L}^{\Psi (i)\ast}f_{n,L}^{\Psi
(j)}\right)=\delta^{ij},~~\int_L^{L^{\prime}} dy\left(f_{n,R}^{\Psi
(i)\ast}f_{n,R}^{\Psi (j)}\right)=\delta^{ij},
\end{eqnarray}
where $\Psi$ stands for $Q$, $\mathscr{U}$ or $\mathscr{D}$. Note
that $n$ is the family index and is not summed.

By the above expanding (\ref{expand-4d-Q-u})-(\ref{expand-4d-d}) and
the normalization conditions (\ref{KK5-1-massive-norm-model}), the
fermion action (\ref{action-psi-5d-3}) becomes
\begin{eqnarray}
\label{action-psi-4d} S_{\psi} &=&\sum_{n,j}\int d^4 x
~\left\{\frac{i}{2}
\left[\bar{\psi}_n^{(j)}\gamma^{\mu}\partial_{\mu}\psi_n^{(j)}-\partial_{\mu}\bar{\psi}_n^{(j)}\gamma^{\mu}\psi_n^{(j)}\right]-i\mathbbm{m}_n^{\psi(j)}\bar{\psi}_n^{(j)}
{\psi}_n^{(j)}\right\},
\end{eqnarray}
where $\psi$ can be $u$ or $d$. This action includes zero modes and
massive modes. The modes $u_n^{(0)}$ and $d_n^{(0)}$ are massless
here. They obtain mass by coupling with Higgs field as in actions
(\ref{yukawa-D-brane-5d-2}) and (\ref{yukawa-U-brane-5d-2}).

For the gauge interaction sector, after the above expanding, we
obtain
\begin{eqnarray}
\label{gauge-int-Q-4d} S_{\mathbb{Q}\mathrm{int}}&=&\int d^4 x
\left\{\bar{Q}_{n,L}^{(0)}\gamma^{\mu}\bigg{[}\hat{g}W^i_\mu(x)\frac{\sigma^i}{2}+g^{\prime}B_\mu(x)\bigg{]}{Q}_{n,L}^{(0)}\right\}\nonumber\\
&+&\sum_{j}\int d^4 x
\left\{\bar{Q}_{n}^{(j)}\gamma^{\mu}\bigg{[}\hat{g}W^i_\mu(x)\frac{\sigma^i}{2}+g^{\prime}B_\mu(x)\bigg{]}{Q}_{n}^{(j)}\right\},~{Q}_{n,L}=\left(\begin{array}{c}
{u}_{n,L}\\
{d}_{n,L}
\end{array}\right),\\
\label{gauge-int-U-4d} S_{\mathcal{U}\mathrm{int}}&=&\int d^4 x
\left\{\bar{u}_{n,R}^{(0)}\gamma^{\mu}[g^{\prime}B_\mu(x)]u_{n,R}^{(0)}\right\}
+\sum_{j}\int d^4 x
\left\{\bar{u}_{n}^{(j)}\gamma^{\mu}[g^{\prime}B_\mu(x)]u_{n}^{(j)}\right\},\\
\label{gauge-int-D-4d} S_{\mathcal{D}\mathrm{int}}&=&\int d^4 x
\left\{\bar{d}_{n,R}^{(0)}\gamma^{\mu}[g^{\prime}B_\mu(x)]d_{n,R}^{(0)}\right\}
+\sum_{j}\int d^4 x
\left\{\bar{d}_{n}^{(j)}\gamma^{\mu}[g^{\prime}B_\mu(x)]d_{n}^{(j)}\right\},
\end{eqnarray}
where $n$ can be regarded as the family index and is summed. In the
above, we have employed the normalization conditions
(\ref{KK5-1-massive-norm-model}). We have omitted three total minus
signs in the above equations. We also isolate zero modes from
massive modes obviously. The gauge interactions of zero modes are
chiral because of the boundary conditions (\ref{KK5-1-bound-model}),
(\ref{KK5-2-bound-model-u}) and (\ref{KK5-2-bound-model-d}); while
the gauge interactions of massive modes are vector-like. We also see
that the gauge interactions are universal for zero modes.

For the Yukawa interaction sector, by the above expanding, we obtain
\begin{eqnarray}
\label{yukawa-D-brane-4d} S_{\mathcal{D}\mathrm{Yukawa}}&=&\int d^4
x dy A\frac{f_{\mathcal{D}}}{\Lambda} [\widehat{{\mathcal
{Y}}}^{D}_{mn}+
i\beta_{\mathcal{D}}\widehat{Y}^{D}_{mn}]\sum_{i,j}\left\{f_{m,L}^{Q
(i)\ast}f_{n,R}^{\mathscr{D}
(j)}\bar{Q}_{m,L}^{(i)}d_{n,R}^{(j)}+f_{m,R}^{Q
(i)\ast}f_{n,L}^{\mathscr{D}
(j)}\bar{Q}_{m,R}^{(i)}d_{n,L}^{(j)}\right\}\nonumber\\
&\times&\Phi(x)\frac{L^{\prime}}{L}\delta(y-L^{\prime})+\mathrm{H.C.},\\
\label{yukawa-U-brane-4d} S_{\mathcal{U}\mathrm{Yukawa}}&=&\int d^4
x dy A\frac{f_{\mathcal{U}}}{\Lambda} [\widehat{{\mathcal
{Y}}}^{U}_{mn}+
i\beta_{\mathcal{U}}\widehat{Y}^{U}_{mn}]\sum_{i,j}\left\{f_{m,L}^{Q
(i)\ast}f_{n,R}^{\mathscr{U}
(j)}\bar{Q}_{m,L}^{(i)}u_{n,R}^{(j)}+f_{m,R}^{Q
(i)\ast}f_{n,L}^{\mathscr{U}
(j)}\bar{Q}_{m,R}^{(i)}u_{n,L}^{(j)}\right\}\nonumber\\
&\times&\tilde{\Phi}(x)\frac{L^{\prime}}{L}\delta(y-L^{\prime})+\mathrm{H.C.},
\end{eqnarray}
where $\widehat{{\mathcal {Y}}}^{D(U)}$ and $\widehat{Y}^{D(U)}$ are
defined as in equations (\ref{yukawa-D-brane-5d-Y-2}) and
(\ref{yukawa-U-brane-5d-Y-2}). Here $m$ and $n$ are the family
indices. They are summed implicitly. These interaction terms include
zero modes and massive modes.

\subsubsection{Mass matrices and mixing matrix}\label{sec:3.2.4}
In the last subsection, we have derived the 4D effective actions
from the 5D ones in subsection \ref{sec:3.2.2}. In this subsection,
we derive the mass matrix for 4D zero modes and their mixing matrix.
Before doing that, we convert the gauge field action and the Higgs
field action to the canonical forms.

For the gauge field, use the metric (\ref{metric-ind-2}), the action
(\ref{action-gauge}) becomes to be
\begin{eqnarray}
\label{action-gauge-cano} S_{gauge}=\int d^4 x
\left\{-{\eta}^{\mu\alpha}{\eta}^{\nu\beta}\bigg{[}\frac{1}{4}\hat{F}^a_{\mu\nu}\hat{F}^a_{\alpha\beta}+\frac{1}{4}\tilde{F}_{\mu\nu}\tilde{F}_{\alpha\beta}\bigg{]}\right\},
\end{eqnarray}
where ${\eta}^{\mu\alpha}=\mathrm{diag}(-1,1,1,1)$ is the 4D Lorentz
metric. Note that we do not need to redefine the gauge fields. So
the gauge interaction actions (\ref{gauge-int-Q-4d}),
(\ref{gauge-int-U-4d}) and (\ref{gauge-int-D-4d}) keep invariant and
still apply in this subsection.

In order to convert the Higgs field action to the canonical form, we
redefine the Higgs field as
\begin{eqnarray}
\label{redef-Higgs}
\varphi(x)=\left(\tilde{s}\frac{L}{L^{\prime}}\right)\Phi(x),~~\tilde{s}=s\frac{e^{\omega
R}+a}{e^{\omega R}+b}.
\end{eqnarray}
By this redefinition and the metric (\ref{metric-ind-2}), the Higgs
action (\ref{higgs-action}) changes to
\begin{eqnarray}
\label{higgs-action-cano} S_{\mathrm{Higgs}}=\int d^4 x
\left\{{\eta}^{\mu\nu}D_{\mu}\varphi^{\dagger}D_{\nu}\varphi
-\mu_0(\varphi^{\dagger}\varphi-v^2)^2\right\},
\end{eqnarray}
where $v=\tilde{s}\frac{L}{L^{\prime}}v_0$. For
$\tilde{s}\frac{L}{L^{\prime}}\ll 1$, it supplies a beautiful
geometrical solution for gauge hierarchy problem suggested by
Randall and Sundrum in \cite{Randall:1999}. Because the gauge fields
do not need to be redefined, the gauge covariance derivative keeps
with the same form as that in (\ref{higgs-action}).

After the redefinition (\ref{redef-Higgs}) for Higgs filed, the
Yukawa interaction terms (\ref{yukawa-D-brane-4d}) and
(\ref{yukawa-U-brane-4d}) change to be
\begin{eqnarray}
\label{yukawa-D-brane-4d-cano} S_{\mathcal{D}\mathrm{Yukawa}}&=&\int
d^4 x dy A\frac{f_{\mathcal{D}}}{\tilde{s}\Lambda}
[\widehat{{\mathcal {Y}}}^{D}_{mn}+
i\beta_{\mathcal{D}}\widehat{Y}^{D}_{mn}]\sum_{i,j}\left\{f_{m,L}^{Q
(i)\ast}f_{n,R}^{\mathscr{D}
(j)}\bar{Q}_{m,L}^{(i)}d_{n,R}^{(j)}+f_{m,R}^{Q
(i)\ast}f_{n,L}^{\mathscr{D}
(j)}\bar{Q}_{m,R}^{(i)}d_{n,L}^{(j)}\right\}\nonumber\\
&\times&\varphi(x)\left(\frac{L^{\prime}}{L}\right)^2\delta(y-L^{\prime})+\mathrm{H.C.},\\
\label{yukawa-U-brane-4d-cano} S_{\mathcal{U}\mathrm{Yukawa}}&=&\int
d^4 x dy A\frac{f_{\mathcal{U}}}{\tilde{s}\Lambda}
[\widehat{{\mathcal {Y}}}^{U}_{mn}+
i\beta_{\mathcal{U}}\widehat{Y}^{U}_{mn}]\sum_{i,j}\left\{f_{m,L}^{Q
(i)\ast}f_{n,R}^{\mathscr{U}
(j)}\bar{Q}_{m,L}^{(i)}u_{n,R}^{(j)}+f_{m,R}^{Q
(i)\ast}f_{n,L}^{\mathscr{U}
(j)}\bar{Q}_{m,R}^{(i)}u_{n,L}^{(j)}\right\}\nonumber\\
&\times&\tilde{\varphi}(x)\left(\frac{L^{\prime}}{L}\right)^2\delta(y-L^{\prime})+\mathrm{H.C.}.
\end{eqnarray}

Following the ZMA approach, we isolate the zero mode terms from
above expressions as follows,
\begin{eqnarray}
\label{yukawa-D-brane-4d-zero} S_{\mathcal{D}\mathrm{Yukawa}}&=&\int
d^4 x dy A \frac{f_{\mathcal{D}}}{\tilde{s}\Lambda}
[\widehat{{\mathcal {Y}}}^{D}_{mn}+
i\beta_{\mathcal{D}}\widehat{Y}^{D}_{mn}]\left\{f_{m,L}^{Q
(0)\ast}f_{n,R}^{\mathscr{D}
(0)}\bar{Q}_{m,L}^{(0)}d_{n,R}^{(0)}\right\}\varphi(x)\left(\frac{L^{\prime}}{L}\right)^2\delta(y-L^{\prime})\nonumber\\
&+&\mathrm{H.C.},\\
\label{yukawa-U-brane-4d-zero} S_{\mathcal{U}\mathrm{Yukawa}}&=&\int
d^4 x dy A\frac{f_{\mathcal{U}}}{\tilde{s}\Lambda}
[\widehat{{\mathcal {Y}}}^{U}_{mn}+
i\beta_{\mathcal{U}}\widehat{Y}^{U}_{mn}]\left\{f_{m,L}^{Q
(0)\ast}f_{n,R}^{\mathscr{U}
(0)}\bar{Q}_{m,L}^{(0)}u_{n,R}^{(0)}\right\}\tilde{\varphi}(x)\left(\frac{L^{\prime}}{L}\right)^2\delta(y-L^{\prime})\nonumber\\
&+&\mathrm{H.C.}.
\end{eqnarray}
As in SM, after that the Higgs filed develops a vacuum expectation
value, the electroweak symmetry breaks. These Yukawa interactions
produce mass terms for fermions. From the above, we see that the
mass terms are related to the fermion zero mode profiles. We have
worked out these profiles in subsection \ref{sec:2.4} and give their
approximation behavior there. By these profiles, we obtain the mass
matrices for quarks as follows
\begin{eqnarray}
\label{mass-d} M^{d}_{mn}&=&\frac{vf_{\mathcal{D}}}{\tilde{s}\Lambda
L} [\widehat{\mathcal {Y}}^{D}_{mn}+
i\beta_{\mathcal{D}}\widehat{Y}^{D}_{mn}]\hat{f}_{m,L}^{Q
(0)\ast}\hat{f}_{n,R}^{\mathscr{D} (0)},\\
\label{mass-u} M^{u}_{mn}&=&\frac{vf_{\mathcal{U}}}{\tilde{s}\Lambda
L} [\widehat{\mathcal {Y}}^{U}_{mn}+
i\beta_{\mathcal{U}}\widehat{Y}^{U}_{mn}]\hat{f}_{m,L}^{Q
(0)\ast}\hat{f}_{n,R}^{\mathscr{U} (0)},
\end{eqnarray}
where the indices $m$ and $n$ are not summed. The term
$Y_{mn}f_mf_n$ stands for the product of three quantities $Y_{mn}$,
$f_m$ and $f_n$. $v$ is the Higgs vacuum expectation value as in
Eq.~(\ref{higgs-action-cano}). $\hat{f}_{n,L}^{Q(0)}$,
$\hat{f}_{n,R}^{\mathscr{U}(0)}$ and
$\hat{f}_{n,R}^{\mathscr{D}(0)}$ are the values of the canonical
zero mode profiles as we defined in equations (\ref{KK5-1-zero-c})
and (\ref{KK5-2-zero-c}) in subsection \ref{sec:2.4}. They are given
by
\begin{eqnarray}
\label{KK5-1-zero-c-q} \hat{f}_{m,L}^{Q
(0)}&=&\sqrt{\frac{2c^{Q}_m-1}{1-\epsilon^{2c^{Q}_m-1}}}\epsilon^{c^{Q}_m-\frac{1}{2}},~~c^{Q}_m={\hat{\lambda}}^{\mathbb{Q}}_m L,\\
\label{KK5-2-zero-c-d} \hat{f}_{n,R}^{\mathscr{D}
(0)}&=&\sqrt{\frac{-2c^{\mathscr{D}}_n-1}{1-\epsilon^{-2c^{\mathscr{D}}_n-1}}}\epsilon^{-c^{\mathscr{D}}_n-\frac{1}{2}},~~c^{\mathscr{D}}_n={\hat{\lambda}}^{\mathcal{D}}_n L,\\
\label{KK5-2-zero-c-u} \hat{f}_{n,R}^{\mathscr{U}
(0)}&=&\sqrt{\frac{-2c^{\mathscr{U}}_n-1}{1-\epsilon^{-2c^{\mathscr{U}}_n-1}}}\epsilon^{-c^{\mathscr{U}}_n-\frac{1}{2}},~~c^{\mathscr{U}}_n={\hat{\lambda}}^{\mathcal{U}}_n
L,
\end{eqnarray}
where ${\hat{\lambda}}^{\mathbb{Q}}_m$,
${\hat{\lambda}}^{\mathcal{D}}_n$ and
${\hat{\lambda}}^{\mathcal{U}}_n$ are determined by equations
(\ref{diagQ-5d-2}), (\ref{diagU-5d-2}) and (\ref{diagD-5d-2}). We
can rewrite above equations with the matrix form as
\begin{eqnarray}
\label{mass-d-martix}
M^{d}&=&\frac{vf_{\mathcal{D}}}{\tilde{s}\Lambda L}P^{Q}_L
[\widehat{\mathcal {Y}}^{D}+
i\beta_{\mathcal{D}}\widehat{Y}^{D}]P^{\mathscr{D}}_R,\\
\label{mass-d-diag} P^{Q}_L&=&\mathrm{diag}(\hat{f}_{1,L}^{Q
(0)},\cdots,\hat{f}_{m,L}^{Q
(0)}),~~P^{\mathscr{D}}_R=\mathrm{diag}(\hat{f}_{1,R}^{\mathscr{D}
(0)},\cdots,\hat{f}_{m,R}^{\mathscr{D}(0)}),\\
\label{mass-u-martix}
M^{u}&=&\frac{vf_{\mathcal{U}}}{\tilde{s}\Lambda L} P^{Q}_L
[\widehat{\mathcal {Y}}^{U}+
i\beta_{\mathcal{U}}\widehat{Y}^{U}]P^{\mathscr{U}}_R,\\
\label{mass-u-diag} P^{Q}_L&=&\mathrm{diag}(\hat{f}_{1,L}^{Q
(0)},\cdots,\hat{f}_{m,L}^{Q
(0)}),~~P^{\mathscr{U}}_R=\mathrm{diag}(\hat{f}_{1,R}^{\mathscr{U}
(0)},\cdots,\hat{f}_{m,R}^{\mathscr{U}(0)}).
\end{eqnarray}
These mass matrices are general complex matrices, and are not
hermitian matrices. We can make the single-value decompositions for
them to derive the mass eigenstates as follows
\begin{eqnarray}
\label{mass-d-svd}
M^{d}&=&V_{dL}^{\dagger}\mathcal{M}_dV_{dR},~~\mathcal{M}_d=\mathrm{diag}(m_{d_1},m_{d_2},\cdots,m_{d_n}),\\
\label{mass-u-svd}
M^{u}&=&V_{uL}^{\dagger}\mathcal{M}_uV_{uR},~~\mathcal{M}_u=\mathrm{diag}(m_{u_1},m_{u_2},\cdots,m_{u_n}),
\end{eqnarray}
where $m_{d_i(u_i)}>0$ for $i=1,2,\cdots,n$ according to the
definition of single-value decomposition.

As we discussed above, the gauge interaction terms keep the same
form with that of equations (\ref{gauge-int-Q-4d}),
(\ref{gauge-int-U-4d}) and (\ref{gauge-int-D-4d}). The zero modes
interact with gauge fields just like that in SM. So we can define
the mixing matrix for quarks like that in SM as
\begin{eqnarray}
\label{ckm} V_{\mathrm{CKM}}&=&V_{uL}V_{dL}^{\dagger}.
\end{eqnarray}

\subsection{Numerical results}\label{sec:3.3}
In subsection \ref{sec:3.1}, we construct our model in 6D bulk. In
subsection \ref{sec:3.2}, we derive 4D effective actions from the 6D
ones by two step KK decompositions. In this subsection, we give
numerical examples to show that the results of our model can be very
close to the experimental data.

From the model in subsection \ref{sec:3.1}, we know that there are
many parameters in this model. These parameters are not determined
by the model. We need to input these parameters by hand to obtain
the numerical results. These parameters can be classified to two
groups: the parameters in the metric and the parameters closely
related to the fermion mass matrices. Before giving the numerical
results, we give some discussions about the permissible extent of
the parameters.

We discuss the parameters in the metric at first. For the extent of
the extra dimension $y$, we let
\begin{eqnarray}
\label{value-y}
\epsilon=\frac{L}{L^{\prime}}=10^{-16},~\frac{1}{L}=10^{19}~\mathrm{GeV},~\frac{1}{L^{\prime}}=\mathrm{TeV},
\end{eqnarray}
which is necessary to interpret the gauge hierarchy as suggested in
\cite{Randall:1999}. From Eq.~(\ref{redef-Higgs}), we know that
$v_0$ relates to $v$ by the factor $\tilde{s}\frac{L}{L^{\prime}}$.
The choice of Eq.~(\ref{value-y}) implies that $\tilde{s}\approx 1$
in order to interpret the gauge hierarchy. The value of $\epsilon$
coincides with that in the paper \cite{Casagrande:2008}. We
designate the boundary value of the dimension $z$ by the equation
\begin{eqnarray}
\label{value-R} \frac{e^{\omega R}}{b}=40.
\end{eqnarray}
We also designate the value of the parameter $\omega$ in the metric
$B(z)$ by the equation
\begin{eqnarray}
\label{value-omega} \omega L=0.15.
\end{eqnarray}
$\omega$ can be regarded as the intrinsic sale of the the dimension
$z$. Eq.~(\ref{value-omega}) implies that $\omega$ is about 10
percents of the Planck scale. We designate $a$ in the metric $B(z)$
by the equation
\begin{eqnarray}
\label{value-a/b} \frac{a}{b}=6.
\end{eqnarray}
The other parameters like $s$ and $b$ are not necessary designated
in the numerical examples as they always emerge in combinations with
other parameters.

In the above, we have designated the necessary parameters in the
metric. We notice that the values of these parameters are not
determined by our model. We expect that they can be determined by
some underlying theories, which are not discussed by the present
paper. We choose them to the above values by hand in this paper,
because we find that these values can make the results of our model
to be very close to the experimental data.

\subsubsection{Numerical results in quark sector}\label{sec:3.3.1}

In this subsection, we give numerical results in the quark sector.
The parameters in the metric have been given above. To obtain the
numerical results, we need to further designate the parameters
related to quark mass matrices. At first, we need to designate the
parameters $m$ and $\mathcal{M}$ in actions (\ref{action-Q}),
(\ref{action-U}) and (\ref{action-D}). From the analysis in
subsection \ref{sec:2.3}, we know that the number of family is
determined by the pair $(\frac{a}{b},\frac{m}{\omega}s)$. So the
parameter $\frac{m}{\omega}s$ is closely related to the number of
family. In subsection \ref{sec:2.3}, we have given a parameter set
which permits the very 3 families. As we analyzed in subsection
\ref{sec:2.5}, these conclusions also apply to the new setup in
subsection \ref{sec:2.5}. This parameter set is given by equations
(\ref{cut1-a-pc}), (\ref{cut1-b-pc}) and (\ref{cut1-c-pc}). We have
designated the value of $\frac{a}{b}$ by Eq.~(\ref{value-a/b}), so
the possible extent for $\frac{m}{\omega}s$ is given by
\begin{eqnarray}
\label{value-m} \frac{m}{\omega}s \in
(\frac{1}{\sqrt{35}+5},~\frac{2}{\sqrt{35}+5})~~\mathrm{and}~~\frac{m}{\omega}s\neq
\frac{1}{10}.
\end{eqnarray}
We exclude the point $\frac{m}{\omega}s=\frac{1}{10}$, because when
$\frac{m}{\omega}s=\frac{1}{10}$, we obtain a massless solution,
which coincides with the zero mode solution as we discussed in
subsection \ref{sec:2.3}. The parameters $m$ in actions
(\ref{action-Q}), (\ref{action-U}) and (\ref{action-D}) should take
values in the intervals Eq.~(\ref{value-m}) to ensure that there are
the very 3 families in our model. While $\mathcal{M}$ is irrelevant
to the family number. Instead it is closely relevant to the values
of quark masses, as it can be seen from the following numerical
example. For $m$ in the intervals Eq.~(\ref{value-m}), the explicit
expressions for 3 family KK modes can be determined as we discussed
in subsection \ref{sec:2.5}. According to those discussions, the
solutions are very similar to that given in appendix \ref{sec:B}.
The differences are that we should replace $\lambda$ with
$\triangle\lambda$ in the expressions in appendix \ref{sec:B}. By
the normalization conditions Eq.~(\ref{norm-app}), these solutions
can be determined completely. The values of $m$ and $\mathcal{M}$
for different fields are different generally. We adjust them by hand
to fit the experimental data. We give the values for them in
Table.~\ref{table1}. Note that the negative real numbers $m$ also
emerge in Table.~\ref{table1}. The solutions for this case are also
discussed in appendix \ref{sec:B}.

The Yukawa couplings $f$ and $\beta$ in equations
(\ref{yukawa-D-brane}) and (\ref{yukawa-U-brane}) also need to be
input by hand. We give their values in Table.~\ref{table1}. In
Table.~\ref{table1}, we have defined that
\begin{eqnarray}
\label{redeef-f}
\tilde{f}_{\mathcal{U}}=\frac{vf_{\mathcal{U}}}{\Lambda
L}\frac{e^{\omega R}+b}{e^{\omega R}+a},~~~
\tilde{f}_{\mathcal{D}}=\frac{vf_{\mathcal{D}}}{\Lambda
L}\frac{e^{\omega R}+b}{e^{\omega R}+a}.
\end{eqnarray}
\begin{table}
\begin{tabular}{ccc}
Table.~3.3.1\label{table1}~~~Parameters in quark sector\\ \hline
Field&$\frac{m}{\omega}s$&$\frac{\mathcal{M}}{\omega}s$ \\
\hline $\mathbb{Q}$&$ 0.152$ & 4.25 \\
$\mathcal{U}$&$-0.173$&$-4.0$\\
$\mathcal{D}$&$-0.093$&$-4.2$\\
$\tilde{f}_{\mathcal{U}}$=15.5~TeV&$\beta_{\mathcal{U}}=0.55-0.43i$\\
$\tilde{f}_{\mathcal{D}}$=12.7~TeV&$\beta_{\mathcal{D}}=-1.2-0.5i$\\
\hline
\end{tabular}
\end{table}

Having designated these parameters, we can obtain the numerical
expressions of kinds of quantities in our model, like the matrices
$K$ and mass matrices $M$ in the fermion actions
(\ref{action-Q-5d-1}), (\ref{action-U-5d-1}) and
(\ref{action-D-5d-1}), the eigenvalues $\hat{\lambda}_n$ in equation
(\ref{action-psi-5d-3}) after two step redefinitions of fermion
fields and so on. In this paper, we omit these intermediate
numerical expressions. We only give the final mass matrices in
equations (\ref{mass-d-martix}) and (\ref{mass-u-martix}). However,
some analytical expressions for these intermediate quantities can be
found in subsection \ref{sec:4.1}.

By the parameter values given above, we obtain the numerical
expressions for mass matrices (\ref{mass-d-martix}) and
(\ref{mass-u-martix}) as follows.
\begin{eqnarray}
\label{numeric-m-u} M^u&=&\left(
\begin{array}{ccc}
 (6.44+0.552i)\cdot10^{-4} & (0.922+1.49 i)\cdot10^{-2} & -0.123-0.158 i \\
 (-1.53-3.25i)\cdot10^{-3} & 0.573& -5.44-4.98 i \\
 (1.29+1.64i)\cdot10^{-2} & 2.23+2.36 i & 135.58-13.4 i
\end{array}
\right)~\mathrm{GeV},\\
\label{numeric-m-d} M^d&=&\left(
\begin{array}{ccc}
 (0.72+2.78i)\cdot10^{-3} & (1.1-0.58i)\cdot10^{-2} & (3.9-7.85i)\cdot10^{-3}\\
 (-6.57+1.8i)\cdot10^{-3} & 4.39\cdot10^{-2}& (-1.36-0.49i)\cdot10^{-2} \\
 0.155-0.46 i & -0.984-0.93 i & -1.30+1.19 i
\end{array}
\right)~\mathrm{GeV}.~~
\end{eqnarray}
Making the single-value decompositions as in equations
(\ref{mass-d-svd}) and (\ref{mass-u-svd}), we obtain the quark
masses as follows
\begin{eqnarray}
m_u&=&0.64~\mathrm{MeV},~~m_c=584.87~\mathrm{MeV},~~m_t=136.48~\mathrm{GeV},\\
m_d&=&2~\mathrm{MeV},~~m_s=36.36~\mathrm{MeV},~~m_b=2.278~\mathrm{GeV}.
\end{eqnarray}
They are consistent with the $\overline{\mathrm{MS}}$ quark masses
evaluated at $1.5$ $\mathrm{TeV}$ in the paper
\cite{Casagrande:2008}. By Eq.~(\ref{ckm}), the mixing matrix and
its absolute value are given by
\begin{eqnarray}
\label{numeric-ckm} V_{\mathrm{CKM}}&=&\left(
\begin{array}{ccc}
 0.974+0.00614 i & 0.224545-0.0293 i & (4.297-0.58i)\cdot10^{-3} \\
 0.1488-0.1699 i & -0.535961+0.811 i & 0.0553-0.0232 i \\
 -0.0011-0.0172 i & 0.01957+0.0542i & -0.249+0.9666 i
\end{array}
\right),\\
\label{numeric-ckm} \mid V_{\mathrm{CKM}}\mid&=&\left(
\begin{array}{ccc}
 0.974014 & 0.226448 & 0.00433647 \\
 0.225832 & 0.972317 & 0.0599961 \\
 0.0172432 & 0.0576282 & 0.998189
\end{array}
\right).
\end{eqnarray}
We also obtain the Jarlskog invariant as
\begin{eqnarray}
\label{numeric-jarlskog}
J=-\mathrm{Im}(V_{ud}V_{cb}V_{ub}^*V_{cd}^*)=3.20416\times10^{-5}.
\end{eqnarray}
They are very close to the experimental data compiled in
\cite{PDG:2008}.

\subsubsection{Numerical results in lepton sector}\label{sec:3.3.2}

In this subsection, we discuss the lepton sector. We suppose that
neutrinos are Dirac ones. In this case, the lepton sector is very
similar to the quark sector. For neutrinos in other scenarios, see
\cite{Huber:2004}.

We introduce the fermion field contents in the 6D bulk as
\begin{eqnarray}
\label{lepton} \mathbb{L}=\left(\begin{array}{c}
   \mathbb{N}\\
   \mathbb{E}
\end{array}\right)
,~~~~\mathcal{N},~~~~\mathcal{E}.
\end{eqnarray}
They transform under the gauge group $\mathrm{SU}(3)_c\times
\mathrm{SU}(2)_L\times\mathrm{U}(1)_Y$ as
$\mathbb{L}=(1,2)_{-1},~\mathcal{N}=(1,1)_{0},~\mathcal{E}=(1,1)_{-2}$.
Note that we also introduce only 1 family lepton in the 6D bulk. The
actions of these fields are the same with that in equations
(\ref{action-Q}), (\ref{action-U}) and  (\ref{action-D}). The model
in subsection \ref{sec:3.1} applies similarly here, other than there
is no color interaction for leptons. The process of deriving 4D
effective actions from the 6D ones in subsection \ref{sec:3.2} also
applies here. While the mixing matrix for leptons is defined by
\begin{eqnarray}
\label{pmns} V_{\mathrm{PMNS}}&=&V_{lL}^{\dagger}V_{\nu L}.
\end{eqnarray}

Now we discuss the parameters in the lepton sector. The parameters
in the metric given by equations (\ref{value-y}), (\ref{value-R}),
(\ref{value-omega}) and (\ref{value-a/b}) still apply in the lepton
sector. The parameters $m$ of leptons should also in the intervals
Eq.~(\ref{value-m}) to ensure that we can obtain the very 3 families
in the lepton sector. Other parameters should also be input by hand
like that in the quark sector. We adjust them by hand to fit the
experimental data. We give their values in Table.~\ref{table2}.
\begin{table}
\begin{tabular}{ccc}
Table.~3.3.2\label{table2}~~~Parameters in lepton sector\\ \hline
Field&$\frac{m}{\omega}s$&$\frac{\mathcal{M}}{\omega}s$ \\
\hline $\mathbb{L}$& $0.16$ &$-9.0$ \\
$\mathcal{N}$&$-0.0944$&$-9.75$\\
$\mathcal{E}$&$-0.147$&$-5.6$\\
$\tilde{f}_{\mathcal{N}}$=23.89~TeV&$\beta_{\mathcal{N}}=1.5+0.45i$\\
$\tilde{f}_{\mathcal{E}}$=21.23~TeV&$\beta_{\mathcal{E}}=0.9752$\\
\hline
\end{tabular}
\end{table}
Having designating these parameters in lepton sector, we can obtain
the numerical expressions for kinds of quantities as in the quark
sector. In this subsection, we also only give the mass matrices for
leptons. They are given by
\begin{eqnarray}
\label{numeric-m-n} M^{\nu}&=&\left(
\begin{array}{ccc}
 3.204-8.405 i & 21.18-13.78 i & 6.37+26.6 i \\
 0.6978+0.632 i & -6.286 & 3.53-2.072 i \\
 -6.48+7.72 i & -12.29-13.11 i & 5.736-26.22 i
\end{array}
\right)\cdot~10^{-3}~\mathrm{eV},\\
\label{numeric-m-u} M^l&=&\left(
\begin{array}{ccc}
 0.287455+3.48 i & -10.01+90.3488 i & 1501.6+154.77 i \\
 -0.4266-6.935 i & 129.213 & 58.594-952.495 i \\
 10.40-1.072 i & 9.527+85.96i & 37.56-454.82 i
\end{array}
\right)~\mathrm{MeV}.
\end{eqnarray}
Making single-value decompositions for these matrices as in
equations (\ref{mass-d-svd}) and (\ref{mass-u-svd}), we obtain the
lepton masses as follows
\begin{eqnarray}
m_e&=&0.511~\mathrm{MeV},~~m_{\mu}=105.229~\mathrm{MeV},~~m_{\tau}=1849.15~\mathrm{MeV},\\
m_1&=&0.0019~\mathrm{eV},~~m_2=0.013~\mathrm{eV},~~m_3=0.05~\mathrm{eV}.
\end{eqnarray}
The masses of electron and muon are close to their experimental
value; while the mass of $\tau$ is moderately large than its
experimental value compiled in \cite{PDG:2008}. The neutrino masses
are of normal hierarchy type. They are close to the experimental
values in \cite{Gonzalez-Garcia:2008}. We can also obtain the mixing
matrix defined in Eq.~(\ref{pmns}) and its absolute value as
\begin{eqnarray}
\label{numeric-pmns} V_{\mathrm{PMNS}}&=&\left(
\begin{array}{ccc}
 0.6799-0.4592 i & 0.5163+0.2442 i & 0.02504+0.003909 i \\
 -0.4619-0.02652 i & 0.2922+0.5664 i & 0.44075+0.43071 i \\
 0.2051-0.2659 i & -0.5136-0.06183 i & 0.7715-0.1561 i
\end{array}
\right),\\
\label{numeric-ckm} \mid V_{\mathrm{PMNS}}\mid&=&\left(
\begin{array}{ccc}
 0.820465 & 0.571134 & 0.02534 \\
 0.462666 & 0.637319 & 0.61625 \\
 0.335822 & 0.517329 & 0.78714
\end{array}
\right).
\end{eqnarray}
They are in the $3\sigma$ extent of the experimental values in
\cite{Gonzalez-Garcia:2008}.

We have supposed neutrinos to be Dirac ones. So as in the quark
sector, we can calculate the Jarlskog invariant for the mixing
matrix as
\begin{eqnarray}
\label{numeric-jarlskog} J=-\mathrm{Im}(V_{e1}V_{\mu3}V_{e3}^*V_{\mu
1}^*)=19.193\times10^{-5}.
\end{eqnarray}
It is lager than that in the quark sector. We note that its size is
not the inevitable result of our model. When we adjust the
parameters, we find that the size of $J$ is very sensitive to the
mass $m$ of the field $\mathcal{N}$. By adjusting
$\frac{m_{\mathcal{N}}}{\omega}s$, we can still keep the mixing
matrix and the neutrino masses to be close to their experimental
values, but $J$ can varies remarkably.

\subsubsection{Brief comments on the numerical results}\label{sec:3.3.3}
We give the numerical results in the last two subsections. We see
that there are still many parameters in our model. The parameters
$\frac{\mathcal{M}}{\omega}s$ are closely relevant to the absolute
size of the fermion masses. The parameter
$\frac{\mathcal{M}}{\omega}s$ of the field $\mathcal{N}$ is more
lager than that of quarks and charged leptons, because the neutrino
masses are remarkably smaller than the masses of quarks and charged
leptons. While the parameters $\frac{m}{\omega}s$ are closely
relevant to the mass hierarchy structure of quarks and charged
leptons. In general, lager absolute value of $\frac{m}{\omega}s$
produces larger mass hierarchy. The parameters $\beta$ are closely
relative to the CP violation measure $J$. $\beta=0$ implies that $J$
vanishes. In the above numerical examples, we see that $\beta\neq 0$
in the lepton sector, so CP violation also emerges in the lepton
sector. We have not found a group of parameters with $\beta=0$ which
can fit the experimental data as the group of parameters in the last
subsection.

We also see that the values of $\frac{\mathcal{M}}{\omega}s$ are
about 100 times larger than the values of $\frac{m}{\omega}s$, while
this little hierarchy is not explained in this paper. In addition,
because there are too many parameters in our model, we only adjust
them by hand to obtain the numerical results. These numerical
results are very close to the experimental values. We have not done
further adjustments to make them in the $1\sigma$ extent permitted
by experiments. In the last two subsections, we only give rough
numerical examples to show that our model can close to the
experimental data in high precision. For the quark masses, we take
the renormalization effects into consideration, and adjust the
parameters to fit the running masses at $1.5$ $\mathrm{TeV}$
compiled in \cite{Casagrande:2008}. While for the leptons masses and
the mixing matrices, we only adjust parameters by hand to fit the
face values compiled in \cite{Gonzalez-Garcia:2008,PDG:2008}, and
the renormalization effects are omitted. So the above numerical
examples are only rough treatments. The renormalization effects of
these quantities should be considered for more detailed comparison
with the experimental data.

\section{Some analytical treatments about the model and more relevant discussions}\label{sec:4}
In section \ref{sec:3}, we introduce our model and give the
numerical results. At first sight, this model seems very
complicated. However, when we give some analytical treatments about
this model, we will see that some quantities in this model have very
concise expressions. We can make some qualitative conclusions from
these concise expressions. These analytical treatments can help us
to understand how our model works more clearly. In subsection
\ref{sec:4.2}, we make more discussions about several relevant
problems.

\subsection{Some analytical treatments about the model}\label{sec:4.1}

For analytical discussions, we take fields $\mathbb{Q}$ and
$\mathcal{D}$ for example. The analytical treatments for
$\mathcal{U}$ are very similar to that of $\mathbb{Q}$ and
$\mathcal{D}$. As we discussed in appendix \ref{sec:B}, the
solutions for KK modes are different for $m$ to be positive or
negative. While the eigenvalues $\lambda$ determined by equations
(\ref{cut1-a}), (\ref{cut1-b}), (\ref{cut2-a}) and (\ref{cut2-b})
can be real or pure imaginary according to the parameter pair
$(\frac{a}{b},~\frac{m}{\omega}s)$. The analytical expressions can
be classified into four classes: (1) $m>0$, $\lambda$ is real;  (2)
$m>0$, $\lambda$ is pure imaginary; (3) $m<0$, $\lambda$ is real;
(4) $m<0$, $\lambda$ is pure imaginary. For the new setup in
subsection \ref{sec:2.5}, we should replace $\lambda$ with
$\triangle\lambda$ as we analyzed in subsection \ref{sec:2.5}. In
the numerical examples for quark sector in subsection
\ref{sec:3.3.1}. The solutions for $\mathbb{Q}$ belong to the
class~(2); while the solutions for $\mathcal{D}$ belong to the
class~(3).  The solutions for $\mathcal{U}$ belong to the class~(4).
In the following discussions, we always suppose that the solutions
for $\mathbb{Q}$ belong to the class~(2) and the solutions for
$\mathcal{D}$ belong to the class~(3). Other situations can be
discussed similarly.

For the analytical treatments, we must discuss the solutions for KK
modes at first. The solutions can be determined according to our
discussions in subsection \ref{sec:2.5}. They have the similar forms
and characters to that given in appendix \ref{sec:B}. By these
solutions, we obtain the expressions for the matrices $K$ and $M$ in
the fermion actions (\ref{action-Q-5d-1}) and (\ref{action-D-5d-1})
as follows
\begin{eqnarray}
\label{ana-K-M-5d-Q} K^{\mathbb{Q}}&=&\left(
\begin{array}{ccc}
 1 &  ia_{\mathbb{Q}} & -ia_{\mathbb{Q}}  \\
 -ia_{\mathbb{Q}}  & 1 & b_{\mathbb{Q}} \\
 ia_{\mathbb{Q}}  & b_{\mathbb{Q}} & 1
\end{array}
\right),M^{\mathbb{Q}}=\mathcal{M}_{\mathbb{Q}}K^{\mathbb{Q}}+\left(
\begin{array}{ccc}
 0 &  \tilde{a}_{\mathbb{Q}} &\tilde{a}_{\mathbb{Q}} \\
 \tilde{a}_{\mathbb{Q}} & 0 & -i\tilde{b}_{\mathbb{Q}}\\
\tilde{a}_{\mathbb{Q}} & i \tilde{b}_{\mathbb{Q}} & 0
\end{array}
\right),\\
\label{ana-K-M-5d-D} K^{\mathcal{D}}&=&\left(
\begin{array}{ccc}
 1 &  a_{\mathcal{D}} & -a_{\mathcal{D}} \\
  a_{\mathcal{D}}  & 1 & b_{\mathcal{D}} \\
  -a_{\mathcal{D}}  & b_{\mathcal{D}} & 1
\end{array}
\right),~~M^{\mathcal{D}}=\mathcal{M}_{\mathcal{D}}K^{\mathcal{D}}+\left(
\begin{array}{ccc}
 0 &  \tilde{a}_{\mathcal{D}}  & \tilde{a}_{\mathcal{D}}\\
 \tilde{a}_{\mathcal{D}} & \triangle\lambda^{\mathcal{D}} & 0\\
\tilde{a}_{\mathcal{D}}& 0 &-\triangle\lambda^{\mathcal{D}}
\end{array}
\right),
\end{eqnarray}
where we have used the normalization conditions in
Eq.~(\ref{norm-app}). We give the expressions for $a_{\mathbb{Q}}$,
$a_{\mathcal{D}}$, $b_{\mathbb{Q}}$ and $b_{\mathcal{D}}$ in
equations (\ref{ana-def-Q}) and (\ref{ana-def-D}) in appendix
\ref{sec:C}. They are all real numbers by definitions. Here and in
the following, we arrange the column and the row indices for the
matrices as $0,1,-1$, which are the indices for KK modes displayed
in appendix \ref{sec:B}. $\triangle\lambda$ are the eigenvalues in
equations (\ref{sol13-new-var}) and (\ref{sol14-new-var}). They can
be calculated according to our discussions in subsection
\ref{sec:2.5}. We have employed Eq.~(\ref{new-var}) to rewrite the
expressions for mass matrices $M$.

Now we discuss the matrices in interaction sectors. For the gauge
interaction sector, from equations (\ref{gauge-int-Q-5d}) and
(\ref{gauge-int-D-5d}), we know that the matrices in this sector are
the same with that in the above. The matrices in the Yukawa
interaction sector are important for our discussions. We obtain
expressions for these matrices in Eq.~(\ref{yukawa-D-brane-5d-M}) as
follows
\begin{eqnarray}
\label{ana-Y} \mathcal{Y}^{D}&=&\left(
\begin{array}{ccr}
 \mathbbm{s} &\mathbbm{a} & -\mathbbm{a} \\
 \mathbbm{b} & \mathbbm{c} & \mathbbm{d} \\
 -\mathbbm{b} & \mathbbm{d} & \mathbbm{c}
\end{array}
\right),~~~Y^{D}=\left(
\begin{array}{ccr}
 0 & \mathbbm{a}^{\prime} & \mathbbm{a}^{\prime} \\
 \mathbbm{b}^{\prime} & \mathbbm{c}^{\prime} & \mathbbm{d}^{\prime} \\
 \mathbbm{b}^{\prime} & -\mathbbm{d}^{\prime} & -\mathbbm{c}^{\prime}
\end{array}
\right).
\end{eqnarray}
The elements in these matrices are defined by equations
(\ref{ana-def-Y-1}) and (\ref{ana-def-Y-2}) in appendix \ref{sec:C}.
These elements can be complex numbers generically.

From the analytical expressions above, we see that these matrices
all have very concise structures. Following the procedures in
subsection \ref{sec:3.2.2}, we need two steps to obtain conventional
5D effective actions.

Step~(I): Making Cholesky decompositions for the matrices $K$, for
the field $\mathbb{Q}$, we obtain
\begin{eqnarray}
\label{ana-K-cho-Q}
K^{\mathbb{Q}}&=&V^{\mathbb{Q}\dagger}\Lambda^{\mathbb{Q}}
V^{\mathbb{Q}}=H^{\mathbb{Q}\dagger}H^{\mathbb{Q}},~H^{\mathbb{Q}}=\sqrt{\Lambda^{\mathbb{Q}}}V^{\mathbb{Q}},\\
\Lambda^{\mathbb{Q}}&=&\mathrm{diag}\left(\Lambda_1^{\mathbb{Q}},\Lambda^{\mathbb{Q}}_2,\Lambda^{\mathbb{Q}}_3\right),
~~\sqrt{\Lambda^{\mathbb{Q}}}=\mathrm{diag}(\sqrt{\Lambda^{\mathbb{Q}}_1},\sqrt{\Lambda^{\mathbb{Q}}_2},\sqrt{\Lambda^{\mathbb{Q}}_3}),\nonumber\\
V^{\mathbb{Q}\dagger}&=&\left(
\begin{array}{ccr}
 -i &0 & 0 \\
 0 & 1& 0 \\
 0 &0 & 1
\end{array}
\right)\left(
\begin{array}{ccr}
 \cos{\theta_{\mathbb{Q}}} &\sin{\theta_{\mathbb{Q}}}  &0 \\
 -\frac{1}{\sqrt{2}}\sin{\theta_{\mathbb{Q}}} &\frac{1}{\sqrt{2}}\cos{\theta_{\mathbb{Q}}} & \frac{1}{\sqrt{2}} \\
 \frac{1}{\sqrt{2}}\sin{\theta_{\mathbb{Q}}} & -\frac{1}{\sqrt{2}}\cos{\theta_{\mathbb{Q}}} & \frac{1}{\sqrt{2}}
\end{array}
\right),~\tan{\theta_{\mathbb{Q}}}=-\frac{a_{\mathbb{Q}}+\sqrt{8a^2_{\mathbb{Q}}+b^2_{\mathbb{Q}}}}{2\sqrt{2}a_{\mathbb{Q}}}.\nonumber
\end{eqnarray}
While for the field $\mathcal{D}$, similarly we obtain
\begin{eqnarray}
\label{ana-K-cho-D}
K^{\mathcal{D}}&=&V^{\mathcal{D}\dagger}\Lambda^{\mathcal{D}}
V^{\mathcal{D}}=H^{\mathcal{D}\dagger}H^{\mathcal{D}},~H^{\mathcal{D}}=\sqrt{\Lambda^{\mathcal{D}}}V^{\mathcal{D}},\\
\Lambda^{\mathcal{D}}&=&\mathrm{diag}\left(\Lambda_1^{\mathcal{D}},\Lambda^{\mathcal{D}}_2,\cdots,\Lambda^{\mathcal{D}}_n\right),
~~\sqrt{\Lambda^{\mathcal{D}}}=\mathrm{diag}(\sqrt{\Lambda^{\mathcal{D}}_1},\sqrt{\Lambda^{\mathcal{D}}_2},\cdots,\sqrt{\Lambda^{\mathcal{D}}_n}),\nonumber\\
V^{\mathcal{D}\dagger}&=&\left(
\begin{array}{ccr}
 \cos{\theta_{\mathcal{D}}} &\sin{\theta_{\mathcal{D}}} &0 \\
 -\frac{1}{\sqrt{2}}\sin{\theta_{\mathcal{D}}} &\frac{1}{\sqrt{2}}\cos{\theta_{\mathcal{D}}} & \frac{1}{\sqrt{2}} \\
 \frac{1}{\sqrt{2}}\sin{\theta_{\mathcal{D}}} & -\frac{1}{\sqrt{2}}\cos{\theta_{\mathcal{D}}} & \frac{1}{\sqrt{2}}
\end{array}
\right),~\tan{\theta_{\mathcal{D}}}=\frac{a_{\mathcal{D}}+\sqrt{8a^2_{\mathcal{D}}+b^2_{\mathcal{D}}}}{2\sqrt{2}a_{\mathcal{D}}}.\nonumber
\end{eqnarray}
The expressions for these eigenvalues in the above are given by
equations (\ref{ana-K-cho-Q-eigenval}) and
(\ref{ana-K-cho-D-eigenval}) in appendix \ref{sec:C}. By the
redefinitions of fermion fields in Eq.~(\ref{redef-5d}), we know
that the matrices in the kinetic terms of fermion actions and gauge
interaction terms become to be the identity matrices. While the
matrices in the mass terms become to be
\begin{eqnarray}
\label{ana-M-Q-tilde}\widetilde{M}^{\mathbb{Q}}&=&(H^{\mathbb{Q}-1})^{\dagger}M^{\mathbb{Q}}
H^{\mathbb{Q}-1}=\mathcal{M}_{\mathbb{Q}}+\left(\begin{array}{ccr}
 0 &0 & i\hat{a}_{\mathbb{Q}} \\
 0 & 0& -i\hat{b}_{\mathbb{Q}} \\
 -i\hat{a}_{\mathbb{Q}} &i\hat{b}_{\mathbb{Q}} &0
\end{array}
\right),\\
\label{ana-M-D-tilde}
\widetilde{M}^{\mathcal{D}}&=&(H^{\mathcal{D}-1})^{\dagger}M^{\mathcal{D}}
H^{\mathcal{D}-1}=\mathcal{M}_{\mathcal{D}}+\left(\begin{array}{ccr}
 0 &0 & \hat{a}_{\mathcal{D}} \\
 0 & 0&\hat{b}_{\mathcal{D}} \\
 \hat{a}_{\mathcal{D}} &\hat{b}_{\mathcal{D}} &0
\end{array}
\right).
\end{eqnarray}
The elements of these matrices are defined by equations
(\ref{ana-M-Q-tilde-element}) and (\ref{ana-M-D-tilde-element}) in
appendix \ref{sec:C}. By definitions, $\hat{a}_{\mathbb{Q}}$,
$\hat{b}_{\mathbb{Q}}$, $\hat{a}_{\mathcal{D}}$ and
$\hat{b}_{\mathcal{D}}$ are all real numbers. We see that the above
expressions are still of very concise structure after this first
field redefinitions. We do not give explicit expressions for the
Yukawa interaction sector here. We will give them in step~(II).

Step~(II): As in equations (\ref{diagQ-5d-2}) and
(\ref{diagD-5d-2}), we diagonalizing the matrices in equations
(\ref{ana-M-Q-tilde})  and (\ref{ana-M-D-tilde}). For the mass
matrix of field $\mathbb{Q}$, we obtain
\begin{eqnarray}
\label{ana-diagQ-5d-2}
\widetilde{M}^{\mathbb{Q}}&=&U^{\mathbb{Q}\dagger}\Delta_{\mathbb{Q}}
U^{\mathbb{Q}},~~
\Delta_{\mathbb{Q}}=\mathrm{diag}(\hat{\lambda}^{\mathbb{Q}}_1,\hat{\lambda}^{\mathbb{Q}}_2,\cdots,\hat{\lambda}^{\mathbb{Q}}_n),\\
U^{\mathbb{Q}\dagger}&=&\left(
\begin{array}{ccr}
 \cos{\vartheta_{\mathbb{Q}}} &i\frac{1}{\sqrt{2}}\sin{\vartheta_{\mathbb{Q}}}&-i\frac{1}{\sqrt{2}}\sin{\vartheta_{\mathbb{Q}}} \\
  -\sin{\vartheta_{\mathbb{Q}}}&i\frac{1}{\sqrt{2}}\cos{\vartheta_{\mathbb{Q}}}&-i\frac{1}{\sqrt{2}}\cos{\vartheta_{\mathbb{Q}}} \\
 0& \frac{1}{\sqrt{2}} & \frac{1}{\sqrt{2}}
\end{array}
\right),~\tan{\vartheta_{\mathbb{Q}}}=-\frac{\hat{a}_{\mathbb{Q}}}{\hat{b}_{\mathbb{Q}}},\nonumber\\
\hat{\lambda}^{\mathbb{Q}}_1&=&\mathcal{M}_{\mathbb{Q}},~\hat{\lambda}^{\mathbb{Q}}_2=\mathcal{M}_{\mathbb{Q}}-\triangle\hat{\lambda}^{\mathbb{Q}},~
\hat{\lambda}^{\mathbb{Q}}_3=\mathcal{M}_{\mathbb{Q}}+\triangle\hat{\lambda}^{\mathbb{Q}},~\triangle\hat{\lambda}^{\mathbb{Q}}=\sqrt{\hat{a}_{\mathbb{Q}}^2+\hat{b}_{\mathbb{Q}}^2}.\nonumber
\end{eqnarray}
While for the mass matrix of field $\mathcal{D}$, we obtain
\begin{eqnarray}
\label{ana-diagD-5d-2}
\widetilde{M}^{\mathcal{D}}&=&U^{\mathcal{D}\dagger}\Delta_{\mathcal{D}}
U^{\mathcal{D}},~~
\Delta_{\mathcal{D}}=\mathrm{diag}(\hat{\lambda}^{\mathcal{D}}_1,\hat{\lambda}^{\mathcal{D}}_2,\cdots,\hat{\lambda}^{\mathcal{D}}_n),\\
U^{\mathcal{D}\dagger}&=&\left(
\begin{array}{ccr}
 \cos{\vartheta_{\mathcal{D}}} &\frac{1}{\sqrt{2}}\sin{\vartheta_{\mathcal{D}}}&-\frac{1}{\sqrt{2}}\sin{\vartheta_{\mathcal{D}}} \\
  -\sin{\vartheta_{\mathcal{D}}}&\frac{1}{\sqrt{2}}\cos{\vartheta_{\mathcal{D}}}&-\frac{1}{\sqrt{2}}\cos{\vartheta_{\mathcal{D}}} \\
 0& \frac{1}{\sqrt{2}} & \frac{1}{\sqrt{2}}
\end{array}
\right),~\tan{\vartheta_{\mathcal{D}}}=\frac{\hat{a}_{\mathcal{D}}}{\hat{b}_{\mathcal{D}}},\nonumber\\
\hat{\lambda}^{\mathcal{D}}_1&=&\mathcal{M}_{\mathcal{D}},~\hat{\lambda}^{\mathcal{D}}_2=\mathcal{M}_{\mathcal{D}}-\triangle\hat{\lambda}^{\mathcal{D}},~
\hat{\lambda}^{\mathcal{D}}_3=\mathcal{M}_{\mathcal{D}}+\triangle\hat{\lambda}^{\mathcal{D}},~\triangle\hat{\lambda}^{\mathcal{D}}=\sqrt{\hat{a}_{\mathcal{D}}^2+\hat{b}_{\mathcal{D}}^2}.\nonumber
\end{eqnarray}
These matrices are still of concise structure. If $\mathcal{M}=0$,
we see that the eigenvalues $\hat{\lambda}$ have the spectrum
$0,\pm\triangle\hat{\lambda}$. This spectrum is similar to
Eq.~(\ref{spectrum1}), which is the spectrum of $\lambda$ before the
redefinitions of fermion fields. While for $\mathcal{M} \neq 0$, the
spectrum is similar to Eq.~(\ref{spectrum2}). These results are
expected in subsection \ref{sec:2.5}.

By the redefinitions of fermion fields as in
Eq.~(\ref{field-redef-5d-2}), we know that the fermion actions
become to be the conventional ones as in
Eq.~(\ref{action-psi-5d-3}); while the gauge interaction terms keep
to be the flavor universal ones as in equations
(\ref{gauge-int-Q-5d-2}) and (\ref{gauge-int-D-5d-2}). For the
Yukawa interaction sector, after the redefinitions, we obtain final
results of these matrices as follows
\begin{eqnarray}
\label{ana-yukawa-D-brane-5d-Y-2} \widehat{\mathcal
{Y}}^{D}&=&\tilde{H}^{\mathbb{Q}\dagger}{\mathcal
{Y}}^{D}\tilde{H}^{\mathcal{D}},~~\widehat{Y}^{D}=
\tilde{H}^{\mathbb{Q}\dagger}Y^{D}\tilde{H}^{\mathcal {D}},\\
\tilde{H}^{\mathbb{Q}\dagger}&=&U^{\mathbb{Q}}(H^{\mathbb{Q}-1})^{\dagger}=U^{\mathbb{Q}}\frac{1}{\sqrt{\Lambda^{\mathbb{Q}}}}V^{\mathbb{Q}}
=\left(
\begin{array}{ccr}
 \mathbbm{s}_{\mathbb{Q}} &\mathbbm{a}_{\mathbb{Q}} & -\mathbbm{a}_{\mathbb{Q}} \\
 \mathbbm{b}_{\mathbb{Q}} & \mathbbm{c}_{\mathbb{Q}} & \mathbbm{d}_{\mathbb{Q}} \\
 -\mathbbm{b}_{\mathbb{Q}} & \mathbbm{d}_{\mathbb{Q}} & \mathbbm{c}_{\mathbb{Q}}
\end{array}
\right),\\
\tilde{H}^{\mathcal{D}}&=&(H^{\mathcal{D}-1})U^{\mathcal{D}\dagger}=V^{\mathcal{D}\dagger}
\frac{1}{\sqrt{\Lambda^{\mathcal{D}}}}U^{\mathcal{D}\dagger}=\left(
\begin{array}{ccr}
 \mathbbm{s}_{\mathcal{D}} &\mathbbm{a}_{\mathcal{D}} & -\mathbbm{a}_{\mathcal{D}} \\
 \mathbbm{b}_{\mathcal{D}} & \mathbbm{c}_{\mathcal{D}} & \mathbbm{d}_{\mathcal{D}} \\
 -\mathbbm{b}_{\mathcal{D}}& \mathbbm{d}_{\mathcal{D}} & \mathbbm{c}_{\mathcal{D}}
\end{array}
\right),
\end{eqnarray}
where $\frac{1}{\sqrt{\Lambda^{\mathbb{Q}}}}$ and
$\frac{1}{\sqrt{\Lambda^{\mathcal{D}}}}$ are the inverse matrices of
$\sqrt{\Lambda^{\mathbb{Q}}}$ and $\sqrt{\Lambda^{\mathcal{D}}}$
respectively. We define the elements of matrices
$\tilde{H}^{\mathbb{Q}}$ and $\tilde{H}^{\mathcal{D}}$ in equations
(\ref{ana-yukawa-D-brane-5d-Y-2-elementQ}),
(\ref{ana-yukawa-D-brane-5d-Y-2-elementQ-cd}),
(\ref{ana-yukawa-D-brane-5d-Y-2-elementD}) and
(\ref{ana-yukawa-D-brane-5d-Y-2-elementD-cd}). From the above, we
see that the transformation matrices $\tilde{H}^{\mathbb{Q}}$ and
$\tilde{H}^{\mathcal{D}}$ have the concise structure similar to that
of matrices ${\mathcal {Y}}^{D}$ and $Y^{D}$ in Eq.~(\ref{ana-Y}).
Working out the product of these matrices in
Eq.~(\ref{ana-yukawa-D-brane-5d-Y-2}), we obtain
\begin{eqnarray}
\label{ana-Y-tran}\widehat{\mathcal{Y}}^{D}&=&\left(
\begin{array}{ccr}
 \hat{\mathbbm{s}} &\hat{\mathbbm{a}} & -\hat{\mathbbm{a}} \\
 \hat{\mathbbm{b}} & \hat{\mathbbm{c}} & \hat{\mathbbm{d}} \\
 -\hat{\mathbbm{b}} & \hat{\mathbbm{d}} & \hat{\mathbbm{c}}
\end{array}
\right),~~~\widehat{Y}^{D}=\left(
\begin{array}{ccr}
 0 & \hat{\mathbbm{a}}^{\prime} & \hat{\mathbbm{a}}^{\prime} \\
 \hat{\mathbbm{b}}^{\prime} & \hat{\mathbbm{c}}^{\prime} & \hat{\mathbbm{d}}^{\prime} \\
 \hat{\mathbbm{b}}^{\prime} & -\hat{\mathbbm{d}}^{\prime} & -\hat{\mathbbm{c}}^{\prime}
\end{array}
\right).
\end{eqnarray}
The elements in these matrices can be expressed with the quantities
in the matrices $\tilde{H}^{\mathbb{Q}}$, $\tilde{H}^{\mathcal{D}}$,
${\mathcal{Y}}^{D}$ and $Y^{D}$ by the matrix multiplication. We
omit their explicit expressions for simplicity. Because of the
special structure of the transformation matrices
$\tilde{H}^{\mathbb{Q}}$ and $\tilde{H}^{\mathcal{D}}$, the
structure of ${\mathcal{Y}}^{D}$ and $Y^{D}$ keep invariant under
these transformations and only their elements are modified to be
different values.

Making single-value decompositions for matrices $\widehat{\mathcal
{Y}}^{D}$ and $\widehat{Y}^{D}$, for $\widehat{\mathcal {Y}}^{D}$,
we obtain
\begin{eqnarray}
\label{ana-calY-tilde-svd}\widehat{\mathcal{Y}}^{D}&=&V_{\mathcal{Y}L}\Sigma^{\mathcal{Y}}
V_{\mathcal{Y}R}^{\dagger},~~V_{\mathcal{Y}R}=\widehat{\mathcal{Y}}^{D\dagger}V_{\mathcal{Y}L}\Sigma^{\mathcal{Y}-1},
~~~\Sigma^{\mathcal{Y}}=\mathrm{diag}(\Sigma^{\mathcal{Y}}_{1},~\Sigma^{\mathcal{Y}}_{2},~\Sigma^{\mathcal{Y}}_{3}),\\
V_{\mathcal{Y}L}&=&\left(
\begin{array}{ccr}
 \exp{i\delta_\mathcal{Y}} &0 & 0 \\
 0 & 1& 0 \\
 0 &0 & 1
\end{array}
\right)\left(
\begin{array}{ccr}
 \cos{\theta_{\mathcal{Y}}} &\sin{\theta_{\mathcal{Y}}}  &0 \\
 -\frac{1}{\sqrt{2}}\sin{\theta_{\mathcal{Y}}} &\frac{1}{\sqrt{2}}\cos{\theta_{\mathcal{Y}}} & \frac{1}{\sqrt{2}} \\
 \frac{1}{\sqrt{2}}\sin{\theta_{\mathcal{Y}}} & -\frac{1}{\sqrt{2}}\cos{\theta_{\mathcal{Y}}} & \frac{1}{\sqrt{2}}
\end{array}
\right).\nonumber
\end{eqnarray}
For $\widehat{Y}^{D}$, similarly we obtain
\begin{eqnarray}
\label{ana-Y-tilde-svd}\widehat{Y}^{D}&=&V_{YL}\Sigma^{Y}
V_{YR}^{\dagger},~~V_{YR}=\widehat{Y}^{D\dagger}V_{YL}\Sigma^{Y-1},
~~~\Sigma^{Y}=\mathrm{diag}(\Sigma^{Y}_{1},~\Sigma^{Y}_{2},~\Sigma^{Y}_{3}),\\
V_{YL}&=&\left(
\begin{array}{ccr}
 \exp{i\delta_Y} &0 & 0 \\
 0 & 1& 0 \\
 0 &0 & 1
\end{array}
\right)\left(
\begin{array}{ccr}
 \cos{\theta_{Y}} &\sin{\theta_{Y}} &0 \\
 -\frac{1}{\sqrt{2}}\sin{\theta_{Y}}&\frac{1}{\sqrt{2}}\cos{\theta_{Y}} & \frac{1}{\sqrt{2}} \\
 \frac{1}{\sqrt{2}}\sin{\theta_{Y}}& -\frac{1}{\sqrt{2}}\cos{\theta_{Y}} & \frac{1}{\sqrt{2}}
\end{array}
\right).\nonumber
\end{eqnarray}
Some quantities in these expressions are defined in equations
(\ref{ana-calY-tilde-svd-mat}), (\ref{ana-calY-tilde-svd-def}) and
(\ref{ana-Y-tilde-svd-def}) in appendix \ref{sec:C}. The unitary
matrices $V_{\mathcal{Y}R}$ and $V_{YR}$ have the similar structure
to that of $V_{\mathcal{Y}L}$ and $V_{YL}$, we do not give them
explicitly here.

In the above, we give some analytical treatment about our model. We
only display the results for down quark sector, but the results for
up quark sector are similar to that in the down quark sector. As we
discussed in subsection \ref{sec:3.2.2}, the lepton sector is
similar to the quark sector. So the above discussions also apply to
the lepton sector. From the above discussions, we see that the
matrices in the 5D effective actions all have very concise
structure. These concise structures are induced by the special
characters of KK modes as we analyzed in subsection \ref{sec:2.3}
and subsection \ref{sec:2.5}. Especially, the unitary transformation
matrices $V_{\mathcal{Y}L}$ and $V_{YL}$ in equations
(\ref{ana-calY-tilde-svd}) and (\ref{ana-Y-tilde-svd}) are very
close to the structure of the experimental PMNS mixing matrix in the
lepton sector. So the Yukawa coupling matrices $\widehat{\mathcal
{Y}}^{D}$ and $\widehat{Y}^{D}$ are appropriate to construct models
for lepton mixing matrices. However, the addition of these two
matrices
$\widehat{\mathcal{Y}}^{D}+i\beta_{\mathcal{D}}\widehat{Y}^{D}$ in
Eq.~(\ref{yukawa-D-brane-5d-2}) becomes to be a general matrix. It
does not have the concise structure like that of matrices
$\widehat{\mathcal {Y}}^{D}$ and $\widehat{Y}^{D}$. Its eigenvectors
are complicated and we do not give them here.

The matrices $\widehat{\mathcal{Y}}^{D}$ and $\widehat{Y}^{D}$ are
the 5D Yukawa couplings. The physical 4D mass matrix is given by
Eq.~(\ref{mass-d-martix}). From Eq.~(\ref{mass-d-martix}), we see
that these matrices are modified further by the 4D zero modes
profiles. The concise structure of $\widehat{\mathcal {Y}}^{D}$ and
$\widehat{Y}^{D}$ are lost and are distorted further by these zero
mode profiles to be a general matrix.

By the above analytical treatments, we can make some qualitative
discussions about how our model works. From Eq.~(\ref{ana-Y}), we
know that the 5D effective Yukawa couplings are determined by the
profiles of KK modes. Due to the special characters of KK modes as
we analyzed in subsection \ref{sec:2.3} and subsection
\ref{sec:2.5}, they have very concise structures. After two step
field redefinitions, the induced Yukawa couplings
Eq.~(\ref{ana-Y-tran}) are still of concise structure. These concise
structures are distorted by the summation in
Eq.~(\ref{yukawa-D-brane-5d-2}). When we reduce further the actions
from 5D to 4D, the 4D zero mode profiles distort the 5d Yukawa
couplings further. The exponential behaviors of 4D zero mode
profiles also induce the hierarchy mass structure in 4D. We see that
the structures of mixing matrices $V_{\mathcal{Y}L}$ and $V_{YL}$
are universal for quark fields. In our model, because the lepton
sector is similar to the quark sector, these structures also apply
to the lepton sector. These concise matrix structures are distorted
by the two sources we discussed above: the summation
$\widehat{\mathcal{Y}}^{D}+i\beta_{\mathcal{D}}\widehat{Y}^{D}$; and
the 4D zero mode profiles in Eq.~(\ref{mass-d-martix}). These two
sources distort these concise matrices to some general matrices.
Their analytical expressions are complicated and are not appropriate
to make qualitative discussions. The above discussions give some
sketchy interpretations about how our model works.

\subsection{More relevant discussions}\label{sec:4.2}
In subsection \ref{sec:2.5}, we suggest a new action
Eq.~(\ref{action-new}) for our model building. This new action
breaks the 6D local Lorentz invariance obviously. In this
subsection, we discuss the possible origin of this action.

The new term $\bar{\Psi}\Gamma^7e^6_a\Gamma^a\Psi$ suggests that we
may introduce the gauge interaction term
\begin{eqnarray}
\label{action-new-1form} A_N\bar{\Psi}\Gamma^7e^N_a\Gamma^a\Psi,
\end{eqnarray}
where $A_N$ is an Abelian gauge field. After that $A_N$ acquires the
background value $A_6=v,~A_i=0,~i=1,2,\cdots,5$, this interaction
term supplies the term $v\bar{\Psi}\Gamma^7e^6_a\Gamma^a\Psi$.
However, we argue that this interaction term is not a proper choice.
Because $\Gamma^7$ emerges in Eq.~(\ref{action-new-1form}), we need
to define the local gauge transformation for $\Psi$ as
\begin{eqnarray}
\label{action-new-1form-gauge} \Psi\rightarrow
e^{i\theta\Gamma^7}\Psi.
\end{eqnarray}
However, under such a gauge transformation, the term
$\bar{\Psi}\Psi$ transforms as
$\bar{\Psi}\Psi\rightarrow-\bar{\Psi}\Psi$. The mass terms
$\bar{\Psi}\Psi$ is prohibited by this gauge transformation. So the
interaction term Eq.~(\ref{action-new-1form}) is not appropriate to
produce the new action Eq.~(\ref{action-new}).

Instead of the above, we suggest another way to produce the action
Eq.~(\ref{action-new}). We introduce a 5-form gauge field strength
as follows
\begin{eqnarray}
\label{action-new-5form-gact} \mathscr{L}_{F}&=&-\frac{1}{2\cdot
5!}F_{MNHKL}F^{MNHKL},\\
F_{MNHKL}&=&\partial_{[M}A_{NHKL]}.\nonumber
\end{eqnarray}
This 5-form gauge field strength is similar to the Maxwell
electromagnetic field strength. We suppose that this 5-form field
interacts with the fermions by the nontrivial interaction term
\begin{eqnarray}
\label{action-new-5form-int} \mathscr{L}_{\mathrm{int}}&\propto&
\bar{\psi}\frac{1}{(F^2)^{\alpha}}F_{MNHKL}\Gamma^M\Gamma^N\Gamma^H\Gamma^K\Gamma^L\psi,\\
F^2&=&-\frac{1}{2\cdot 5!}F_{MNHKL}F^{MNHKL},\nonumber
\end{eqnarray}
where $\alpha$ is a constant real number. We have defined that
$\Gamma^N=e^N_a\Gamma^a$, in which $\Gamma^a$ are given by
Eq.~(\ref{Eq3}) and the index $a$ is summed.

In order to make the 5-form field to produce the appropriate
background value, we construct the following interaction system
\begin{eqnarray}
\label{gra-system} S&=&\int d^4 x dy dz \sqrt{-g} \{2M^4 R\} + \int
d^4 x dy dz \sqrt{-g}
\left\{\frac{1}{2}g^{KL}\nabla_{K}\phi\nabla_{L}\phi+V(\phi)\right\}\nonumber\\
&+&\int d^4 x dy dz \sqrt{-g} \left\{-\frac{1}{2\cdot
5!}F_{MNHKL}F^{MNHKL}\right\}.
\end{eqnarray}
in which $\phi$ is a scalar field and $V(\phi)$ is its potential
term. We solve this system supposing the metric Ansatz
Eq.~(\ref{introduce}). Suppose that
\begin{eqnarray}
\label{gra-system-5form}
F^{\mu\nu\alpha\beta\gamma}&=&\frac{1}{\sqrt{-g}}f(z)\epsilon^{\mu\nu\alpha\beta\gamma},~~f(z)=vB^{-4}(z),\\
\mu,\nu,\alpha,\beta,\gamma&=&0,1,2,3,5,~~\epsilon^{01235}=1,\nonumber
\end{eqnarray}
where $v$ is a constant. Then the equation of motion of this 5-form
field strength and the Bianchi identities for it are both satisfied.
We also suppose that the scalar field $\phi$ only depends on the
coordinate $z$. By the metric Ansatz Eq.~(\ref{introduce}), we
obtain the following equations
\begin{eqnarray}
\label{einstein2}
4B^{-1}B_{zz}+2B^{-2}B^2_{z}+3A^{-3}A_{yy}&=&\frac{1}{4M^4}\left[-B^2\left(\frac{1}{2}B^{-2}
\phi^2_{z}+V(\phi)\right)\right]-\frac{v^2}{8M^4}B^{-8},\\
\label{einstein3}
4B^{-1}B_{zz}+2B^{-2}B^2_{z}+6A^{-4}A^2_{y}&=&\frac{1}{4M^4}\left[-B^2\left(\frac{1}{2}B^{-2}
\phi^2_{z}+V(\phi)\right)\right]-\frac{v^2}{8M^4}B^{-8},\\
\label{einstein4}
10B^{-2}B^2_{z}+4A^{-3}A_{yy}+2A^{-4}A^2_{y}&=&\frac{1}{4M^4}\left[\phi^2_{z}-B^2\left(\frac{1}{2}B^{-2}
\phi^2_{z}+V(\phi)\right)\right]+\frac{v^2}{8M^4}B^{-8},~~\\
\label{einstein5} B^{-2}\phi_{zz}+4B^{-3}B_{z} \phi_z-\frac{d
V(\phi)}{d \phi}&=&0,
\end{eqnarray}
in which $A_y=\frac{dA}{dy},~B_z=\frac{dB}{dz},~\phi_z=\frac{d
\phi}{dz}$. Obviously, $A(y)$ should be of the form
\begin{eqnarray}
\label{gra-5d} A(y)=\frac{1}{ky+c},
\end{eqnarray}
in which $k,~c$ are constants. This system is similar to that in our
previous paper \cite{Guo:2008}. The conclusions for that system
apply here. For any $B(z)$, there exists an appropriate $V(\phi)$,
which makes Eqs. (\ref{einstein2})-(\ref{einstein5}) to be
satisfied. So we can choose an appropriate $V(\phi)$ to make the
metric Eq.~(\ref{suppose1}) to be our background solutions. This
system is also complicated, and we can not find a superpotential to
express the general solutions.

By the above discussions, if we suppose $\alpha=\frac{2}{5}$ in
Eq.~(\ref{action-new-5form-int}), then we obtain
\begin{eqnarray}
\label{action-new-5form-int-1} \mathscr{L}_{int}&\propto&
\bar{\psi}vB^{-1}\Gamma^7\Gamma^6\psi.
\end{eqnarray}
This term is exactly that we expect in Eq.~(\ref{action-new}). So
the above system can give the new action Eq.~(\ref{action-new}) and
give the background solutions Eq.~(\ref{suppose1-model}) at the same
time. However, note that the term Eq.~(\ref{action-new-5form-int})
has a very nontrivial form for $\alpha=\frac{2}{5}$, which implies
that it will be intractable when we treat it as a quantum theory. We
have not found a more simple term to replace it yet.

\subsection{Gauge fields in the bulk and a spontaneously broken framework for CP violation}\label{sec:4.3}

In our model building in subsection \ref{sec:3.1}, we have supposed
that the gauge fields are confined on the 3-brane sited at
$(z=R,~y=L^{\prime})$. By this assumption, the gauge fields only
propagate in the physical 4D spacetime, and we do not need to make
the conventional KK decompositions for them. In this subsection, we
consider the possibility that the gauge fields propagate in higher
dimensions.

When we consider gauge fields in higher dimensions, a natural choice
is that gauge fields propagate in the 6D bulk. However, gauge fields
propagating in the 6D bulk induces intractable problems. Because the
5th space dimension $z$ is intrinsically semi-infinite in the metric
Ansatz (\ref{ansatz-model}), the bounded KK modes of gauge fields
must be non-constant. Such non-constant profiles of gauge fields
would break the unitarity of mixing matrix in the ZMA approach. This
situation differs from the conventional flavor models in RS
spacetime. In RS flavor models like
\cite{Agashe:2005,Casagrande:2008}, the 5th dimension is a finite
interval and the zero modes of gauge fields have constant profiles.
These constant zero modes profiles keep the unitarity of mixing
matrix in the ZMA approach. So in our model building in subsection
\ref{sec:3.1}, we consider the situation that the gauge fields only
propagate in the 4D spacetime. This choice makes the numerical
results of our model to be close to the experimental data.

As we just discussed above, gauge fields in the 6D bulk induce
intractable problems. However, the gauge fields can propagate in the
5D spacetime. We can consider the situation that the gauge fields
are confined on the 4-brane sited at $z=R$. Because the zero mode
profiles of gauge fields can be constant, the unitarity of mixing
matrix can be kept in the ZMA approach. By some modifications, the
model in subsection \ref{sec:3.1} can apply similarly in this
situation.

Moreover, the gauge fields propagating in 5D can bring interesting
influence on the mechanics for CP violation. In our model in
subsection \ref{sec:3.1}, CP violations originate from the terms
after the coefficients $\beta$, which are put in by hand. If we
consider that gauge fields propagate in 5D, we may have a new
mechanics for CP violation. According to the mechanics suggested in
\cite{Cosme:2004}, the 5th component of gauge field can develop a
vacuum expectation value through the gauge invariant line integral
\begin{eqnarray}
\label{spon-cp} \langle A_y\rangle=\int dy A_y.
\end{eqnarray}
This vacuum expectation value can break the non-Abelian gauge group
and also supply an origin for CP violation. In order to make a
realistic model, we may embed the electroweak unification group
$SU(2)\times U(1)$ into a larger unification group $SO(5)\times
U(1)$. The vacuum expectation value in Eq.~(\ref{spon-cp}) has also
been discussed in gauge-Higgs unification framework
\cite{Hosotani:2008}. So instead of putting in CP violation by hand
as we did in subsection \ref{sec:3.1}, the gauge fields in 5D can
supply a spontaneously broken framework for CP violation according
to the above mechanics.

\section{Conclusions}\label{sec:5}

In warped extra dimensional RS model, the fermion mass hierarchies
can be produced by the 5D bulk mass parameters of the same order. In
our previous paper \cite{Guo:2008}, we suggest that these 5D mass
parameters can be interpreted in a two-layer warped 6D model, and
such an approach also supply a solution for family problem. In this
paper, we combine these suggestions and construct a specific model
to address the fermion mass hierarchy problem and the family
problems at the same time. We give numerical examples in subsection
\ref{sec:3.3} to show that the numerical results of this model can
be very close to the experimental data in both the quark sector and
the lepton sector. However, because there still exist many
parameters in our model, we only make rough numerical treatments
about the model, and do not further adjust parameters to fit the
experimental data in higher precision.

We further make some analytical treatments for our model in
subsection \ref{sec:4.1}. These analytical treatments show that some
very concise structures exist in this model. They imply some common
features shared by quarks and leptons. However, the breaking of
these concise structures makes the model to be a complicated one,
and we do not make more analytical discussions. Some approximate
treatments may be helpful to illuminate this model more clearly. In
addition, a natural question is that whether we can interpret the
parameters in Table.~\ref{table1} and Table.~\ref{table2}. It seems
that it is appropriate to interpret the origin of those parameters
in a grand unification framework like in \cite{Goldberger:2003}.

{\bf Acknowledgement} This work is partially supported by National
Natural Science Foundation of China (No.~10721063), by the Key Grant
Project of Chinese Ministry of Education (No.~305001), by the
Research Fund for the Doctoral Program of Higher Education (China).

\appendix

\section{Analysis about the equations determining the eigenvalues}\label{sec:A}
In this appendix, we analyze how equations (\ref{cut1-a}),
(\ref{cut1-b}), (\ref{cut2-a}) and (\ref{cut2-b}) restrict the
number of eigenvalues to be finite.

Solving equations (\ref{cut1-a}), (\ref{cut1-b}), (\ref{cut2-a}) and
(\ref{cut2-b}) for $\lambda$, we obtain
\begin{eqnarray}
\label{cut1-a-s}
\left(\frac{\lambda}{\omega}\right)^2&=&\left(\frac{m}{\omega}s\right)^2-\left\{\frac{(\frac{m}{\omega}s)^2[(\frac{a}{b})^2-1]-(n+\rho)^2}{2(n+\rho)}\right\}^2~~~~~~~\mathrm{for}~~(\ref{cut1-a}),\\
\label{cut1-b-s}
\left(\frac{\lambda}{\omega}\right)^2&=&\left(\frac{m}{\omega}s\right)^2-\left\{\frac{(\frac{m}{\omega}s)^2[(\frac{a}{b})^2-1]-(n+1-\rho)^2}{2(n+1-\rho)}\right\}^2~~\mathrm{for}~~(\ref{cut1-b}),\\
\label{cut2-a-s}
\left(\frac{\lambda}{\omega}\right)^2&=&\left(\frac{m}{\omega}s\right)^2-\left\{\frac{(\frac{m}{\omega}s)^2[(\frac{a}{b})^2-1]-(n+\rho)^2}{2(n+\rho)}\right\}^2~~~~~~~\mathrm{for}~~(\ref{cut2-a}),\\
\label{cut2-b-s}
\left(\frac{\lambda}{\omega}\right)^2&=&\left(\frac{m}{\omega}s\right)^2-\left\{\frac{(\frac{m}{\omega}s)^2[(\frac{a}{b})^2-1]-(n+1-\rho)^2}{2(n+1-\rho)}\right\}^2~~\mathrm{for}~~(\ref{cut2-b}).
\end{eqnarray}
From these solutions, because we suppose that $\frac{m}{\omega}s$,
$\frac{a}{b}$ are real and $n$ is a non-negative integer, we can
infer that $\left(\frac{\lambda}{\omega}\right)^2$ must be real, in
other words, $\frac{\lambda}{\omega}$ must be real or pure
imaginary. For simplicity, define
$t=\left(\frac{\lambda}{\omega}\right)^2$, then $t$ is a real
number.

In the following analysis, we suppose that $m>0$. For the case
$m<0$, we discuss it in appendix \ref{sec:B}. Now we discuss these
solutions in two cases:

Case~(1): In this case, let $\frac{a}{b}>1$, so
$\rho=\frac{m}{\omega}s(1-\frac{a}{b})<0$. Rewrite equations
(\ref{cut1-a}), (\ref{cut1-b}), (\ref{cut2-a}) and (\ref{cut2-b}) as
\begin{eqnarray}
\label{cut1-a-r}
n+\rho&=&\sqrt{\left(\frac{m}{\omega}s\frac{a}{b}\right)^2-t}-\sqrt{\left(\frac{m}{\omega}s\right)^2-t},\\
\label{cut1-b-r}
n+1-\rho&=&\sqrt{\left(\frac{m}{\omega}s\frac{a}{b}\right)^2-t}-\sqrt{\left(\frac{m}{\omega}s\right)^2-t},\\
\label{cut2-a-r}
n+\rho&=&-\sqrt{\left(\frac{m}{\omega}s\frac{a}{b}\right)^2-t}-\sqrt{\left(\frac{m}{\omega}s\right)^2-t},\\
\label{cut2-b-r}
n+1-\rho&=&-\sqrt{\left(\frac{m}{\omega}s\frac{a}{b}\right)^2-t}-\sqrt{\left(\frac{m}{\omega}s\right)^2-t}.
\end{eqnarray}
Because $\rho<0$ and $n\geq 0$, Eq.~(\ref{cut2-b-r}) has no
solutions for any $t$ obviously; while equations (\ref{cut1-a-r}),
(\ref{cut1-b-r}) and (\ref{cut2-a-r}) might have solutions only when
$t<\left(\frac{m}{\omega}s\right)^2$. Define
\begin{eqnarray}
\label{cut1-a-f}
f(t)&=&\sqrt{\left(\frac{m}{\omega}s\frac{a}{b}\right)^2-t}-\sqrt{\left(\frac{m}{\omega}s\right)^2-t},\\
\label{cut2-a-g}
g(t)&=&-\sqrt{\left(\frac{m}{\omega}s\frac{a}{b}\right)^2-t}-\sqrt{\left(\frac{m}{\omega}s\right)^2-t}.
\end{eqnarray}
$f(t)$ and $g(t)$ are both increasing functions about $t$ when
$t<\left(\frac{m}{\omega}s\right)^2$. By this feature, we obtain
\begin{eqnarray}
\label{cut1-a-ex}
0<f(t)\leq \frac{m}{\omega}s\sqrt{\left(\frac{a}{b}\right)^2-1},\\
\label{cut2-a-ex}g(t)\leq
-\frac{m}{\omega}s\sqrt{\left(\frac{a}{b}\right)^2-1}.
\end{eqnarray}
By equations (\ref{cut1-a-ex}) and (\ref{cut2-a-ex}), we can
determine the extent of $n$ as
\begin{eqnarray}
\label{cut1-a-n-ex}
\frac{m}{\omega}s\left(\frac{a}{b}-1\right)<&n&\leq
\frac{m}{\omega}s\left[\sqrt{\left(\frac{a}{b}\right)^2-1}+\left(\frac{a}{b}-1\right)\right]~~~~\mathrm{for~~(\ref{cut1-a-r})},\\
\label{cut1-b-n-ex}
-\frac{m}{\omega}s\left(\frac{a}{b}-1\right)<&n+1&\leq
\frac{m}{\omega}s\left[\sqrt{\left(\frac{a}{b}\right)^2-1}-\left(\frac{a}{b}-1\right)\right]~~~~\mathrm{for~~(\ref{cut1-b-r})},\\
\label{cut2-a-n-ex} &n&\leq
\frac{m}{\omega}s\left[-\sqrt{\left(\frac{a}{b}\right)^2-1}+\left(\frac{a}{b}-1\right)\right]~~\mathrm{for~~(\ref{cut2-a-r})}.
\end{eqnarray}
When $\frac{a}{b}>1$,
$\frac{m}{\omega}s\left[-\sqrt{\left(\frac{a}{b}\right)^2-1}+\left(\frac{a}{b}-1\right)\right]<0$.
So Eq.~(\ref{cut2-a-n-ex}) is impossible, and the corresponding
equation (\ref{cut2-a-r}) has no solutions.

In summaries, in the case, the extent of $n$ is bounded by equations
(\ref{cut1-a-n-ex}) and (\ref{cut1-b-n-ex}). Note that it does not
imply that the two equations should be satisfied at the same time.
They mean we can cut off the series in two different ways; while
each cut off of the series provides a kind of solutions for the
equation (\ref{solution17}). We might have two kinds of solutions in
this case.

Case~(2): In this case, let $\frac{a}{b}<1$, so
$\rho=\frac{m}{\omega}s(1-\frac{a}{b})>0$. From equations
(\ref{cut1-a-r}), (\ref{cut1-b-r}), (\ref{cut2-a-r}) and
(\ref{cut2-b-r}), we can infer that Eq.~(\ref{cut2-a-r}) has no
solutions for any $t$; while equations (\ref{cut1-a-r}),
(\ref{cut1-b-r}) and (\ref{cut2-b-r}) might have solutions when
$t<\left(\frac{m}{\omega}s\frac{a}{b}\right)^2$.

In this case, $f(t)$ is a decreasing function about $t$ when
$t<\left(\frac{m}{\omega}s\frac{a}{b}\right)^2$; while $g(t)$ is
still a increasing function about $t$ when
$t<\left(\frac{m}{\omega}s\frac{a}{b}\right)^2$. We obtain
\begin{eqnarray}
\label{cut1-a-ex-2}
-\frac{m}{\omega}s\sqrt{1-\left(\frac{a}{b}\right)^2}&\leq & f(t)<0,\\
\label{cut2-a-ex-2}g(t)&\leq &
-\frac{m}{\omega}s\sqrt{1-\left(\frac{a}{b}\right)^2}.
\end{eqnarray}
By these equations, we determine the extent of $n$ to be
\begin{eqnarray}
\label{cut1-a-n-ex-2}
-\frac{m}{\omega}s\left[\sqrt{1-\left(\frac{a}{b}\right)^2}+\left(1-\frac{a}{b}\right)\right]&\leq & n<-\frac{m}{\omega}s\left(1-\frac{a}{b}\right)
~~~~~~~~~~~~~~~~~~~~\mathrm{for~~(\ref{cut1-a-r})},\\
\label{cut1-b-n-ex-2}
\frac{m}{\omega}s\left[\sqrt{1-\left(\frac{a}{b}\right)^2}+\left(1-\frac{a}{b}\right)\right]&\leq
& n+1<\frac{m}{\omega}s\left(1-\frac{a}{b}\right)
~~~~~~~~~~~~~~~~~\mathrm{for~~(\ref{cut1-b-r})},\\
\label{cut2-b-n-ex-2} n+1&\leq &
\frac{m}{\omega}s\left[-\sqrt{1-\left(\frac{a}{b}\right)^2}+\left(1-\frac{a}{b}\right)\right]
~~~\mathrm{for~~(\ref{cut2-b-r})}.
\end{eqnarray}
For $\frac{a}{b}<1$,
$\frac{m}{\omega}s\left[-\sqrt{1-\left(\frac{a}{b}\right)^2}+\left(1-\frac{a}{b}\right)\right]<0$.
So Eq.~(\ref{cut2-b-n-ex-2}) is impossible, and the corresponding
equation (\ref{cut2-b-r}) has no solutions. While
Eq.~(\ref{cut1-a-n-ex-2}) is also impossible obviously, and the
corresponding equation (\ref{cut1-a-r}) has no solutions.

In summaries, in this case, only Eq.~(\ref{cut1-b-n-ex-2}) is
possible, and the corresponding equation (\ref{cut1-b-r}) has
solutions. We only have a way to cut off the series, and we can have
one kind of solutions corresponding to this cut off when
$\frac{a}{b}<1$.

In addition, we make more discussions about a special case. We
analyze whether equations (\ref{cut1-a-r}), (\ref{cut1-b-r}),
(\ref{cut2-a-r}) and (\ref{cut2-b-r}) can have the solution $t=0$.
Replacing $t$ with $0$ in these equations, we find that only
Eq.~(\ref{cut1-a-r}) is possible. We obtain the condition
\begin{eqnarray}
\label{lambda-zero} n=2\frac{m}{\omega}s\left(\frac{a}{b}-1\right).
\end{eqnarray}
For $\frac{a}{b}>1$, this condition can be satisfied. When the
condition Eq.~(\ref{lambda-zero}) is satisfied, the $n^{th}$ massive
solutions will coincide with the zero mode solutions given by
Eq.~(\ref{ZeroMode-a}) in appendix \ref{sec:B}.

\section{Explicit solutions for zero modes and massive modes}\label{sec:B}
In this appendix, we give the solutions for zero modes and massive
modes explicitly. In the following we discuss two cases: $m>0$ and
$m<0$.

Case~(1): For the positive bulk mass parameters, that is, $m>0$. The
normalizable zero mode for the metric (\ref{suppose1}) is given by
\begin{eqnarray}
\label{ZeroMode-a}
F_{0}(x)=0,~G_{0}(x)=\frac{\sqrt{\omega}}{\sqrt{N_0}}x^{-\frac{m}{\omega}s\frac{a}{b}}(x+1)^{\frac{m}{\omega}s(\frac{a}{b}-1)},
\end{eqnarray}
where we have defined $x=\frac{e^{\omega z}}{b}$.

For the massive modes, the solution determined by the parameter set
(\ref{cut1-a-pc}), (\ref{cut1-b-pc}) and (\ref{cut1-c-pc}) is given
by
\begin{eqnarray}
\label{Massive-2}
F_1(x)=\frac{\sqrt{\omega}}{\sqrt{N_1}}x^{-\mu_1}(x+1)^{\mu_1-\nu_1}\left[1-\frac{\beta_1}{\gamma_1}\frac{x}{x+1}\right],
\end{eqnarray}
in which
\begin{eqnarray}
\label{Massive-1a}\gamma_n&=&1-2\mu_n,~~n=1,2,3,\cdots\\
\label{Massive-1b}\beta_n&=&1-\rho-\mu_n+\nu_n,~\rho=\frac{m}{\omega}s\left(1-\frac{a}{b}\right),\\
\label{Massive-1c}\nu_n&=&\sqrt{\left(\frac{m}{\omega}s\right)^2-\left(\frac{\lambda_n}{\omega}\right)^2},\\
\label{Massive-1d}\mu_n&=&\sqrt{\left(\frac{m}{\omega}s\right)^2\left(\frac{a}{b}\right)^2-\left(\frac{\lambda_n}{\omega}\right)^2}.
\end{eqnarray}
The solution for $G_1(x)$ is determined by the equation
\begin{eqnarray}
\label{Massive-G}
G_{n}(x)=\frac{\omega}{\lambda_n}\left[\frac{m}{\omega}s\frac{x+\frac{a}{b}}{x+1}F_{n}(x)-x\frac{d}{dx}F_{n}(x)\right],~~n=1,2,3,\cdots.
\end{eqnarray}

According to the discussions in subsection \ref{sec:2.3}, the
massive modes emerge in pairs. The other solution in pairs with the
solution $(F_1,~G_1)$, that is, the solution corresponding to the
eigenvalue $-\lambda_1$, is given by
\begin{eqnarray}
\label{Massive-2-pair}
F_{-1}(x)=F_1(x),~~G_{-1}(x)=-G_1(x).
\end{eqnarray}

Case~(2): Now we discuss the case $m<0$. Redefine $m=-\tilde{m}$,
then $\tilde{m}>0$. The action (\ref{action1}) becomes to be
\begin{eqnarray}
\label{action1-nem} S = \int d^4 x dy dz \sqrt{-g} \left\{
\frac{i}{2} \left[\bar{\Psi} \, e_a^M \Gamma^a \nabla_M \Psi -
\nabla_M \bar{\Psi} \, e_a^M \Gamma^a \Psi \right] +i \, \tilde{m}
\bar{\Psi} \Psi\right\}.
\end{eqnarray}
After KK decompositions like in subsection \ref{sec:2.1}, the
equations (\ref{solution13}) and (\ref{solution14}) change to
\begin{eqnarray}
\label{solution13-nem}
\left(\frac{d}{d z}+\tilde{m} B\right)F_n(z) + \lambda_n G_n(z)=0,\\
\label{solution14-nem} \left(\frac{d}{d z}-\tilde{m} B \right)G_n(z)
- \lambda_n F_n(z)=0.
\end{eqnarray}
The induced second equations also change correspondingly to be
\begin{eqnarray}
\label{solution17-nem} -\frac{d^2}{d z^2}F_n(z)
+V(z)F_n(z)&=&\lambda_n^2 F_n(z),\\
\label{solution18-nem} -\frac{d^2}{d z^2}G_n(z)+\widetilde{V}(z)
G_n(z) &=&\lambda_n^2 G_n(z),
\end{eqnarray}
with potentials
\begin{eqnarray}
\label{solution19-nem} V(z)=-\tilde{m} B_z+\tilde{m}^2
B^2,~\widetilde{V}(z)=\tilde{m} B_z+\tilde{m}^2 B^2,
\end{eqnarray}
From equations (\ref{solution18-nem}) and (\ref{solution19-nem}), we
see that $G_n(z)$ conform to the similar equation like $F_n(z)$ in
equations (\ref{solution17}) and (\ref{solution18}). The massive
solutions for $G_n(z)$ are given by
\begin{eqnarray}
\label{suppose1-2-nem} G(z)&=&C_1 e^{-\mu \omega z}(e^{\omega
z}+b)^{\mu-\nu}
\mathrm{hypergeom}\left(\rho-\mu+\nu,1-\rho-\mu+\nu;1-2\mu,\frac{e^{\omega
z}}{e^{\omega z}+b}\right) \nonumber \\&+&C_2 e^{\mu \omega
z}(e^{\omega z}+b)^{-\mu-\nu}
\mathrm{hypergeom}\left(\rho+\mu+\nu,1-\rho+\mu+\nu;1+2\mu,\frac{e^{\omega
z}}{e^{\omega z}+b}\right),~~~
\end{eqnarray}
where $\rho=\frac{\tilde{m}}{\omega}s(1-\frac{a}{b})$,
$\mu=\sqrt{\left(\frac{\tilde{m}}{\omega}s\right)^2\left(\frac{a}{b}\right)^2-\left(\frac{\lambda}{\omega}\right)^2}$,
and
$\nu=\sqrt{\left(\frac{\tilde{m}}{\omega}s\right)^2-\left(\frac{\lambda}{\omega}\right)^2}$.
Therefore, the conclusions about the extent of $n$ in subsection
\ref{sec:2.3} and appendix \ref{sec:A} applies here. We just need
replace $m$ with $\tilde{m}$, because we assume $m>0$ in those
analysis. However, the solutions of zero mode and massive modes
change.

The zero mode solution is given by
\begin{eqnarray}
\label{ZeroMode-a-nem}
F_{0}(x)=\frac{\sqrt{\omega}}{\sqrt{N_0}}x^{-\frac{\tilde{m}}{\omega}s\frac{a}{b}}(x+1)^{\frac{\tilde{m}}{\omega}s(\frac{a}{b}-1)},~G_{0}(x)=0.
\end{eqnarray}
The massive mode for $G_1(z)$ is given by
\begin{eqnarray}
\label{Massive-2-nem}
G_1(x)=\frac{\sqrt{\omega}}{\sqrt{N_1}}x^{-\mu_1}(x+1)^{\mu_1-\nu_1}\left[1-\frac{\beta_1}{\gamma_1}\frac{x}{x+1}\right],\\
\end{eqnarray}
in which
\begin{eqnarray}
\label{Massive-1a-nem}\gamma_n&=&1-2\mu_n,~~n=1,2,3,\cdots\\
\label{Massive-1b-nem}\beta_n&=&1-\rho-\mu_n+\nu_n,~\rho=\frac{\tilde{m}}{\omega}s\left(1-\frac{a}{b}\right),\\
\label{Massive-1c-nem}\nu_n&=&\sqrt{\left(\frac{\tilde{m}}{\omega}s\right)^2-\left(\frac{\lambda_n}{\omega}\right)^2},\\
\label{Massive-1d-nem}\mu_n&=&\sqrt{\left(\frac{\tilde{m}}{\omega}s\right)^2\left(\frac{a}{b}\right)^2-\left(\frac{\lambda_n}{\omega}\right)^2}.
\end{eqnarray}
While the solution for $F_1(x)$ is determined by the equation
\begin{eqnarray}
\label{Massive-G}
F_{n}(x)=\frac{\omega}{\lambda_n}\left[-\frac{\tilde{m}}{\omega}s\frac{x+\frac{a}{b}}{x+1}G_{n}(x)+x\frac{d}{dx}G_{n}(x)\right],~~n=1,2,3,\cdots.
\end{eqnarray}
For the other solution in pairs with the solution $(F_1,~G_1)$, now
we obtain
\begin{eqnarray}
\label{Massive-2-pair-nem} F_{-1}(x)=-F_1(x),~~G_{-1}(x)=G_1(x).
\end{eqnarray}

In order to determine the normalization constants $N_0$ and $N_1$ in
the above equations, we designate the normalization conditions as
\begin{eqnarray}
\label{norm-app} \int dz
\left({F}_{n}^{\ast}{F}_n+{G}_{n}^{\ast}{G}_n\right)=\delta_{nn},~~n=0,1,-1,
\end{eqnarray}
where $\delta_{nn}=1$ and $n$ is not be summed.

\section{Definitions for quantities}\label{sec:C}

We define the quantities in equations (\ref{ana-K-M-5d-Q}) and
(\ref{ana-K-M-5d-D}) as follows. For the field $\mathbb{Q}$, we
define that
\begin{eqnarray}
\label{ana-def-Q} a_{\mathbb{Q}}&=&-i\int dz
(F_0^{\mathbb{Q}\ast}F_1^{\mathbb{Q}}+G_0^{\mathbb{Q}\ast}G_1^{\mathbb{Q}}),~~b_{\mathbb{Q}}=\int
dz
(F_1^{\mathbb{Q}\ast}F_{-1}^{\mathbb{Q}}+G_1^{\mathbb{Q}\ast}G_{-1}^{\mathbb{Q}}),\\
\tilde{a}_{\mathbb{Q}}&=&\frac{i}{2}a_{\mathbb{Q}}\triangle\lambda^{\mathbb{Q}},~~\tilde{b}_{\mathbb{Q}}=-ib_{\mathbb{Q}}\triangle\lambda^{\mathbb{Q}}.\nonumber
\end{eqnarray}
Note that $\triangle\lambda^{\mathbb{Q}}$ is pure imaginary
according to our designation in subsection \ref{sec:4.1}, so
$\tilde{a}_{\mathbb{Q}}$ and $\tilde{b}_{\mathbb{Q}}$ are real. For
the field $\mathcal{D}$, we define that
\begin{eqnarray}
\label{ana-def-D} a_{\mathcal{D}}&=&\int dz
(F_0^{\mathcal{D}\ast}F_1^{\mathcal{D}}+G_0^{\mathcal{D}\ast}G_1^{\mathcal{D}}),~~b_{\mathcal{D}}=\int
dz
(F_1^{\mathcal{D}\ast}F_{-1}^{\mathcal{D}}+G_1^{\mathcal{D}\ast}G_{-1}^{\mathcal{D}}),\\
\tilde{a}_{\mathcal{D}}&=&\frac{1}{2}a_{\mathcal{D}}\triangle\lambda^{\mathcal{D}}.\nonumber
\end{eqnarray}
Note that here $\triangle\lambda^{\mathcal{D}}$ is real according to
our designation in subsection \ref{sec:4.1}. Functions $F_n(z)$ and
$G_n(z)$ are determined by equations (\ref{sol13-new-var}) and
(\ref{sol14-new-var}). They can be determined according to our
discussions in subsection \ref{sec:2.5}. Obviously they have the
similar forms and characters to the solutions given in appendix
\ref{sec:B}. Due to the special characters of these solutions
discussed in subsection \ref{sec:2.3}, the matrix $K$ and other
matrices in subsection \ref{sec:4.1} all have very concise
structures.

For the quantities in Eq.~(\ref{ana-Y}), we have defined them as
follows. For $\mathcal{Y}^{D}$, we define that
\begin{eqnarray}
\label{ana-def-Y-1} \mathbbm{s}&=&\int dz
B(z)(F_0^{\mathbb{Q}\ast}G_0^{\mathcal
{D}}+G_0^{\mathbb{Q}\ast}F_0^{\mathcal {D}}),~~\mathbbm{a}=\int dz
B(z)(F_0^{\mathbb{Q}\ast}G_1^{\mathcal
{D}}+G_0^{\mathbb{Q}\ast}F_1^{\mathcal {D}}),\\
\mathbbm{b}&=&\int dz B(z)(F_1^{\mathbb{Q}\ast}G_0^{\mathcal
{D}}+G_1^{\mathbb{Q}\ast}F_0^{\mathcal {D}}),~~\mathbbm{c}=\int dz
B(z)(F_1^{\mathbb{Q}\ast}G_1^{\mathcal
{D}}+G_1^{\mathbb{Q}\ast}F_1^{\mathcal
{D}}),\nonumber\\
\mathbbm{d}&=&\int dz B(z)(F_1^{\mathbb{Q}\ast}G_{-1}^{\mathcal
{D}}+G_1^{\mathbb{Q}\ast}F_{-1}^{\mathcal {D}}).\nonumber
\end{eqnarray}
For $Y^{D}$, we define that
\begin{eqnarray}
\label{ana-def-Y-2} \mathbbm{a}^{\prime}&=&\int dz
B(z)(F_0^{\mathbb{Q}\ast}F_1^{\mathcal
{D}}-G_0^{\mathbb{Q}\ast}G_1^{\mathcal
{D}}),~~\mathbbm{b}^{\prime}=\int dz
B(z)(F_1^{\mathbb{Q}\ast}F_0^{\mathcal
{D}}-G_1^{\mathbb{Q}\ast}G_0^{\mathcal {D}}),\\
\mathbbm{c}^{\prime}&=&\int dz
B(z)(F_1^{\mathbb{Q}\ast}F_1^{\mathcal
{D}}-G_1^{\mathbb{Q}\ast}G_1^{\mathcal
{D}}),~~\mathbbm{d}^{\prime}=\int dz
B(z)(F_1^{\mathbb{Q}\ast}F_{-1}^{\mathcal
{D}}-G_1^{\mathbb{Q}\ast}G_{-1}^{\mathcal {D}}).\nonumber
\end{eqnarray}
The elements in equations (\ref{ana-def-Y-1}) and
(\ref{ana-def-Y-2}) can be complex numbers generally.

The eigenvalues in equations (\ref{ana-K-cho-Q}) and
(\ref{ana-K-cho-D}) are given by
\begin{eqnarray}
\label{ana-K-cho-Q-eigenval} \Lambda_1^{\mathbb{Q}}&=&1-\frac{
b_{\mathbb{Q}}}{2}-\frac{1}{2}\sqrt{8a^2_{\mathbb{Q}}+b^2_{\mathbb{Q}}},~~\Lambda_2^{\mathbb{Q}}=1-\frac{
b_{\mathbb{Q}}}{2}+\frac{1}{2}\sqrt{8a^2_{\mathbb{Q}}+b^2_{\mathbb{Q}}},~~\Lambda^{\mathbb{Q}}_3=1+b_{\mathbb{Q}},\\
\label{ana-K-cho-D-eigenval} \Lambda_1^{\mathcal{D}}&=&1-\frac{
b_{\mathcal{D}}}{2}-\frac{1}{2}\sqrt{8a^2_{\mathcal{D}}+b^2_{\mathcal{D}}},~~\Lambda_2^{\mathcal{D}}=1-\frac{
b_{\mathcal{D}}}{2}+\frac{1}{2}\sqrt{8a^2_{\mathcal{D}}+b^2_{\mathcal{D}}},~~\Lambda^{\mathcal{D}}_3=1+b_{\mathcal{D}}.
\end{eqnarray}

The elements in equations (\ref{ana-M-Q-tilde}) and
(\ref{ana-M-D-tilde}) are given as follows. For the field
$\mathbb{Q}$, we define that
\begin{eqnarray}
\label{ana-M-Q-tilde-element}
\hat{a}_{\mathbb{Q}}&=&\left(-\frac{1}{\sqrt{2}}a_{\mathbb{Q}}\triangle\lambda^{\mathbb{Q}}\cos{\theta_{\mathbb{Q}}}+
b_{\mathbb{Q}}\triangle\lambda^{\mathbb{Q}}\sin{\theta_{\mathbb{Q}}}
\right)\frac{1}{\sqrt{\Lambda^{\mathbb{Q}}_1}}\frac{1}{\sqrt{\Lambda^{\mathbb{Q}}_3}},\\
\hat{b}_{\mathbb{Q}}&=&\left(\frac{1}{\sqrt{2}}a_{\mathbb{Q}}\triangle\lambda^{\mathbb{Q}}\sin{\theta_{\mathbb{Q}}}+
b_{\mathbb{Q}}\triangle\lambda^{\mathbb{Q}}\cos{\theta_{\mathbb{Q}}}
\right)\frac{1}{\sqrt{\Lambda^{\mathbb{Q}}_2}}\frac{1}{\sqrt{\Lambda^{\mathbb{Q}}_3}}.\nonumber
\end{eqnarray}
For the field $\mathcal{D}$, we define that
\begin{eqnarray}
\label{ana-M-D-tilde-element}
\hat{a}_{\mathcal{D}}&=&\left(\frac{1}{\sqrt{2}}a_{\mathcal{D}}\triangle\lambda^{\mathcal{D}}\cos{\theta_{\mathcal{D}}}-
\triangle\lambda^{\mathcal{D}}\sin{\theta_{\mathcal{D}}}
\right)\frac{1}{\sqrt{\Lambda^{\mathcal{D}}_1}}\frac{1}{\sqrt{\Lambda^{\mathcal{D}}_3}},\\
\hat{b}_{\mathcal{D}}&=&\left(\frac{1}{\sqrt{2}}a_{\mathcal{D}}\triangle\lambda^{\mathcal{D}}\sin{\theta_{\mathcal{D}}}+
\triangle\lambda^{\mathcal{D}}\cos{\theta_{\mathcal{D}}}
\right)\frac{1}{\sqrt{\Lambda^{\mathcal{D}}_2}}\frac{1}{\sqrt{\Lambda^{\mathcal{D}}_3}}.\nonumber
\end{eqnarray}

The elements of $\tilde{H}^{\mathbb{Q}\dagger}$ in
Eq.~(\ref{ana-yukawa-D-brane-5d-Y-2}) are given as follows.
$\mathbbm{s}_{\mathbb{Q}}$, $\mathbbm{a}_{\mathbb{Q}}$ and
$\mathbbm{b}_{\mathbb{Q}}$ are given by
\begin{eqnarray}
\label{ana-yukawa-D-brane-5d-Y-2-elementQ}
\mathbbm{s}_{\mathbb{Q}}&=&i\left(\frac{1}{\sqrt{\Lambda^{\mathbb{Q}}_1}}\cos{\theta_{\mathbb{Q}}}\cos{\vartheta_{\mathbb{Q}}}-
\frac{1}{\sqrt{\Lambda^{\mathbb{Q}}_2}}\sin{\theta_{\mathbb{Q}}}\sin{\vartheta_{\mathbb{Q}}}\right),\\
\mathbbm{a}_{\mathbb{Q}}&=&-\frac{1}{\sqrt{2}}\left(\frac{1}{\sqrt{\Lambda^{\mathbb{Q}}_1}}\sin{\theta_{\mathbb{Q}}}\cos{\vartheta_{\mathbb{Q}}}+
\frac{1}{\sqrt{\Lambda^{\mathbb{Q}}_2}}\cos{\theta_{\mathbb{Q}}}\sin{\vartheta_{\mathbb{Q}}}\right),\nonumber\\
\mathbbm{b}_{\mathbb{Q}}&=&\frac{1}{\sqrt{2}}\left(\frac{1}{\sqrt{\Lambda^{\mathbb{Q}}_1}}\cos{\theta_{\mathbb{Q}}}\sin{\vartheta_{\mathbb{Q}}}+
\frac{1}{\sqrt{\Lambda^{\mathbb{Q}}_2}}\sin{\theta_{\mathbb{Q}}}\cos{\vartheta_{\mathbb{Q}}}\right).\nonumber
\end{eqnarray}
While $\mathbbm{c}_{\mathbb{Q}}$ and $\mathbbm{d}_{\mathbb{Q}}$ are
given by
\begin{eqnarray}
\label{ana-yukawa-D-brane-5d-Y-2-elementQ-cd}
\mathbbm{c}_{\mathbb{Q}}&=&\frac{1}{2}\left[\frac{1}{\sqrt{\Lambda^{\mathbb{Q}}_3}}+i\left(\frac{1}{\sqrt{\Lambda^{\mathbb{Q}}_1}}\sin{\theta_{\mathbb{Q}}}\sin{\vartheta_{\mathbb{Q}}}+
\frac{1}{\sqrt{\Lambda^{\mathbb{Q}}_2}}\cos{\theta_{\mathbb{Q}}}\cos{\vartheta_{\mathbb{Q}}}\right)\right],\\
\mathbbm{d}_{\mathbb{Q}}&=&\frac{1}{2}\left[\frac{1}{\sqrt{\Lambda^{\mathbb{Q}}_3}}-i\left(\frac{1}{\sqrt{\Lambda^{\mathbb{Q}}_1}}\sin{\theta_{\mathbb{Q}}}\sin{\vartheta_{\mathbb{Q}}}+
\frac{1}{\sqrt{\Lambda^{\mathbb{Q}}_2}}\cos{\theta_{\mathbb{Q}}}\cos{\vartheta_{\mathbb{Q}}}\right)\right].\nonumber
\end{eqnarray}

The elements of $\tilde{H}^{\mathcal{D}}$ in
Eq.~(\ref{ana-yukawa-D-brane-5d-Y-2}) are given as follows.
$\mathbbm{s}_{\mathcal{D}}$, $\mathbbm{a}_{\mathcal{D}}$ and
$\mathbbm{b}_{\mathcal{D}}$ are given by
\begin{eqnarray}
\label{ana-yukawa-D-brane-5d-Y-2-elementD}
\mathbbm{s}_{\mathcal{D}}&=&\left(\frac{1}{\sqrt{\Lambda^{\mathcal{D}}_1}}\cos{\theta_{\mathcal{D}}}\cos{\vartheta_{\mathcal{D}}}-
\frac{1}{\sqrt{\Lambda^{\mathcal{D}}_2}}\sin{\theta_{\mathcal{D}}}\sin{\vartheta_{\mathcal{D}}}\right),\\
\mathbbm{a}_{\mathcal{D}}&=&\frac{1}{\sqrt{2}}\left(\frac{1}{\sqrt{\Lambda^{\mathcal{D}}_1}}\cos{\theta_{\mathcal{D}}}\sin{\vartheta_{\mathcal{D}}}+
\frac{1}{\sqrt{\Lambda^{\mathcal {D}}_2}}\sin{\theta_{\mathcal {D}}}\cos{\vartheta_{\mathcal{D}}}\right),\nonumber\\
\mathbbm{b}_{\mathcal{D}}&=&-\frac{1}{\sqrt{2}}\left(\frac{1}{\sqrt{\Lambda^{\mathcal{D}}_1}}\sin{\theta_{\mathcal{D}}}\cos{\vartheta_{\mathcal{D}}}+
\frac{1}{\sqrt{\Lambda^{\mathcal
{D}}_2}}\cos{\theta_{\mathcal{D}}}\sin{\vartheta_{\mathcal
{D}}}\right).\nonumber
\end{eqnarray}
While $\mathbbm{c}_{\mathcal{D}}$ and $\mathbbm{d}_{\mathcal{D}}$
are given by
\begin{eqnarray}
\label{ana-yukawa-D-brane-5d-Y-2-elementD-cd}
\mathbbm{c}_{\mathcal{D}}&=&\frac{1}{2}\left[\frac{1}{\sqrt{\Lambda^{\mathcal{D}}_3}}-\left(\frac{1}{\sqrt{\Lambda^{\mathcal{D}}_1}}\sin{\theta_{\mathcal
{D}}}\sin{\vartheta_{\mathcal{D}}}-
\frac{1}{\sqrt{\Lambda^{\mathcal{D}}_2}}\cos{\theta_{\mathcal{D}}}\cos{\vartheta_{\mathcal{D}}}\right)\right],\\
\mathbbm{d}_{\mathcal
{D}}&=&\frac{1}{2}\left[\frac{1}{\sqrt{\Lambda^{\mathcal{D}}_3}}+\left(\frac{1}{\sqrt{\Lambda^{\mathcal{D}}_1}}\sin{\theta_{\mathcal{D}}}\sin{\vartheta_{\mathcal{D}}}-
\frac{1}{\sqrt{\Lambda^{\mathcal{D}}_2}}\cos{\theta_{\mathcal{D}}}\cos{\vartheta_{\mathcal{D}}}\right)\right].\nonumber
\end{eqnarray}

In order to make single-value decomposition for
$\hat{\mathcal{Y}}^{D}$ and $\hat{Y}^{D}$, we define two matrices as
follows
\begin{eqnarray}
\label{ana-calY-tilde-svd-mat}
\hat{\mathbb{M}}=\hat{\mathcal{Y}}^{D}\hat{\mathcal{Y}}^{D\dagger}=\left(
\begin{array}{ccr}
 \mathcal{S} &\mathcal{A} &-\mathcal{A} \\
\mathcal{A}^{\ast} & \mathcal{B}& \mathcal{C} \\
 -\mathcal{A}^{\ast} &\mathcal{C} & \mathcal{B}
\end{array}\right),~\hat{\mathbb{M}}^{\prime}=\hat{Y}^{D}\hat{Y}^{D\dagger}=\left(
\begin{array}{ccr}
 \mathcal{S}^{\prime} &\mathcal{A}^{\prime} &-\mathcal{A}^{\prime} \\
\mathcal{A}^{\prime\ast} & \mathcal{B}^{\prime}& \mathcal{C}^{\prime} \\
 -\mathcal{A}^{\prime\ast} &\mathcal{C}^{\prime} & \mathcal{B}^{\prime}
\end{array}\right).
\end{eqnarray}
The matrices $\hat{\mathbb{M}}$ and $\hat{\mathbb{M}}^{\prime}$ are
defined by the above matrix multiplication. Their elements can be
expressed with the elements of $\hat{\mathcal{Y}}^{D}$ and
$\hat{Y}^{D}$ respectively. We omit their explicit expressions here.
By the definitions in Eq.~(\ref{ana-calY-tilde-svd-mat}), the
elements $\mathcal{S}$, $\mathcal{B}$, $\mathcal{C}$,
$\mathcal{S}^{\prime}$, $\mathcal{B}^{\prime}$ and
$\mathcal{C}^{\prime}$ are all real numbers. By the elements of
these two new matrices, the quantities in equations
(\ref{ana-calY-tilde-svd}) and (\ref{ana-Y-tilde-svd}) can be
defined as follows. For $\hat{\mathcal{Y}}^{D}$, we define that
\begin{eqnarray}
\label{ana-calY-tilde-svd-def}
\Sigma^{\mathcal{Y}}_{1}&=&\frac{1}{2}\left[(\mathcal{B}-\mathcal{C}+\mathcal{S})+\sqrt{(\mathcal{B}-\mathcal{C}-\mathcal{S})^2+8\mid\mathcal{A}\mid^2}\right],\\
\Sigma^{\mathcal{Y}}_{2}&=&\frac{1}{2}\left[(\mathcal{B}-\mathcal{C}+\mathcal{S})-\sqrt{(\mathcal{B}-\mathcal{C}-\mathcal{S})^2+8\mid\mathcal{A}\mid^2}\right],~
\Sigma^{\mathcal{Y}}_{3}=\mathcal{B}+\mathcal{C},\nonumber\\
\mathcal{A}&=&\mid\mathcal{A}\mid\exp{i\delta_\mathcal{Y}},~
\tan{\theta_{\mathcal{Y}}}=-\frac{(\mathcal{B}-\mathcal{C}-\mathcal{S})+\sqrt{(\mathcal{B}-\mathcal{C}-\mathcal{S})^2+8\mid\mathcal{A}\mid^2}}{2\sqrt{2}\mid\mathcal{A}\mid}.\nonumber
\end{eqnarray}
For $\hat{Y}^{D}$, we define that
\begin{eqnarray}
\label{ana-Y-tilde-svd-def}
\Sigma^{Y}_{1}&=&\frac{1}{2}\left[(\mathcal{B}^{\prime}-\mathcal{C}^{\prime}+\mathcal{S}^{\prime})+\sqrt{(\mathcal{B}^{\prime}-\mathcal{C}^{\prime}-\mathcal{S}^{\prime})^2+8\mid\mathcal{A}^{\prime}\mid^2}\right],\\
\Sigma^{Y}_{2}&=&\frac{1}{2}\left[(\mathcal{B}^{\prime}-\mathcal{C}^{\prime}+\mathcal{S}^{\prime})-\sqrt{(\mathcal{B}^{\prime}-\mathcal{C}^{\prime}-\mathcal{S}^{\prime})^2+8\mid\mathcal{A}^{\prime}\mid^2}\right],
~\Sigma^{Y}_{3}=\mathcal{B}^{\prime}+\mathcal{C}^{\prime},\nonumber\\
\mathcal{A}^{\prime}&=&\mid\mathcal{A}^{\prime}\mid\exp{i\delta_Y},~
\tan{\theta_{Y}}=-\frac{(\mathcal{B}^{\prime}-\mathcal{C}^{\prime}-\mathcal{S}^{\prime})+\sqrt{(\mathcal{B}^{\prime}-\mathcal{C}^{\prime}-\mathcal{S}^{\prime})^2+8\mid\mathcal{A}^{\prime}\mid^2}}{2\sqrt{2}\mid\mathcal{A}^{\prime}\mid}.\nonumber
\end{eqnarray}

%%%%%%%%%%%%%%%%%%%%%%%

\end{document}